\documentclass[twocolumn]{aastex6}

\usepackage{color}
\usepackage{amsmath}

\begin{document}

\title{Identifying inflated super-Earths and photo-evaporated cores}

\author{Daniel Carrera\altaffilmark{1,2}, Eric B. Ford\altaffilmark{1,2,3}, Andre Izidoro\altaffilmark{4}, Daniel Jontof-Hutter\altaffilmark{1,2}, Sean N. Raymond\altaffilmark{5}, Angie Wolfgang\altaffilmark{1,2}}

\noaffiliation

\altaffiltext{1}{Center for Exoplanets and Habitable Worlds, 525 Davey Laboratory, The Pennsylvania State University, University Park, PA, 16802, USA.}
\altaffiltext{2}{Department of Astronomy and Astrophysics, The Pennsylvania State University, 525 Davey Laboratory, University Park, PA 16802, USA}
\altaffiltext{3}{Institute for CyberScience, The Pennsylvania State University, University Park, PA, 16802, USA}
\altaffiltext{4}{UNESP, Univ. Estadual Paulista - Grupo de Din\^amica Orbital \& Planetologia, Guaratinguet\'a, CEP 12516-410 S\~ao Paulo, Brazil}
\altaffiltext{5}{Laboratoire d$'$astrophysique de Bordeaux, Univ. Bordeaux, CNRS, B18N, all\'ee Geoffroy Saint-Hilaire, 33615 Pessac, France}

%
%
\newcommand{\K}{{\rm K}}
\newcommand{\cs}{c_{\rm s}}
\newcommand{\kb}{k_{\rm B}}
\newcommand{\Fe}{F_\oplus}
\newcommand{\Rb}{R_{\rm B}}
\newcommand{\Rc}{R_{\rm c}}
\newcommand{\Rh}{R_{\rm H}}
\newcommand{\Td}{T_{\rm d}}
\newcommand{\Mc}{M_{\rm c}}
\newcommand{\Ms}{M_\star}
\newcommand{\Me}{M_\oplus}
\newcommand{\au}{{\rm au}}
\newcommand{\yr}{{\rm yr}}
\newcommand{\rcb}{{\rm rcb}}
\newcommand{\Myr}{{\rm Myr}}
\newcommand{\tpe}{t_{\rm pe}}
\newcommand{\Rtr}{R_{\rm tr}}
\newcommand{\Ptr}{P_{\rm tr}}
\newcommand{\Teq}{T_{\rm eq}}
\newcommand{\Teff}{T_{\rm eff}}
\newcommand{\rout}{r_{\rm out}}
\newcommand{\Matm}{M_{\rm atm}}
\newcommand{\rhod}{\rho_{\rm d}}
\newcommand{\tdisk}{t_{\rm disk}}
\newcommand{\Mdisk}{M_{\rm disk}}
\newcommand{\Omegak}{\Omega_{\rm k}}

\renewcommand{\Re}{R_\oplus}

\begin{abstract}
We present empirical evidence, supported by a planet formation model, to show that the curve $R/R_\oplus = 1.05\,(F/F_\oplus)^{0.11}$ approximates the location of the so-called photo-evaporation valley. Planets below that curve are likely to have experienced complete photo-evaporation, and planets just above it appear to have inflated radii; thus we identify a new population of inflated super-Earths and mini-Neptunes. Our N-body simulations are set within an evolving protoplanetary disk and include prescriptions for orbital migration, gas accretion, and atmospheric loss due to giant impacts. Our simulated systems broadly match the sizes and periods of super-Earths in the Kepler catalog. They also reproduce the relative sizes of adjacent planets in the same system, with the exception of planet pairs that straddle the photo-evaporation valley. This latter group is populated by planet pairs with either very large or very small size ratios ($R_{\rm out} / R_{\rm in} \gg 1$ or $R_{\rm out} / R_{\rm in} \ll 1$) and a dearth of size ratios near unity. It appears that this feature could be reproduced if the planet outside the photo-evaporation valley (typically the outer planet, but some times not) has its atmosphere significantly expanded by stellar irradiation. This new population of planets may be ideal targets for future transit spectroscopy observations with the upcoming James Webb Space Telescope.
\end{abstract}


%
%
\section{Introduction}
The Kepler mission has lead to the discovery of thousands of transiting exoplanets, and exoplanet candidates \citep{Batalha_2013}. This includes a large number of super-Earths and mini-Neptunes --- planets with radii between 1 and 4 $R_\oplus$ that are usually observed in multiple-planet systems with compact close-in orbits. While this type of planet is absent in the solar system, it may be the most common class of planet in the Galaxy \citep{Batalha_2013,Howard_2013,Petigura_2013,Mullally_2015}.

There is evidence that many super-Earths have been stripped of their atmospheres by photo-evaporation \citep{Lopez_2012,Lopez_2014}. Furthermore, \citet{Fulton_2017} have argued for the presence of a local minimum in the marginal density of radii of super-Earths and mini-Neptunes that is potentially the result of photo-evaporation. In this view, the planets on one side of the radius valley would be the photo-evaporated cores of previously gas-rich planets, while the planets on the other side of the valley would be those that have retained their gaseous envelopes.

In the traditional formation model, planet formation begins with the formation of planetesimals, which are small bodies with size $R \sim 1-100$ km \citep{Chiang_2010,Johansen_2014}. These planetesimals collide and grow to form more massive bodies. The first phase is that of runaway growth, in which the mass of the planetesimal grows has $\dot{M}/M \propto M^{1/3}$ \citep{Greenberg_1978,Wetherill_1989,Kokubo_1996}. At some point the velocity dispersion of the planetesimals becomes dominated by a small number of oligarchs. Growth then proceeds more slowly ($\dot{M}/M \propto M^{-1/3}$) until the oligarchs reach their isolation mass \citep{Kokubo_1998,Kokubo_2000,Thommes_2003,Chambers_2006}. More recent work suggests that giant planet cores may form through the rapid accretion of small centimetre-sized icy ``pebbles'' \citep{Lambrechts_2012,Johansen_2017}. This process may also play a role in the formation of super-Earths.

Numerous mechanisms have been proposed for the formation of hot super-Earths; most of these have serious challenges based on theoretical or observational grounds \citep[for a review, see][]{Raymond_2008,Raymond_2014b}. For example, various authors have shown that strict in-situ formation \citep{Hansen_2012,Hansen_2013,Chiang_2013} requires protoplanetary disks that are inconsistent with hydrostatic equilibrium \citep{Raymond_2014} and possibly gravitationally unstable \citep{Schlichting_2014}. Furthermore, \citet{Ogihara_2015} showed that the high surface densities required by in-situ formation lead to very rapid planet formation; since the planets form before the disk dispersal, they should experience rapid inward migration, so that the formation is no longer in-situ. This leaves two formation scenarios that merit further investigation and comparison to observational constraints:

\begin{itemize}
\item Drift model: One possibility is that small rocks or pebbles form in the outer disk, drift inward through aerodynamic drag, and pile up inside a pressure bump \citep{Boley_2013,Chatterjee_2014,Chatterjee_2015}. The pile-up of solids leads to the formation of planetesimals, and then planets.

\item Migration model: Another possibility is that hot super-Earths and Neptunes form by mergers of inward migrating planetary embryos. In this scenario, embryos form at large orbital periods. The embryos migrate inward, and as they do so they experience mergers, and get captured into mean motion resonances \citep{Terquem_2007,Ogihara_2009,McNeil_2010,Cossou_2014}.
\end{itemize}

These formation scenarios may not be mutually exclusive -- a planet could form from the pile up of pebbles in a pressure bump and subsequently migrate inward. In our investigation we will explore the migration model. We have developed a comprehensive planet formation model that simulates the formation of planetary systems starting from the formation of planetary embryos, through the process of gas accretion, N-body dynamics, planetary mergers, disk migration, dynamical instabilities after the dissipation of the protoplanetary disk, and ending with the thermal evolution of the planets and the photo-evaporation of their atmospheres.

A related work by \citet{Jin_2014} also combined planet formation with subsequent planet evolution including atmospheric escape. One critical difference between our work and theirs is that they only modelled one planetary embryo per disk. We believe that ours is the first model that combines full N-body dynamics in an evolving protoplanetary disk with the subsequent evolution of planetary atmospheres. We also take the novel step of testing our model against the radius ratios of adjacent planets, which are both more precisely and more accurately measured than the individual planet radii.

This paper is organized as follows. In section \ref{sec:obs} we use a novel approach to empirically constrain the location of the photo-evaporation valley. We also present supporting evidence in the form of constraints from transit timing variations, and identify five potentially inflated super-Earths. In section \ref{sec:model} we describe our planet formation model and initial conditions. In section \ref{sec:radius} we outline the calculation of the planet radius, and then present an example simulation in section \ref{sec:example}. We present our results in section \ref{sec:results}, where we show that our planet formation model successfully reproduces many of the features of the observed population of super-Earths, but planets just outside the photo-evaporation valley must have extended atmospheres. In section \ref{sec:discussion} we discuss our results. Finally, in section \ref{sec:conclusions} we summarize our results and draw conclusions.

%
%
\section{Observed radii and TTVs}
\label{sec:obs}

We begin by looking at the observational evidence for a transition radius between super-Earths with primordial atmospheres accreted directly from the disc, and photo-evaporated cores. Previous models have attempted to predict the shape of the photo-evaporation valley. \citet{Lopez_2016} used an atmosphere evolution model to estimate that the transition radius between rocky and non-rocky planets should scale as

\begin{equation}\label{eqn:R_LR}
	R_{\rm LR} \propto F^{0.11}
\end{equation}
where $R$ is the planet radius and $F$ is the incident stellar flux. \citet{Owen_2017} used an analytic derivation to argue that the transition radius scales as $R_{\rm trans} \propto P^{-0.25}$, where $P$ is the orbital period. In terms of incident flux, the transition radius would be

\begin{equation}\label{eqn:R_OW}
	R_{\rm OW} \propto F^{3/16} \approx F^{0.19}
\end{equation}

Finally, \citet{Van_Eylen_2018} used a small sample of stars for which stellar parameters could be derived from asteroseismology to estimate that $R_{\rm trans} \propto P^{-0.09}$. In terms of flux, this corresponds to

\begin{equation}
	R_{\rm V+} \propto F^{0.0675}
\end{equation}

Unfortunately, the sample of stars asteroseismic constraints is biased toward more massive stars relative to the Kepler planet search sample. For this reason, in this investigation we will focus on the model predictions of \citet{Lopez_2016} and \citet{Owen_2017}.

Our first goal is to estimate the proportionality constants for Equations \ref{eqn:R_LR} and \ref{eqn:R_OW}. To do this, we plot the ratios of planetary radii, $R_{\rm out} / R_{\rm in}$, between adjacent super-Earths in the Kepler field. Planet pairs that straddle the photo-evaporation valley should have larger $R_{\rm out} / R_{\rm in}$ than the rest, since one planet would have an atmosphere and the other would not. Our reliance on radius ratios is an important novel approach because it eliminates systematic uncertainties resulting from the assumed properties of the host star.

\subsection{Target star and planet selection}

We started with the cumulative catalogue of Kepler exoplanets and planet candidates (updated for DR25), which we downloaded from the NASA Exoplanet Archive \footnote{https://exoplanetarchive.ipac.caltech.edu/
\newline \hspace*{2em}(retrieved on Nov 2, 2017)}. We will refer to both confirmed planets and planet candidates as simply ``planets''. As a first pass, we applied the following selection criteria:

\begin{itemize}
\item We removed the likely false positives (disposition score $\le$ 0.5).

\item We selected FGK stars ($3900 \K < \Teff < 7600 \K$). This gives stars that are comparable to the Sun-like star that we used in our simulations.

\item We removed planets with period $P > 200$d to reduce concerns about false alarms in the data.

\item We selected super-Earth-size planets ($1\Re < R < 4\Re$), since those are the focus of our simulations.
\end{itemize}

\subsection{Photo-evaporation limit}
\label{sec:obs:photo}

\begin{figure*}[th!]
  \centering
  \includegraphics[width=0.45\textwidth]{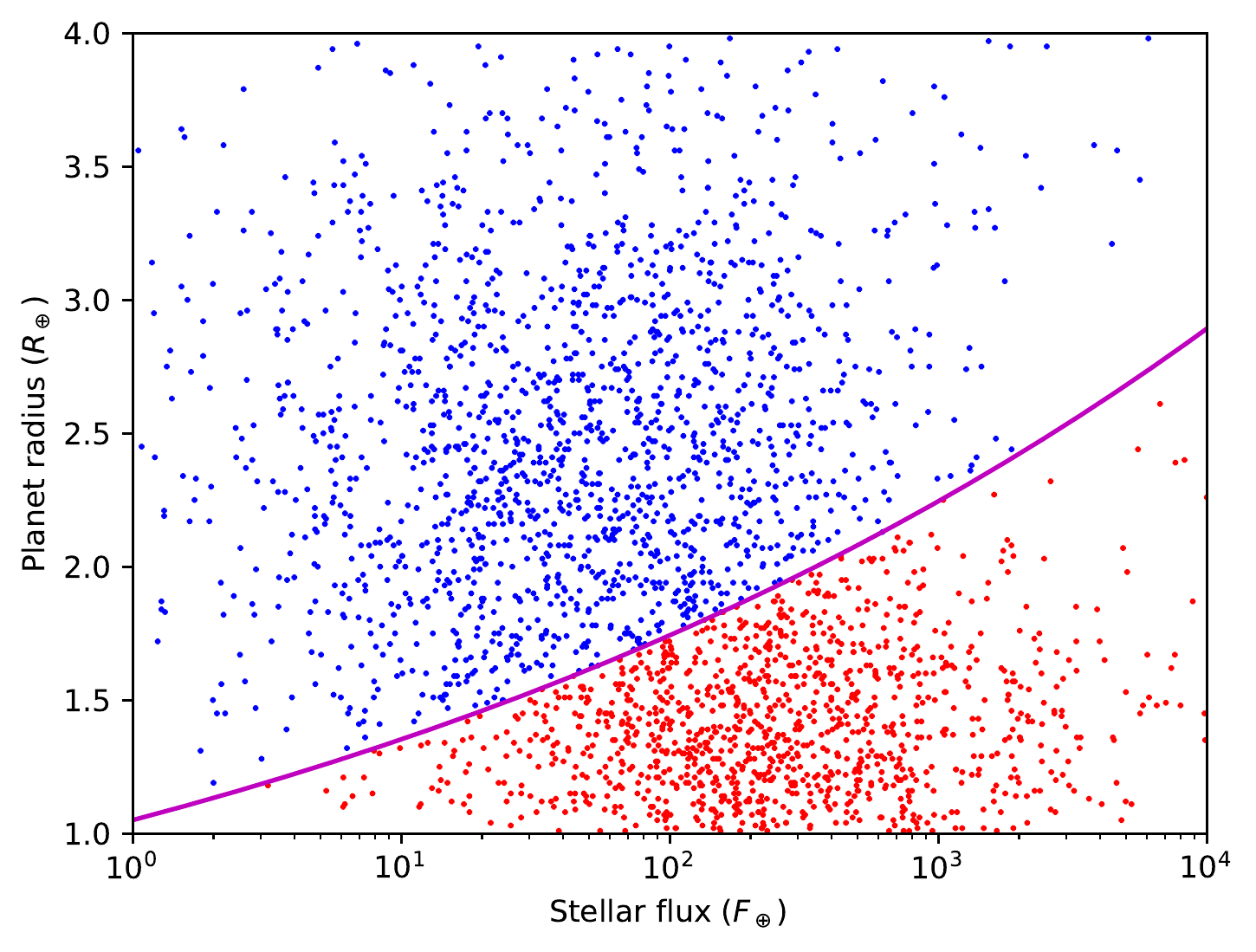}
  \includegraphics[width=0.45\textwidth]{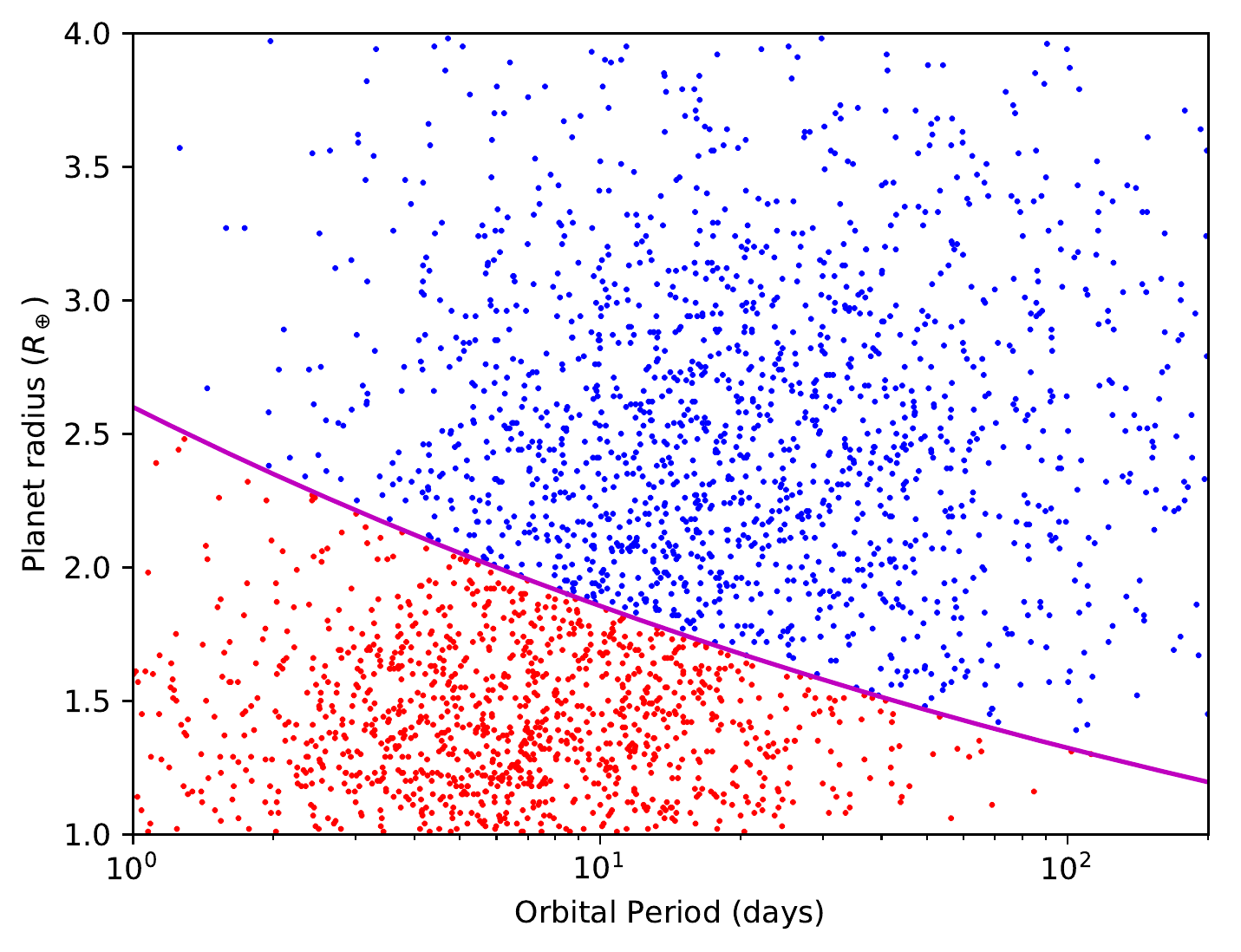} \\
  \includegraphics[width=0.45\textwidth]{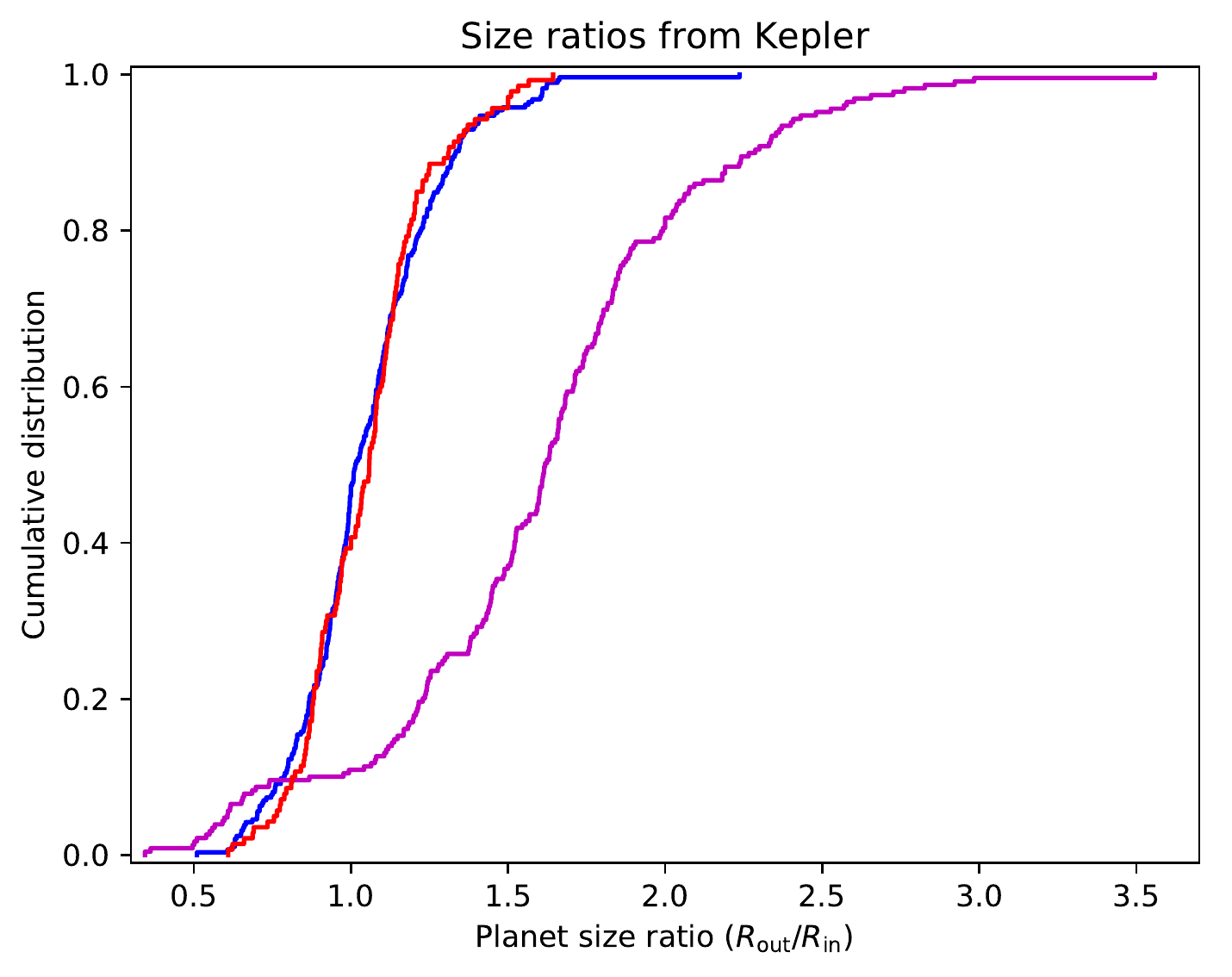}
  \includegraphics[width=0.45\textwidth]{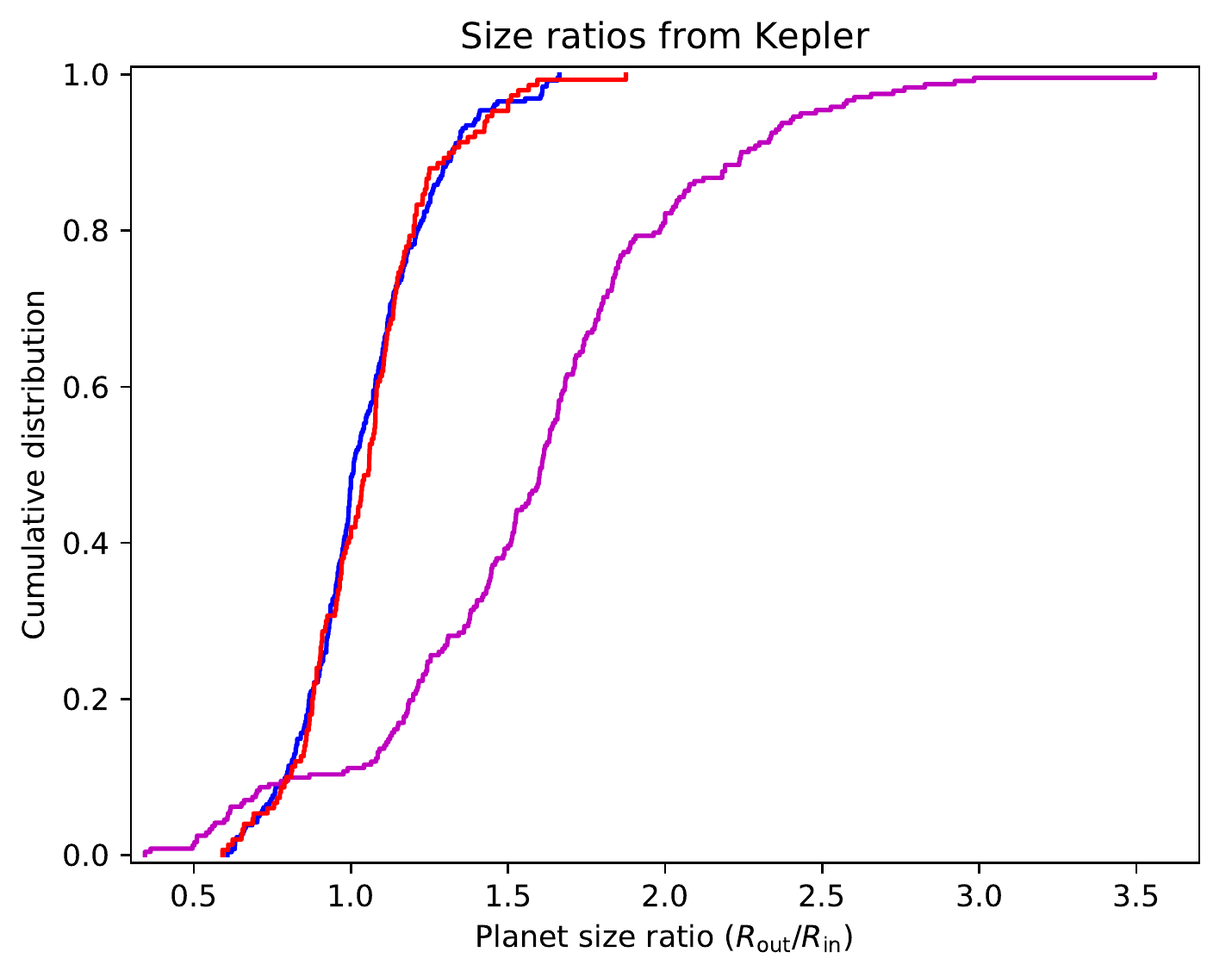} \\
  \caption{
  \textit{Top:} Scatter plot of Kepler exoplanets, separated by the line $R_{\rm trans}/\Re = 1.05\, (F/\Fe)^{0.11}$ (left) or $R_{\rm trans}/\Re = 2.6\, (P/{\rm yr})^{-0.1467}$ (right). \textbf{Bottom:} Cumulative distribution of size ratios ($R_{\rm out}/R_{\rm in}$). Planet pairs that are entirely above (blue) or entirely below (red) $R_{\rm trans}$ have similar radii. But when the planets straddle the line, the outer planet is typically significantly larger than the inner planet.}
  \label{fig:kepler}
\end{figure*}

Our data selection produces a catalogue of 2,770 Kepler planets. Figure \ref{fig:kepler} (top) shows the radius, incident flux, and periods of these planets, as reported in the archive. On the left plots we have divided the planets into two groups, split across the curve

\begin{equation}\label{eqn:R_vs_F}
	\frac{R_{\rm trans}}{\Re} = 1.05 \left( \frac{F}{\Fe} \right)^{0.11},
\end{equation}

and on the right side we use

\begin{equation}\label{eqn:R_vs_P}
	\frac{R_{\rm trans}}{\Re} = 2.6 \left( \frac{P}{P_\oplus} \right)^{-0.1467}.
\end{equation}

The bottom plots of Figure \ref{fig:kepler} show the cumulative distributions of $R_{\rm out}/R_{\rm in}$ for planet pairs that are both above $R_{\rm trans}$ (blue), both below $R_{\rm trans}$ (red), and include one planet on either side of the curve (magenta). The two cuts are approximately equivalent. In this investigation we use Equation \ref{eqn:R_vs_F}, but we have verified that an analysis with Equation \ref{eqn:R_vs_P} produces the same results.

The cumulative plots show that planet pairs that straddle $R_{\rm trans}$ typically have significantly larger $R_{\rm out} / R_{\rm in}$, sometimes have significantly smaller $R_{\rm out} / R_{\rm in}$ and rarely have $R_{\rm out} / R_{\rm in} \sim 1$. To make this general observation more concrete, we compute the olmogorov-Smirnov (KS) distance between the blue and magenta lines in Figure \ref{fig:kepler} ($D_{\rm bm}$), between the red and magenta lines ($D_{\rm rm}$), and between the blue and red lines ($D_{\rm br}$). Then we aggregate these values into a single score,

\begin{equation}\label{eqn:score}
	{\rm score} = D_{\rm bm} + D_{\rm rm} - D_{\rm br}.
\end{equation}

This score quantifies the extent to which the blue and red curves are close to each other, while simultaneously being distant to the magenta curve. Then, we write the general expression for the transition radius

\begin{equation}\label{eqn:R_trans_Cb}
	\frac{R_{\rm trans}}{\Re} = C \left( \frac{F}{\Fe} \right)^{b}.
\end{equation}

\begin{figure}[th!]
  \centering
  \includegraphics[width=0.45\textwidth]{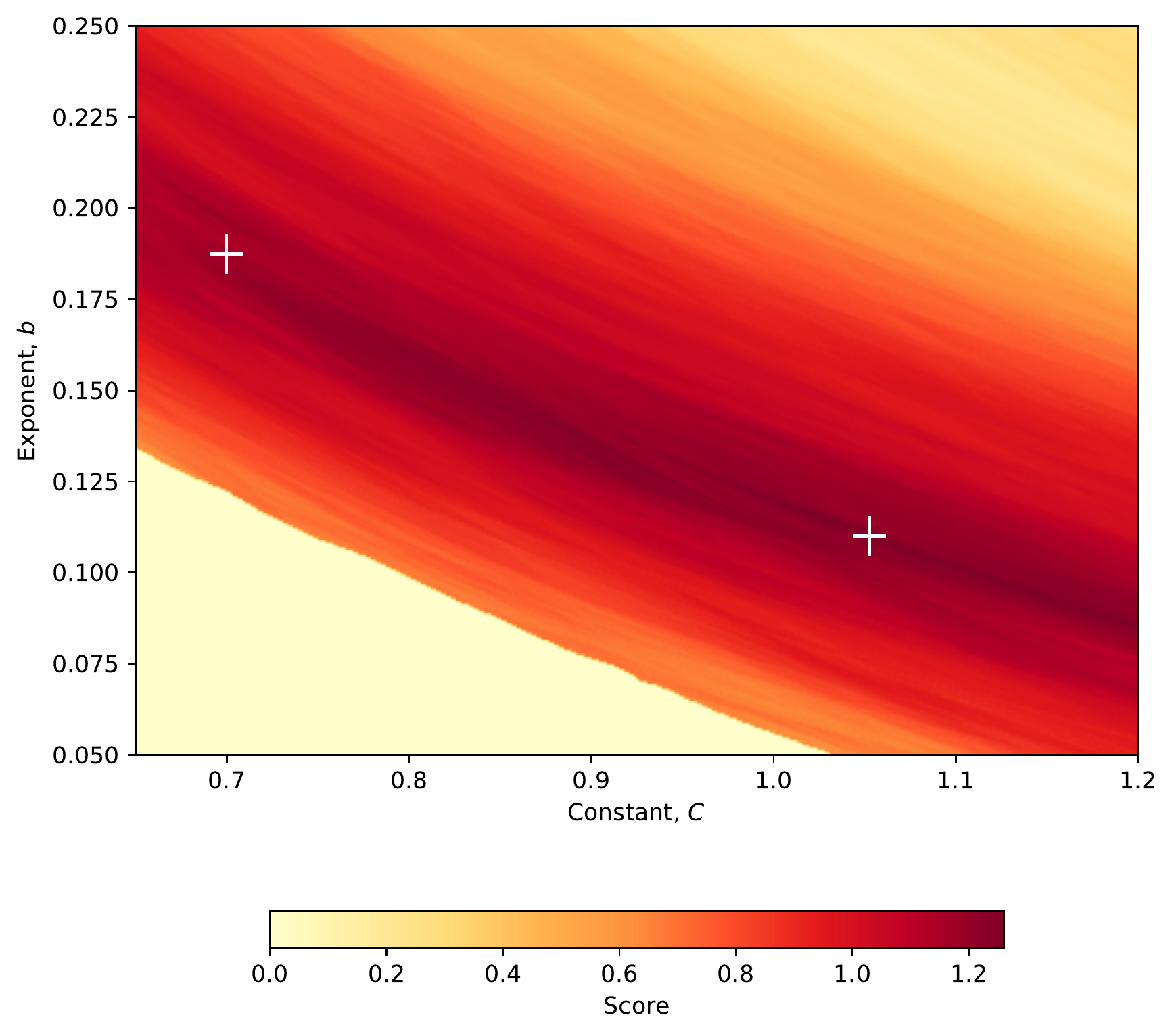}\\
  \caption{Score values (Equation \ref{eqn:score}) for each line $R/\Re = C (F/\Fe)^b$. Lines with high scores mark the location of a large transition in planet radius, which may indicate the location of the photo-evaporation valley. The two white crosses mark the best fits (Equations \ref{eqn:R_trans_Lopez} and \ref{eqn:R_trans_Owen}) for the photo-evaporation power laws suggested by \citet{Owen_2017} (top) and  \citet{Lopez_2016} (bottom)}.
  \label{fig:photoevap-map}
\end{figure}

Intuitively, $C$ is the transition radius for a planet receiving the same incident flux as the Earth. Figure \ref{fig:photoevap-map} shows the score values for a range of $(C,b)$ values. Previous studies have proposed power law indices for a photo-evaporation threshold. We obtain a slightly better fit for the $R \sim F^{0.1}$ relation proposed by \citet{Lopez_2016} than the $R \sim F^{3/16}$ from \citet{Owen_2017} (${\rm score} = 1.26$ vs 1.19). The best-fit formulas are, respectively,

\begin{eqnarray}
	\label{eqn:R_trans_Lopez}
	\frac{R_{\rm trans}}{\Re} &=& 1.05 \left( \frac{F}{\Fe} \right)^{0.11} \\
	\label{eqn:R_trans_Owen}
	\frac{R_{\rm trans}}{\Re} &=& 0.70 \left( \frac{F}{\Fe} \right)^{3/16}.
\end{eqnarray}

These fits are shown in Figure \ref{fig:photoevap-map} as white crosses. Since the $R \sim F^{0.11}$ scaling of \citet{Lopez_2016} and the region near it in Figure \ref{fig:photoevap-map} have higher scores, we adopt Equation \ref{eqn:R_trans_Lopez} for the rest of this investigation. We have verified that the choice between Equation \ref{eqn:R_trans_Lopez} and \ref{eqn:R_trans_Owen} has very little impact in our subsequent results.

\begin{table*}[ht!]
  \caption{We found six TTV planet pairs where at least 95\% of the TTV fits give a density ratio $\rho_{\rm out}/\rho_{\rm in} < 0.12$. For reference, Kepler-36 b and c have a density ratio of $\rho_{\rm c}/\rho_{\rm b} = 0.119$. $P_{\rm in}$ and $P_{\rm out}$ are the periods of the two planets, and $q$ is the fraction of TTV fits with $\rho_{\rm out}/\rho_{\rm in} < 0.12$. The reported error bars contain 68\% of the TTV fits. Kepler names are also shown when available. Except for KOI-593, all these pairs straddle the photo-evaporation valley (Equation \ref{eqn:R_trans_Lopez}). Planet properties are shown in Table \ref{tab:ttv-all}. All stars except KOI-593 had updated parameters from the California-Kepler Survey (CKS).}
  \label{tab:ttv-pairs}
  \centering
  \begin{tabular}{rlcrrccc}
KOI\hspace{2.5em} & \multicolumn{2}{c}{Kepler Name} & $P_{\rm in}$ (day) & 
$P_{\rm out}$ (day) & $\rho_{\rm out}/\rho_{\rm in}$ & $q$ & CKS data? \\
\hline
0115.03 and .01 & Kepler-105 & - \& b &  3.4364 & 5.4118 & $0.018^{+0.0277}_{-0.0110}$ & 0.9657 & Yes \\
0377.03 and .01 & Kepler-9   & d \& b &  1.5930 &19.2478 & $0.003^{+0.0029}_{-0.0009}$ & 0.9963 & Yes \\
0520.02 and .01 & Kepler-176 & b \& c &  5.4331 &12.7624 & $0.004^{+0.0256}_{-0.0035}$ & 0.9556 & Yes \\
0593.01 and .03 & Kepler-616 & b \& - &  9.9976 &51.0869 & $0.014^{+0.0278}_{-0.0087}$ & 0.9586 & No  \\
1831.03 and .01 & Kepler-324 & - \& c & 34.1936 &51.8238 & $0.013^{+0.0237}_{-0.0101}$ & 0.9916 & Yes \\
1955.03 and .01 & Kepler-342 & e \& b &  1.6442 &15.1693 & $0.005^{+0.0168}_{-0.0036}$ & 0.9696 & Yes \\
  \end{tabular}
\end{table*}


\subsection{Transit timing variations}
\label{sec:obs:ttv}

We were able to obtain transit timing variations (TTVs) for one hundred planet pairs in the Kepler sample. For each pair we obtained 10,000 samples from the posterior distribution of planet masses and orbits given the measured TTVs and computed the density ratio of the two planets for each. We identified six pairs for which more than 95\% of the samples imply an extreme mean planet density ratio, which we define as $\rho_{\rm out}/\rho_{\rm in} < 0.12$. For reference, the density ratio of Kepler-36b and c, which is possibly the best known example of an extreme density ratio, is $\rho_{\rm c}/\rho_{\rm b} = 0.119$. Out of the six planet pairs with extreme density ratios, five pairs straddle the photo-evaporation line. The one pair that does not straddle the photo-evaporation line, KOI-593, also has the largest uncertainties in the planet radii. Therefore, our TTV analysis strongly supports the finding, in section \ref{sec:obs:photo}, that at least some planets just outside the photo-evaporation valley have extreme, possibly inflated planet radii.

To do this analysis we used the transiting times of \citet{Rowe_2015}. We performed N-body simulations fitting presumed transit times to the data. We assumed the orbits have negligible mutual inclinations and that no planets other than those with measured transit times are required to explain the TTVs. Our free parameters were the mass ratio of each planet to the host star, and their orbital periods, phases and eccentricity vector components at the epoch BJD=2,455,680. To identify the region parameter space worth of detailed investigation, we performed many Levenberg-Marquardt minimizations with the eccentricity vector components initialized at each point on a grid. Then, we computed a posterior sample using the  Differential Evolution Markov Chain Monte Carlo (DEMCMC) algorithm \citep{TerBraak_2006,Nelson_2014}. The results of the Levenberg-Marquardt minimizations were used to construct the initial population of parameters for the DEMCMC sampler \citep{Jontof-Hutter_2015,Jontof-Hutter_2016}. For each planet pair we obtained 10,000 posterior samples, and thus, 10,000 mass ratios.

We computed the planet radii and stellar flux directly, using the stellar data from the California-Kepler Survey (CKS) whenever possible \citep{Petigura_2017}. If CKS data is not available, we use the stellar parameters published in the Kepler DR25 catalog. This allows us to incorporate the uncertainties in all the stellar parameters throughout our calculation.  We model all of the stellar properties ($R_\star$, $M_\star$, and $T_{\rm eff}$) as Gaussian distributions with the reported values as the means of those distributions. Note that these symmetric distributions become skewed when converted into derived quantities like planet radius and stellar flux. For this reason, our reported error bars are usually not symmetric. We compute the planet radius directly from the transit depth

\begin{equation}
    \frac{ R_{\rm p} }{R_\oplus}
	 = 109.2921 \; \frac{R_\star}{R_\odot} \; \sqrt{ \frac{\Delta F}{F_\star}},
\end{equation}
where $R_{\rm p}$ is the planet radius, $R_\star$ is the star radius, and $\Delta F/F_\star$ is the transit depth. We do not account for limb darkening.

\begin{table}[ht!]
  \caption{We found six TTV planet pairs where at least 95\% of the TTV fits give a density ratio $\rho_{\rm out}/\rho_{\rm in} < 0.12$ (see Table \ref{tab:ttv-pairs}). The reported error bars contain 68\% of the distribution. Except for KOI-593, all these pairs straddle the photo-evaporation valley ($R_t = R_{\rm trans}$ in Equation \ref{eqn:R_trans_Lopez}).}
	\label{tab:ttv-all}
	\centering
  \begin{tabular}{crrrc}
KOI    & $P$ (day)    & $R_p/R_\oplus$  &  $F_p/F_\oplus$  &  $R_p < R_t$ \\
\hline
0115.03 & 3.436  &  $0.556^{+0.0673}_{-0.0639}$  & $623.58^{+121.92}_{-113.04}$ & Yes \\
0115.01 & 5.412  &  $2.848^{+0.2535}_{-0.259 }$  & $340.35^{+66.492}_{-61.717}$ & No  \\
        &        &                               &                              &     \\
0377.03 & 1.593  &  $1.752^{+0.1393}_{-0.1383}$  & $1438.4^{+250.13}_{-228.60}$ & Yes \\
0377.01 & 19.248 &  $9.066^{+0.721 }_{-0.7113}$  & $51.877^{+9.0206}_{-8.2443}$ & No  \\
        &        &                               &                              &     \\
0520.02 & 5.433  &  $1.621^{+0.0806}_{-0.0782}$  & $125.37^{+13.962}_{-12.946}$ & Yes \\
0520.01 & 12.76  &  $2.659^{+0.1261}_{-0.1238}$  & $40.145^{+4.476 }_{-4.1355}$ & No  \\
        &        &                               &                              &     \\
0593.01 & 9.998  &  $2.694^{+0.5587}_{-0.5403}$  & $128.01^{+38.976}_{-31.102}$ & No  \\
0593.03 & 51.084 &  $2.611^{+0.5427}_{-0.5358}$  & $14.547^{+4.4249}_{-3.5327}$ & No  \\
        &        &                               &                              &     \\
1831.03 & 34.195 &  $1.304^{+0.0804}_{-0.0797}$  & $13.544^{+1.7329}_{-1.5856}$ & Yes \\
1831.01 & 51.823 &  $3.083^{+0.1775}_{-0.1733}$  & $7.7790^{+0.9948}_{-0.9093}$ & No  \\
        &        &                               &                              &     \\
1955.03 & 1.644  &  $1.027^{+0.135 }_{-0.136 }$  & $3004.3^{+855.29}_{-753.61}$ & Yes \\
1955.01 & 15.17  &  $2.228^{+0.2907}_{-0.2955}$  & $155.27^{+44.181}_{-38.956}$ & No  \\
  \end{tabular}
\end{table}

Tables \ref{tab:ttv-pairs} and \ref{tab:ttv-all} show the properties of the six planet pairs where at least 95\% of TTV fits gave a density ratio $\rho_{\rm out}/\rho_{\rm in} < 0.12$. We use $\rho_{\rm out}/\rho_{\rm in} = 0.12$ as an empirical cut-off because Kepler-36 b and c, which is possibly the best known planet system with an extreme density ratio, has a density ratio of $\rho_{\rm out}/\rho_{\rm in} = 0.119$. Five of the six pairs straddle the photo-evaporation valley (Equation \ref{eqn:R_trans_Lopez}). The exception is KOI-593, which is one of the least well characterized systems. The errors in planet radii are large, and this is the only system in the table without updated stellar parameters from the CKS. That said, straddling the line in Equation \ref{eqn:R_trans_Lopez} is excluded by the estimated error bars.

%
%
\section{Planet Formation Model}
\label{sec:model}

In addition to the observational argument of the previous section, theoretical modelling provides another line of evidence that super-Earths just outside the photo-evaporation valley have inflated (or ``puffy'') atmospheres. Here we describe our planet formation model. In section \ref{sec:radius} we show how we compute the radii of simulated planets, and in sections \ref{sec:example}-\ref{sec:discussion} we present and discuss our results.

\subsection{Overview}

\label{sec:model:nbody}
We use N-body simulations to model the dynamical evolution of planetary embryos embedded in an evolving protoplanetary disk. The planetary embryos experience mutual gravitational interactions, disk torques, and collisions. We keep track of the water mass fraction of each planet, along with gas accretion, and gas loss through giant impacts. Our simulations begin shortly before the embryos reach their isolation masses. After the disk dissipates, we continue to model the dynamical evolution of the planetary system up to an age of 25 Myr.

We use the hybrid N-body integrator in \textsc{mercury} \citep{Chambers_1999} along with the modifications of \citet{Izidoro_2017}. Most importantly, \citet{Izidoro_2017} added a user-defined force that computes disk torques \citep{Paardekooper_2010,Paardekooper_2011}, applied to the disk model of \citet{Bitsch_2015}, as well as eccentricity and inclination damping following \citet{Cresswell_2006,Cresswell_2008}. We have implemented gas accretion and gas loss after giant impacts as a post-processing step that occurs after the N-body simulation is complete. We have modified \textsc{mercury} to report the location and speed of each impact. We use that information to calculate the amount of atmosphere loss after each collision (section \ref{sec:model:impacts}).

\subsection{Disk Structure}
\label{sec:model:disk}

All our simulations take place around a Sun-like star, in a protoplanetary disk with the structure described by \citet{Bitsch_2015}. They conducted 3D hydrodynamic simulations of protoplanetary disks and fit 1D formulas to their simulation results. The key model parameter is the disk age, from which they approximate the disk accretion rate

\begin{equation}
	\log_{\rm 10}\left( \frac{\dot{M}_{\rm disk}}{M_\odot / \yr} \right)
    = - 8 - 1.4\,\log_{\rm 10}\left( \frac{\tdisk + 10^5 \yr}{10^6 \yr} \right).
\end{equation}

This equation is based on the correlation found by \citet{Hartmann_1998}, and modified by \citeauthor{Bitsch_2015} so that $\tdisk = 0$ corresponds to a stellar age of $t_\star = 10^5 \yr$. A steady-state accretion disk has a constant mass flux $\dot{M}_{\rm disk} = 2\pi r \Sigma v_r$ at each point $r$, where $\Sigma$ is the disk surface density and $v_r$ is the radial velocity. Following the $\alpha$-viscosity model of \citet{Shakura_1973} we write

\begin{equation}\label{eqn:Mdot_disk}
	\dot{M}_{\rm disk} = 3\pi \, \nu \, \Sigma = 3 \pi \alpha H^2 \Omegak \Sigma
\end{equation}
where $\nu = \alpha H \cs$ is the disk viscosity, $H$ is the disk scale height, $\cs = H \Omegak$ is the sound speed, and $\Omegak$ is the Keplerian frequency. The disk model of \citet{Bitsch_2015} has $\alpha = 0.0054$. The disk midplane density is given by $\rhod = \Sigma / H \sqrt{2}$. The scale height (and therefore also the sound speed) is determined by the local temperature and the condition of hydrostatic equilibrium,

\begin{equation}\label{eqn:T_disk}
	T = \left( \frac{H}{r} \right)^2 \frac{G M_\star}{r} \frac{\mu}{\mathcal{R}},
\end{equation}
where $G$ is the gravitational constant, $M_\star$ is the stellar mass, $\mu = 2.3$ is the molecular mass, and $\mathcal{R}$ is the gas constant. Therefore, given a disk temperature profile $T(r)$ and an accretion rate $\dot{M}_{\rm disk}$, Equations \ref{eqn:Mdot_disk} and \ref{eqn:T_disk} uniquely determine the disk structure --- $\Sigma(r)$, $\rhod(r)$, $H(r)$. The disk temperature profile is provided by the model of \citet{Bitsch_2015}, as a function of disk metallicity and $\dot{M}_{\rm disk}$.

The simulations of \citet{Bitsch_2015} did not consider the effect of disk photo-evaporation. Different models of photo-evaporation seem to agree that once it starts, the inner $\sim$2 AU of the disk are cleared rapidly, on a time scale of around $10^5 \yr$, as the inner disk drains on a viscous timescale \citep[e.g.][]{Gorti_2009}. Since we are mainly interested in this inner region, we add a parameter to the disk model: $\tpe$ is the time when photo-evaporation begins to carve cavity in the inner disk. Between $\tpe$ and $\tpe + 10^5 \yr$ the disk temperature profile is held constant, but $\Sigma$ is reduced exponentially with an e-folding timescale of $10^4 \yr$. At that time, the disk effects are removed entirely from the simulation.

\subsection{Disk Torques}
\label{sec:model:torque}

Super-Earths and mini-Neptunes migrate through the disk through Type-I migration. The total torque exerted by the disk on the planet has two main components,

\begin{equation}
	\Gamma_{\rm tot} = \Gamma_{\rm L} \Delta_{\rm L}
    				 + \Gamma_{\rm C} \Delta_{\rm C}
\end{equation}
where $\Gamma_{\rm L}$ is the Lindblad torque and $\Gamma_{\rm C}$ is the co-rotation torque; $\Delta_{\rm L}$ and $\Delta_{\rm C}$ are factors of order unity that are equal to one for circular orbits on the plane of the disk. The Lindblad torque is usually negative and the co-rotation torque is positive. The sum of the two torques can lead to either inward migration (negative torque) or outward migration (positive torque). The expressions for $\Gamma_{\rm L}$ and $\Delta_{\rm L}$ are derived by \citet{Paardekooper_2010,Paardekooper_2011}, while $\Delta_{\rm L}$ and $\Delta_{\rm C}$ were calculated by \citet{Cresswell_2008,Coleman_2014,Fendyke_2014}. The full set of equations was gathered together by \citet{Izidoro_2017} and are reproduced again here in Appendix \ref{appendix:torques}.

\subsection{Gas accretion}
\label{sec:model:accretion}

The gas accretion model is described in Appendix \ref{appendix:gas-accretion}. In summary, the planet's initial atmosphere is the volume of gas inside its Bondi radius,

\begin{equation}
  \Rb = \frac{G \Mc \mu}{\kb \Td}.
\end{equation}

As the atmosphere cools, it contracts, allowing more gas to enter the Bondi radius. Therefore, the planet's accretion rate is set by its cooling rate, and the amount of energy present in the atmosphere,

\begin{eqnarray}
	L &\sim& \sigma \Td^4 \, \Rb, \\
	E &\sim& \frac{G\,\Mc\,\Matm}{\Rc}
\end{eqnarray}
where $\Mc$ is the core mass, $\Matm$ is the atmosphere mass, $\Rc$ is the core radius, $\sigma$ is the Stephan-Boltzmann constant, and $\Td$ is the local disk temperature. Setting $L = - \dot{E}$ gives the accretion rate. The detailed derivation is in Appendix \ref{appendix:gas-accretion}.

\subsection{Giant impacts}
\label{sec:model:impacts}

Giant impacts are a common occurrence in the planet formation process. In this model the rock and water component of each planet is retained after each merger, but planets experience atmospheric mass loss. Let $m_{\rm imp}$ be the mass of the impactor, and $v_{\rm imp}$ be the impact speed. \citet{Schlichting_2015} estimate that, for an adiabatic atmosphere, the global atmospheric mass loss resulting from a giant impact is

\begin{equation}
	X_{\rm loss} =
	\left\{
		\begin{array}{ll}
			1  & \mbox{if } x > 1 \\
			0.4x + 1.8x^2 - 1.2x^3 & \mbox{if } x \leq 1
		\end{array}
	\right.
\end{equation}
where $x = (v_{\rm imp} m_{\rm imp}) / (v_{\rm esc} \Mc)$, and $v_{\rm esc}$ is the planet's escape speed. If the planet's atmospheric mass drops below $\sim \rhod \; \Rb^3$ (Equation \ref{eqn:Matm}), the planet will quickly re-accrete a new initial atmosphere. Therefore, after each collision we set the the atmosphere mass of the new planet to

\begin{equation}
	\Matm =
	\left\{
		\begin{array}{ll}
			\Matm^*(1 - X_{\rm loss})  & \mbox{if } t > \tdisk \\
			{\rm max}\left(\Matm^*(1 - X_{\rm loss}),
            	\rhod \; \Rb^3\right) & \mbox{if } t \leq \tdisk
		\end{array}
	\right.
\end{equation}
where $\Matm^*$ is the atmospheric mass of the more massive planet. In practice, if the two planets have equal mass, then $X_{\rm loss} \approx 1$.

\subsection{Initial conditions}
\label{sec:model:init}

We assume that planetesimals form very early and do not experience significant migration due to either aerodynamic drag or disk torques. Therefore, the surface density of embryos reflects the surface density of solids at the beginning of the disk phase:

\begin{equation}\label{eqn:Sigma_solid}
	\Sigma_{\rm solid} = Z \; \Sigma_{\rm gas}(t = 0).
\end{equation}
We follow the recommendation of \citet{Bitsch_2015} and define $t = 0$ as the moment when the disk becomes gravitationally stable; this corresponds to a stellar age of 100 kyr. The timescale for embryo formation is $\sim 1\,\Myr$ at 1 AU and increases with distance \citep{Kokubo_2000}. To approximate the first 1 Myr of evolution, we insert 125-250 planetary embryos, each with a mass of $0.4 M_\oplus$, spaced so as to follow the surface density $\Sigma_{\rm solid}$ (Equation \ref{eqn:Sigma_solid}). Our N-body simulations begin at 1 Myr when planetary embryos have become massive enough that disk migration starts to become an important process. Our runs capture the final stages of the formation of isolation masses. Starting at semimajor axis $a_0$, the disk is divided into 125-250 radial bins such that the total solid mass inside each bin is $0.4 M_\oplus$. Each embryo is placed in the middle of its respective radial bin.

At $t = 0$ the snow line is located at around 5 AU. Therefore, we assume that planetary embryos that form inside 5 AU are dry, and those beyond 5 AU are icy. The formation of water ice significantly increases the surface density beyond the snow line, which leads to the formation of more massive embryos \citep{Morbidelli_2015}. In the solar system, the most ice-rich bodies near the orbit of Jupiter, like the Galilean moon Callisto, are approximately equal parts rock and ice \citep{Kuskov_2005}. Therefore, we double the solid surface density beyond 5 AU and model the embryos that form there as 50\% rock and 50\% ice. Over the course of the simulation we track the ice fraction in the forming planets.

The baseline model has $Z = 1\%$ and begins with 125 embryos for a total mass of $50 M_\oplus$. The embryos span from $a_0 = 1$ AU to 5.96 AU, and the separation between embryos ranges from $7.3 R_{\rm Hill}$ for the two innermost embryos, to $0.4 R_{\rm Hill}$ for the two outermost embryos. The outer embryos are highly collisional and quickly merge to form larger bodies.

The embryos start out with low but non-zero inclinations and eccentricities. All embryos begin with eccentricity $e = 0.002$ and inclination $I = 0.10^\circ$. Each embryo is given random mean anomaly, argument of pericentre, and longitude of ascending node, all chosen uniformly between $0^\circ$ and $360^\circ$. Therefore, these last three orbital angles are the only parameters that varies between different instances of each model. For each model we perform 200 simulations unless otherwise indicated.

The disk lifetime is 5 Myr. From 1 to 5 Myr, planets experience migration, eccentricity and inclination damping, gas accretion, N-body gravitational interactions, and atmosphere loss from giant impacts. At 5 Myr, we hold the temperature profile constant and allow the surface density to drop exponentially over the course of 0.1 Myr. The simulation then proceeds as a pure N-body simulation until it reaches 25 Myr. In addition to the baseline model, we have investigated several alternative models. Our full set of models are shown in Table \ref{tab:models}.

\begin{table}[h]
  \caption{We examined six planet formation models. In this table, $a$ is the range of semimajor axes of the initial planetary embryos, $Z$ is the disk metallicity, and $M_{\rm tot}$ is the total mass in embryos.}
  \label{tab:models}
  \begin{tabular}{l|c|c|r}
  Model & $a_0/{\rm AU}$  & $Z$ & $M_{\rm tot}/M_\oplus$ \\
  \hline
  Baseline   & 1.0 - 6.0          & 1.0\%          &  50 \\
  Ice-rich   & \textbf{5.0 - 7.6} & 1.0\%          &  50 \\
  Metal-rich & 1.0 - 4.9          & \textbf{2.0\%} &  50 \\
  Metal-poor & 1.0 - 7.6          & \textbf{0.5\%} &  50 \\
  High-mass  & 1.0 - 8.5          & 1.0\%          & \textbf{100} \\
  Low-mass   & 1.0 - 4.3          & 1.0\%          & \textbf{ 25} \\
  \end{tabular}
\end{table}

%
%
\section{Computing the planet radius}
\label{sec:radius}

Each N-body simulation produces a planetary system (see Figure \ref{fig:example}). In sections \ref{sec:model:accretion} and \ref{sec:model:impacts} we discuss how we compute the water mass fraction and the mass of the atmosphere. To estimate each planet radius we separately compute the size of the rock-water core, and the height of the atmosphere. \citet{Zeng_2016} have published a table of planet core radii for a range of planet masses and water mass fractions\footnote{https://www.cfa.harvard.edu/~lzeng/planetmodels.html\#mrtables}. We interpolate this table to compute the core radii of our simulated planets. Finally, to compute the height of the gaseous envelope we use Equation 4 of \citet{Lopez_2014}

\begin{equation}\label{eqn:adiabatic}
	\frac{R_{\rm env}}{R_\oplus}  \sim 2.06 \, f^{0.59} \,
    		\left( \frac{\Mc}{\Me} \right)^{-0.21} \,
            \left( \frac{F_{\rm p}}{F_\oplus} \right)^{0.044}
\end{equation}
where $f = \Matm / \Mc$ is the atmosphere mass fraction, and $F_{\rm p} \propto a^{-2}$ is the incident stellar flux. The exact value of $R_{\rm env}$ also depends on the opacity law \citep[e.g.][]{Ginzburg_2016}, and there is a $R_{\rm env} \propto t^{-0.18}$ dependence on the planet age \citep{Lopez_2014}. In this work we assume an age of 5 Gyr for a typical star in the Kepler catalog. For example, for an Earth-mass planet with $f = 10\%$ and an Earth-like stellar flux, a factor-of-two error in the age of the star would cause a $\lesssim 13\%$ error in the estimated density of the planet.

\subsection{Inflated atmospheres}
\label{sec:radius:inflated}

In this section we develop a very simple model of how a highly irradiated atmosphere might become inflated, of ``puffy'', due to very high temperatures. We use this model to examine whether this kind of process could potentially explain the extreme size ratios of planets that straddle the photo-evaporation threshold. We begin with the formula for hydrostatic equilibrium,

\begin{equation}\label{eqn:hydrostatic}
	\frac{1}{\rho} \frac{{\rm d} P}{{\rm d}R} = - \frac{G \Mc}{R^2}
\end{equation}
where $P$ is the gas pressure, and $\rho$ is the gas density. We assume that the formula for the adiabatic lower layer of the atmosphere from \citet{Lopez_2014} continues to be valid, and instead we focus only on the upper isothermal layer. For an isothermal atmosphere the equation of state is $P = \cs^2 \rho$, where $\cs$ is the isothermal sound speed

\begin{equation}
	\cs \sim \sqrt{ \frac{\kb T}{\mu} },
\end{equation}
where $\kb$ is the Boltzmann constant, $T$ is the temperature, and $\mu$ is the molecular weight. Hence, we rewrite Equation \ref{eqn:hydrostatic} for an isothermal atmosphere as 

\begin{equation}
	\frac{1}{P} \frac{{\rm d} P}{{\rm d}R} = - \frac{\Rb}{R^2},
\end{equation}
where $\Rb = G \Mc \mu / (\kb T)$ is the Bondi radius. Let $P_\rcb$ and $R_\rcb$ be the pressure and radius at the radiative-convective boundary. Let $\Rtr$ be the planet's transit radius (i.e.\,the point where the atmosphere becomes optically transparent) and let $\Ptr$ be the pressure at $R = \Rtr$. Integrating from $P_\rcb$ to $\Ptr$ we obtain

\begin{equation}\label{eqn:R_fin}
	\Rtr = \frac{\Rb \; R_\rcb}{\Rb + R_\rcb \log(\Ptr / P_\rcb) }.
\end{equation}

According to \citet{Lopez_2014}, $P_\rcb \sim $100 to 1000 bar and $\Ptr \sim 20$ mbar; therefore, we adopt $\log(\Ptr / P_\rcb) = - 9$. But notice that, as $R_\rcb \rightarrow \Rb/9$, the planet radius diverges. This is a reminder that a fixed $\Ptr / P_\rcb$ may not capture the complexity of highly inflated atmospheres. We checked that none of the simulated planets in section \ref{sec:results:inflated} have $R_\rcb \ge \Rb/9$.

%
%
\section{Example simulation}
\label{sec:example}

\begin{figure*}[p!]
  \centering
  \includegraphics[width=1\textwidth]{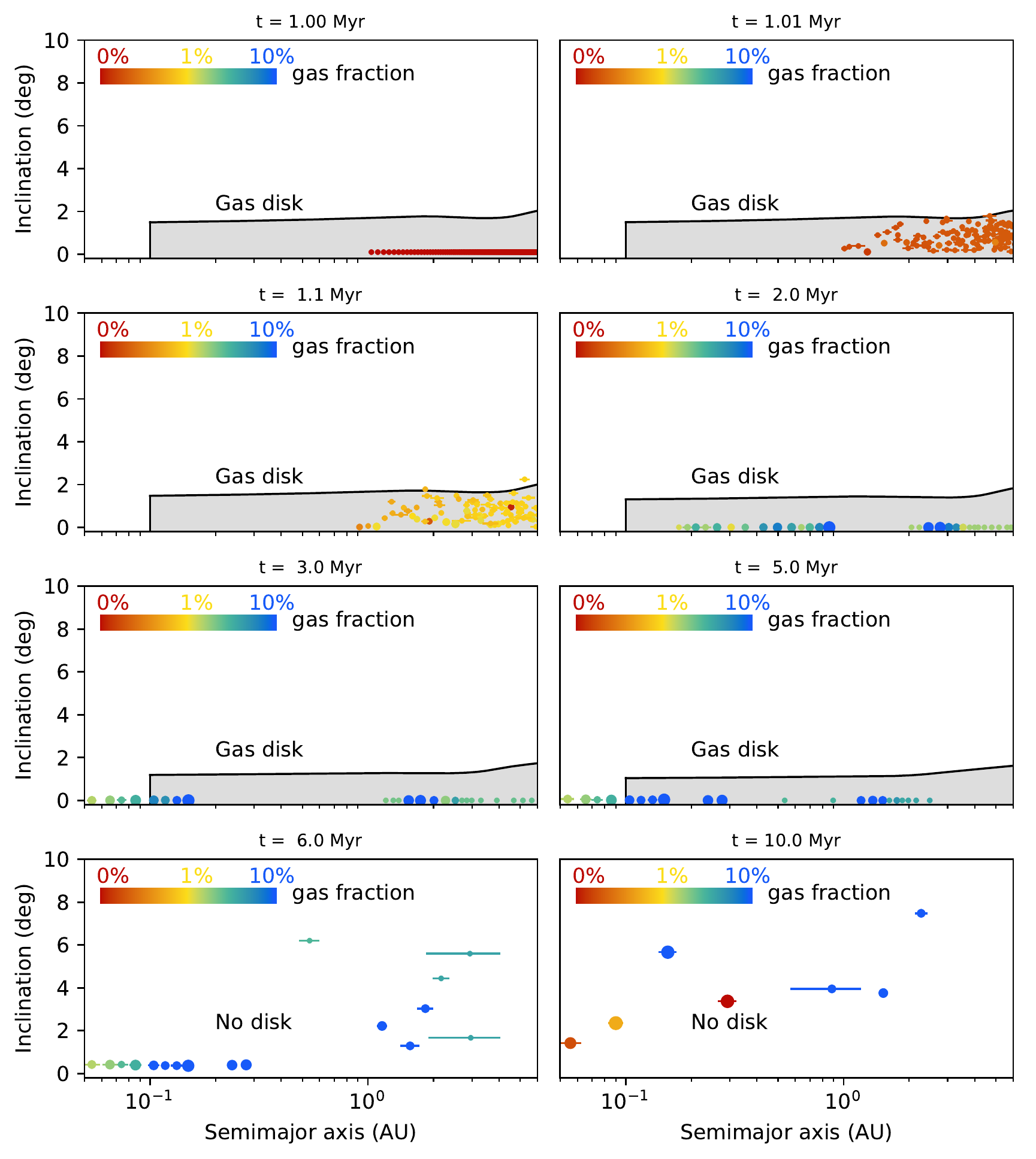}
  \caption{Eight snapshots of one of the simulations in the baseline model (Table \ref{tab:models}). Each planet is represented by a circle of size $r \propto m^{1/3}$. The color of each planet indicates the fraction of the planet's mass that is in the atmosphere; it follows a bi-linear scale, first from 0\% (red) to 1\%(green), and then to 10\% (blue). Each planet also has a horizontal line that goes from apastron to periastron. The vertical axis is the orbital inclination for each planet. The grey region shows the height of the disk as an inclination, $I = \rm{tan}^{-1}(H/r)$, where $H$ is the disk scale height, and $r$ is the orbital separation. The snapshot at $t = 1$ Myr shows the initial conditions of the N-body simulation: 125 embryos with a mass of $0.4 M_\oplus$ each, spaced according to the solid surface density at the time when planetesimals form ($t = 0$; see main text). The snapshot at $t = 5$ Myr occurs just before the disk dissipates. (An animation is available at \url{https://youtu.be/14OqAkn_OR0} and \url{https://zenodo.org/record/1401675}).}
  \label{fig:example}
\end{figure*}

Figure \ref{fig:example} shows eight snapshots for one of our simulations in the baseline model. The simulation begins at $t = 1$ Myr, when the protoplanetary disk is one million years old. The simulation begins with 125 embryos, each with a mass of $M = 0.4 M_\oplus$, starting at $a_0 = 1$ AU (see section \ref{sec:model:init}). Each planet is shown as a colored circle, with a color scale that indicates the atmosphere mass fraction. At the beginning of the simulation, the embryos have no atmosphere. Though not visible at $t = 1$ Myr, each planet also has a horizontal line that goes from apastron to periastron.

In the figure, the grey region marks the scale height $H$ of the disk. For planets, the vertical axis gives the orbital inclination. For the disk, we use $I = \rm{tan}^{-1}(H/r)$ to convert the disk scale height into an inclination; where $r$ is the orbital separation. Over the course of the simulation, the scale height of the disk drops slightly as the disk temperature decreases.

The snapshots at $t = 1.01$ Myr and 1.1 Myr show the early dynamical evolution of the system. The embryos are initially densely packed and quickly interact dynamically. In this way, we simulate the final assembly of isolation-mass bodies. This might be an important difference between this investigation and that of \citet{Izidoro_2017}, who began their simulations with more widely spaced and more massive bodies, applying the isolation-mass results of \citet{Kokubo_2000}. During this early formation period the planetary embryos experience strong gravitational scatterings and acquire inclinations that reach close to or even above one scale height. The eccentricities and inclinations result from a balance between gravitational scatterings, and the dampening effect of the disk.

At $t = 2$ Myr, the embryos have merged and formed a smaller number of more massive planets. The planets are locked in compact resonant chains that migrate as a group while the planets continue to accrete gas. Occasionally, planet neighbors collide again to form a more massive planet. At $t = 2$ Myr there is also a planet trap (edge of the outward migration zone) at $a \sim 2$ AU that has allowed the nascent planets to separate into two different resonant chains.

\begin{table}[h]
  \caption{20 sub-Neptune ($R < 4 R_\oplus$) planet pairs with $R_{\rm out} < R_{\rm in}$ and $M_{\rm out} > M_{\rm in}$. These planets cannot be explained by standard formation models plus photo-evaporation, and their presence points to late-stage collisions that removed a significant portion of the planet's atmosphere. The error bars correspond to the 68\% credible interval.}
  \label{tab:collisions}
  \begin{tabular}{l|rrrr}
KOI & $P_{\rm in}$ (day) & $P_{\rm out}$ (day)
& $R_{\rm out} / R_{\rm in}$ & $M_{\rm out} / M_{\rm in}$ \\
  \hline
KOI-0085  & 5.8597  & 8.1319  & $0.588^{+0.006}_{-0.006}$  &  $1.398^{+3.634}_{-0.802}$ \\
KOI-0115  & 5.4118  & 7.1262  & $0.566^{+0.006}_{-0.006}$  &  $1.161^{+1.012}_{-0.398}$ \\
KOI-0250  & 12.283  & 17.2517 & $0.858^{+0.007}_{-0.007}$  &  $1.318^{+0.144}_{-0.121}$ \\
KOI-0250  & 17.2517 & 46.8304 & $0.865^{+0.015}_{-0.015}$  &  $1.333^{+1.782}_{-0.922}$ \\
KOI-0520  & 12.7624 & 25.7561 & $0.915^{+0.013}_{-0.013}$  &  $1.549^{+0.335}_{-0.275}$ \\
KOI-0654  & 8.6051  & 10.2179 & $0.841^{+0.036}_{-0.038}$  &  $2.677^{+3.906}_{-1.183}$ \\
KOI-0701  & 18.1646 & 122.383 & $0.868^{+0.009}_{-0.010}$  &  $1.103^{+3.440}_{-0.818}$ \\
KOI-0707  & 21.7725 & 31.7881 & $0.746^{+0.007}_{-0.007}$  &  $1.787^{+1.667}_{-0.689}$ \\
KOI-0870  & 5.9124  & 8.9858  & $0.957^{+0.010}_{-0.010}$  &  $1.018^{+0.192}_{-0.155}$ \\
KOI-0877  & 12.0424 & 20.8364 & $0.593^{+0.020}_{-0.021}$  &  $1.504^{+1.787}_{-0.943}$ \\
KOI-1598  & 56.4756 & 92.8834 & $0.777^{+0.011}_{-0.011}$  &  $1.206^{+1.546}_{-0.539}$ \\
KOI-1955  & 15.1693 & 26.2381 & $0.936^{+0.018}_{-0.019}$  &  $1.09 ^{+2.070}_{-0.628}$ \\
KOI-2086  & 8.9187  & 11.8981 & $0.852^{+0.039}_{-0.041}$  &  $1.084^{+0.395}_{-0.272}$ \\
KOI-2195  & 20.0537 & 30.0958 & $0.892^{+0.027}_{-0.028}$  &  $1.279^{+1.267}_{-0.511}$ \\
KOI-0232  & 37.9861 & 56.2619 & $0.929^{+0.016}_{-0.016}$  &  $23.20^{+68.72}_{-14.33}$ \\
KOI-0285  & 26.7242 & 49.357  & $0.817^{+0.012}_{-0.012}$  &  $3.236^{+10.21}_{-2.664}$ \\
KOI-0904  & 2.211   & 4.6166  & $0.891^{+0.016}_{-0.017}$  &  $1.79 ^{+3.896}_{-1.151}$ \\
KOI-0904  & 27.9647 & 42.1311 & $0.974^{+0.024}_{-0.024}$  &  $1.379^{+0.097}_{-0.096}$ \\
KOI-1781  & 7.8345  & 58.0196 & $0.799^{+0.010}_{-0.011}$  &  $1.686^{+4.712}_{-1.263}$ \\
KOI-2038  & 17.9129 & 25.2176 & $0.938^{+0.062}_{-0.066}$  &  $1.35 ^{+3.417}_{-0.753}$ \\
  \end{tabular}
\end{table}

At $t = 3$ Myr, the inner resonant chain has reached the edge of the disk ($a = 0.1$ AU) and pushed past it, as the combined inward torque of the outer planets overwhelms the outward torque due to the pressure bump at the disk edge. Farther out, the second chain of planets remains behind the planet trap. However, as the disk evolves, the planet trap has moved to $\sim$ 1 AU.

The snapshot at $t = 5$ Myr is taken just before the disk dissipates. The planet trap has evolved, and allowed the outer planets to move inward. At $t = 6$ Myr, or one million years after the disk dissipates, the inner planets are still locked into a resonant chain, but without the dampening effect of the disk, the inclinations and eccentricities have grown. Finally, by $t = 10$ Myr, the inner resonant chain has also broken apart. Most of the planets in the chain have collided and formed a system with a smaller number of more massive planets with higher eccentricities and inclinations. This formation story, in which sub-Neptunes form compact resonant chains which then break after the disk dissipates, was previously identified by \citet{Izidoro_2017}.

One new wrinkle in the story is that the final phase of post-disk giant impacts can lead to a significant loss in the planet's volatile budget. \citet{Inamdar_2016} have shown that late giant impacts can reproduce much of the diversity in the densities of super-Earths. In the snapshot at $t = 10$ Myr we see one planet that is completely depleted of volatiles (red) sitting in between two gas-rich planets (blue). This type of architecture is a prediction of the ``breaking chains'' formation scenario, and is not easily replicated by other processes like photo-evaporation. In our study of transit timing variations we found 20 sub-Neptune ($R < 4 R_\oplus$) planet pairs, shown in Table \ref{tab:collisions}, where the outer planet has a larger mass and a smaller radius than the inner planet. This type of architecture cannot be the result of photo-evaporation, because photo-evaporation is strongest on either the inner planet or the less massive planet. The existence of so many planet pairs where the outer planet is more massive and evidently has a smaller gaseous envelope is strong evidence that, similar to Figure \ref{fig:example}, the outer planet experienced late-stage giant impacts that removed the planet's atmosphere.

%
%
\section{Simulation results}
\label{sec:results}

In this section we present our results. Broadly speaking, we find that our simulations produce planetary systems that broadly resemble the population of super-Earths in the Kepler field. This includes the size ratios of most planet pairs, with the crucial exception of planet pairs that straddle the photo-evaporation valley. Among those planet pairs, we find that the planet that has not lost its atmosphere must typically be inflated relative to our model.

\subsection{Period ratios}
\begin{figure}[th!]
  \centering
  \includegraphics[width=0.45\textwidth]{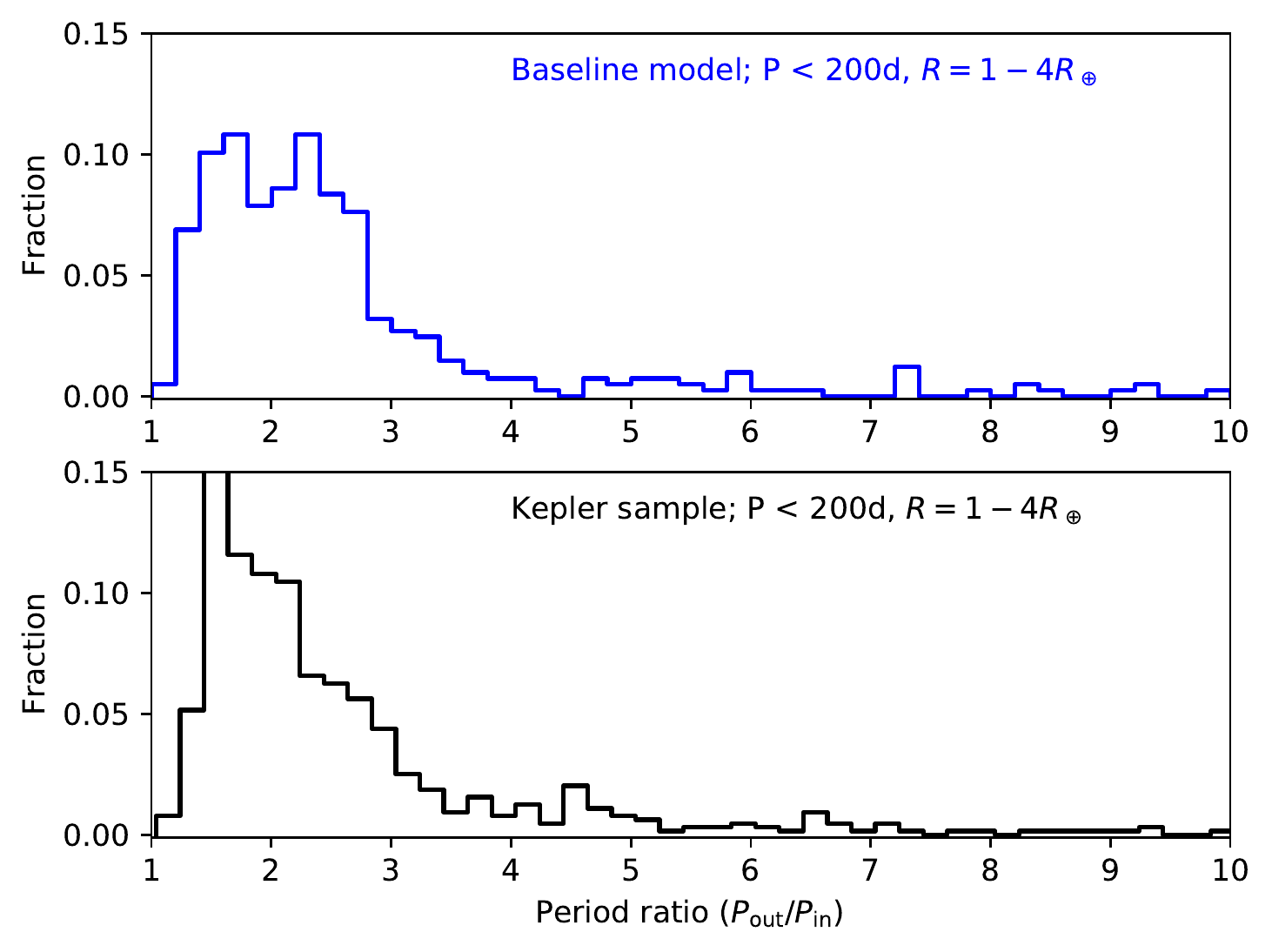}
  \caption{Distribution of period ratios ($P_{\rm out} / P_{\rm in}$) between neighboring planets. Our simulations (top) produce planets in orbital configurations that are broadly similar to those in the Kepler sample (bottom).}
  \label{fig:period-ratios}
\end{figure}

Figure \ref{fig:period-ratios} shows a histogram of the period ratios for the planetary systems produced by our baseline simulations, as well as the period ratios of planet pairs in our sample of Kepler planets. For the simulated population, we have plotted only a subset of the period ratios, so as to mimic the primary detection bias due to viewing geometry. We do not account for the conditional detection efficiency given that both planet transit, as this would depends on stellar properties. Broadly speaking, the two distributions are similar: Both distributions show a small tail for ratios greater than 4, and an increase in frequency from $P_{\rm out} / P_{\rm in} \sim 4$ to $\sim$ 2 or 3, and a peak near $P_{\rm out} / P_{\rm in} \sim 2$. However, the planets in the Kepler sample peak at smaller period ratios. This discrepancy may point to a limitation of the planet formation model, or might be caused by our incomplete implementation of observational biases. For example, since close-in planets transit more often than planets with large semi-major axes, the integrated transit signal-to-noise is greater for the inner planet than the outer planet for a fixed planet size and given star. Thus, accounting for the detection efficiency would be expected to result in further reducing the frequency of planet pairs with large period ratios in our simulated sample. In an up-coming work we will investigate how the results of our simulations change once observational biases are fully taken into account.

\subsection{Periods and radii}

\begin{figure*}[p!]
  \centering
  \includegraphics[width=0.48\textwidth]{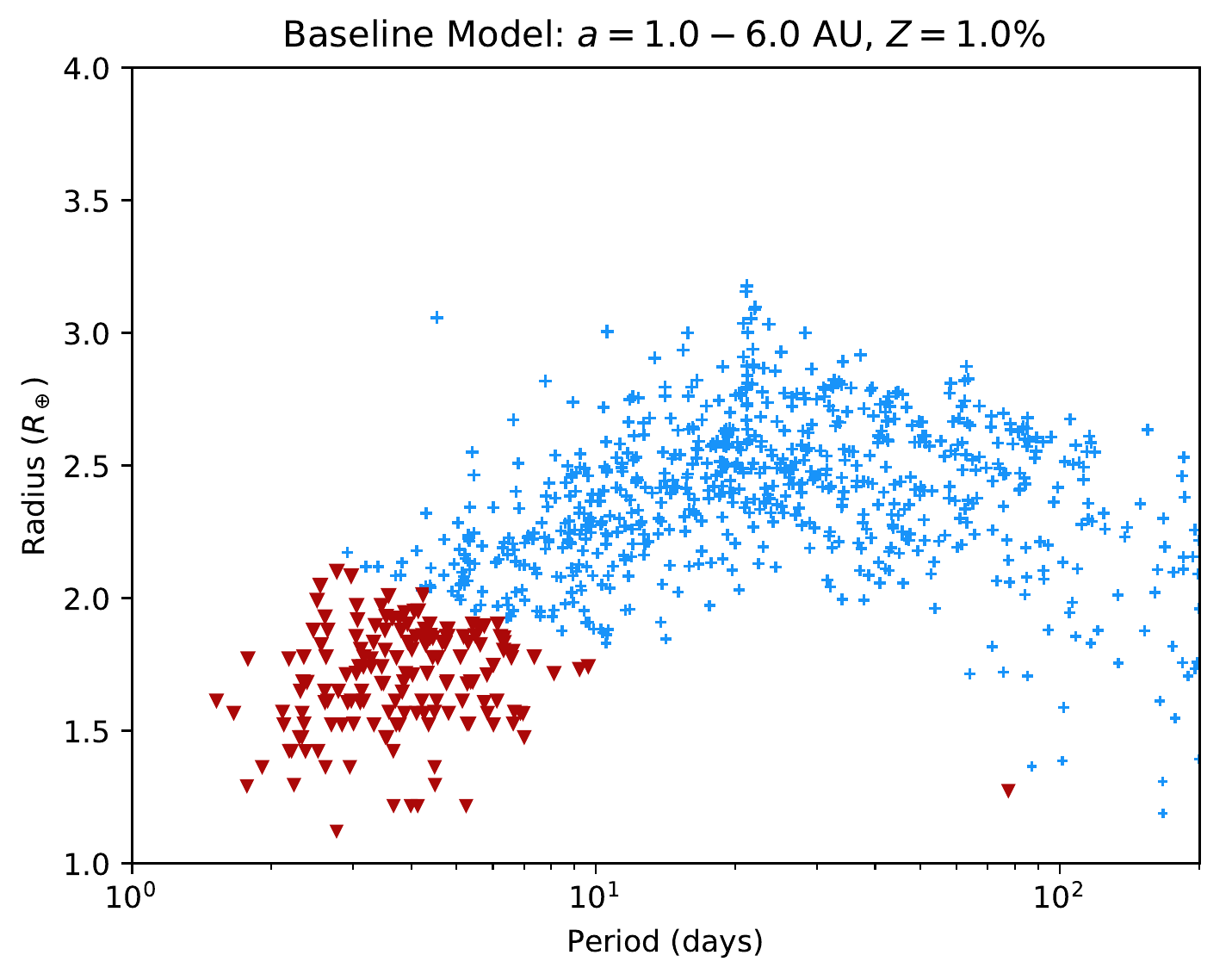}
  \includegraphics[width=0.48\textwidth]{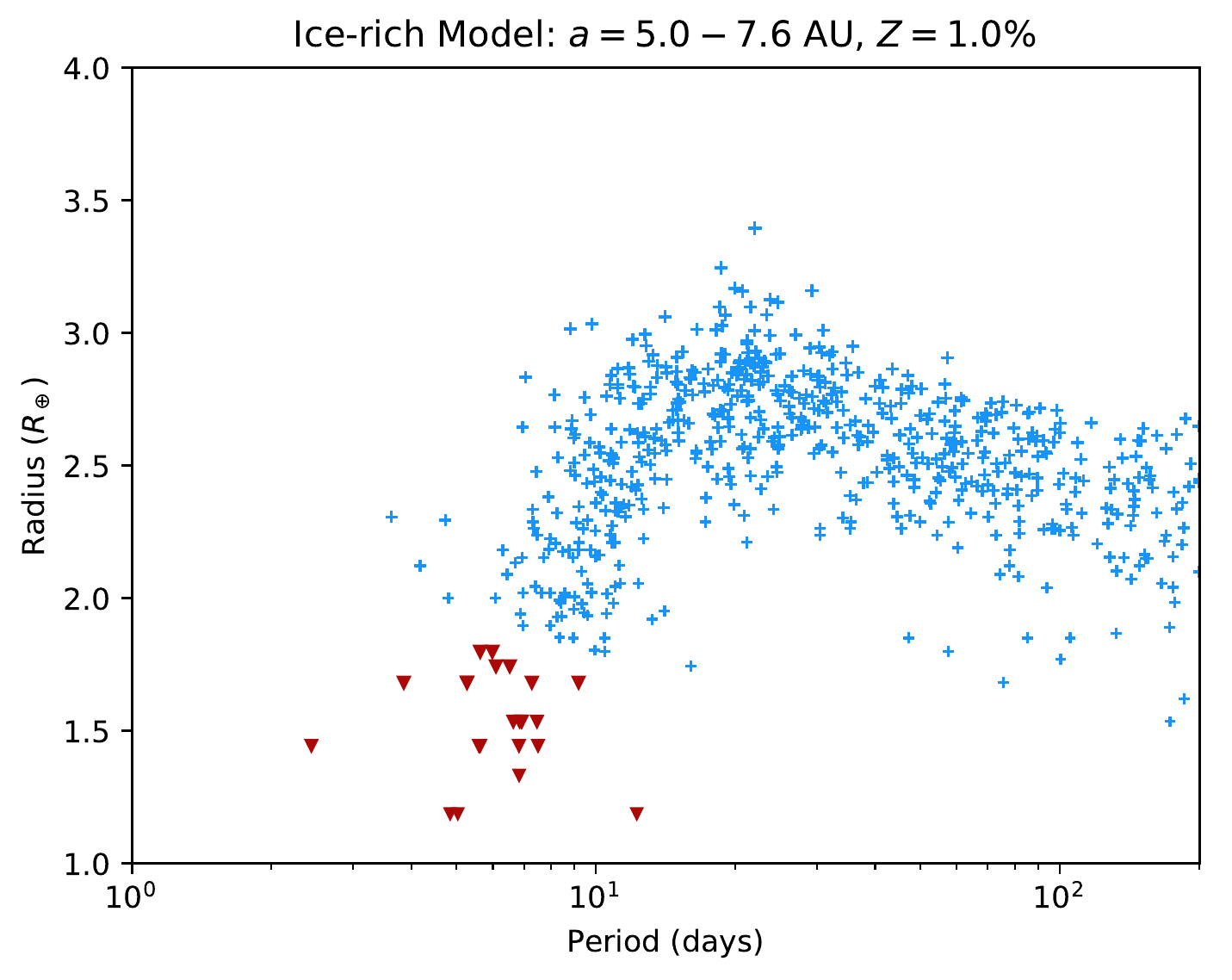}\\
  \includegraphics[width=0.48\textwidth]{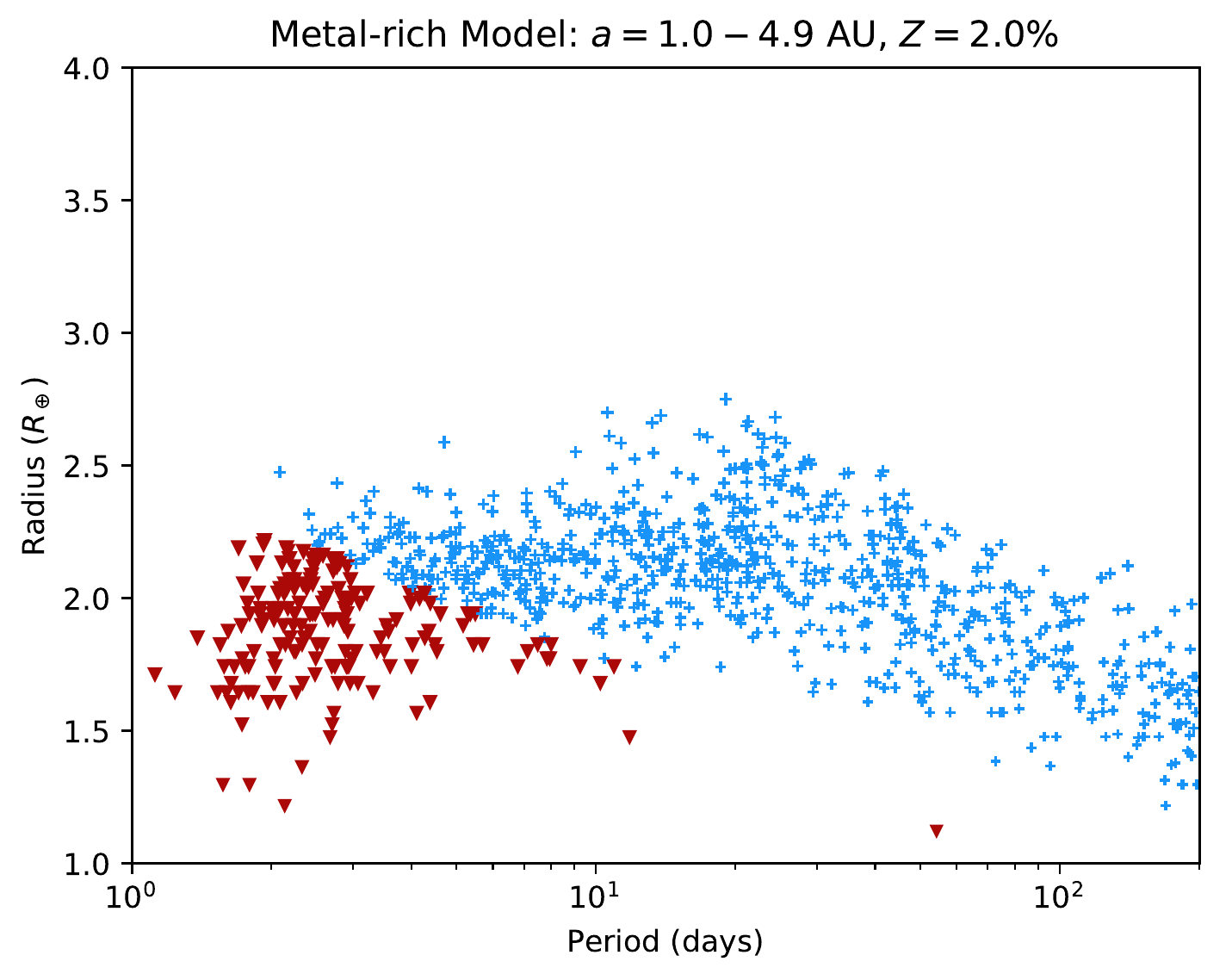}
  \includegraphics[width=0.48\textwidth]{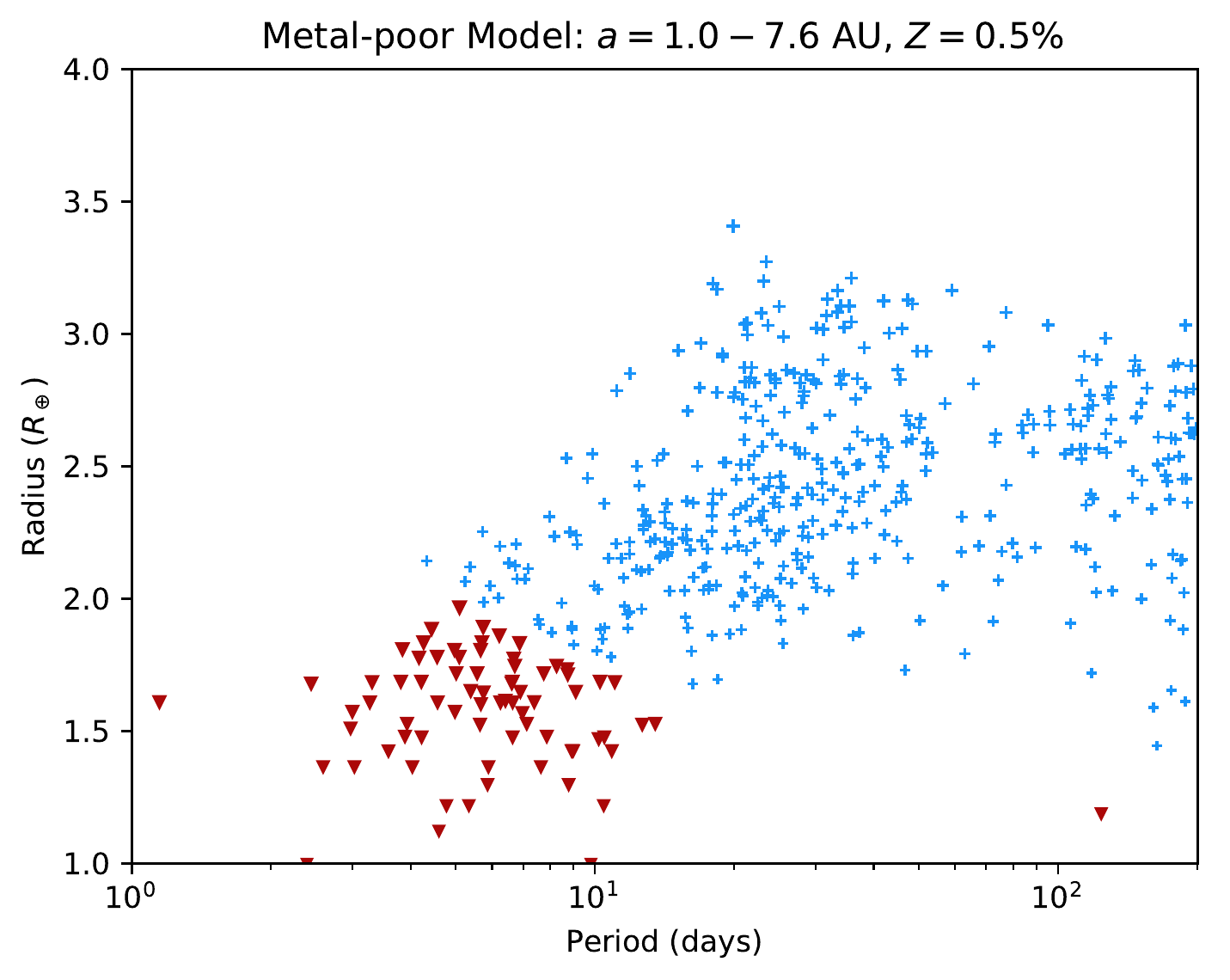}\\
  \includegraphics[width=0.48\textwidth]{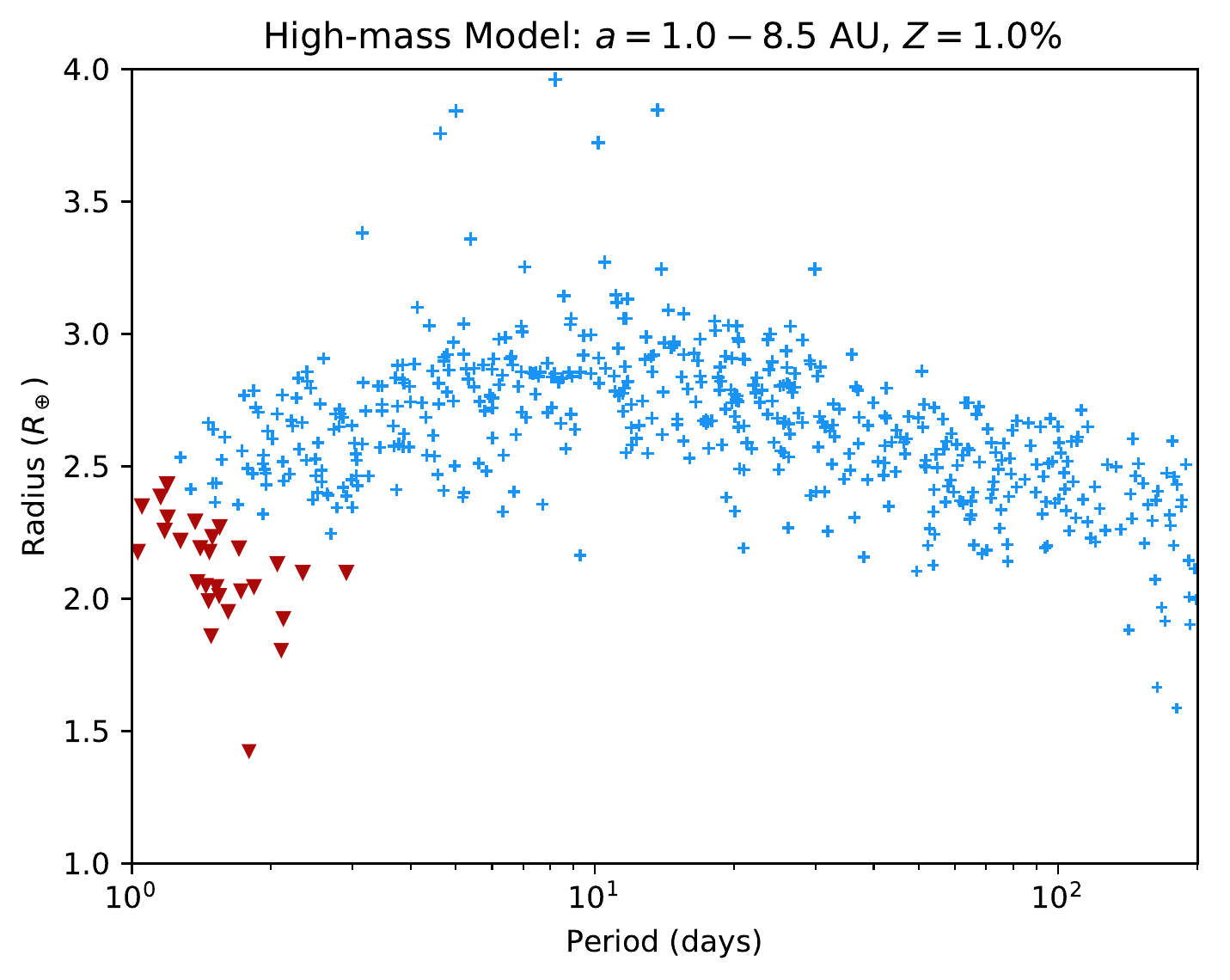}
  \includegraphics[width=0.48\textwidth]{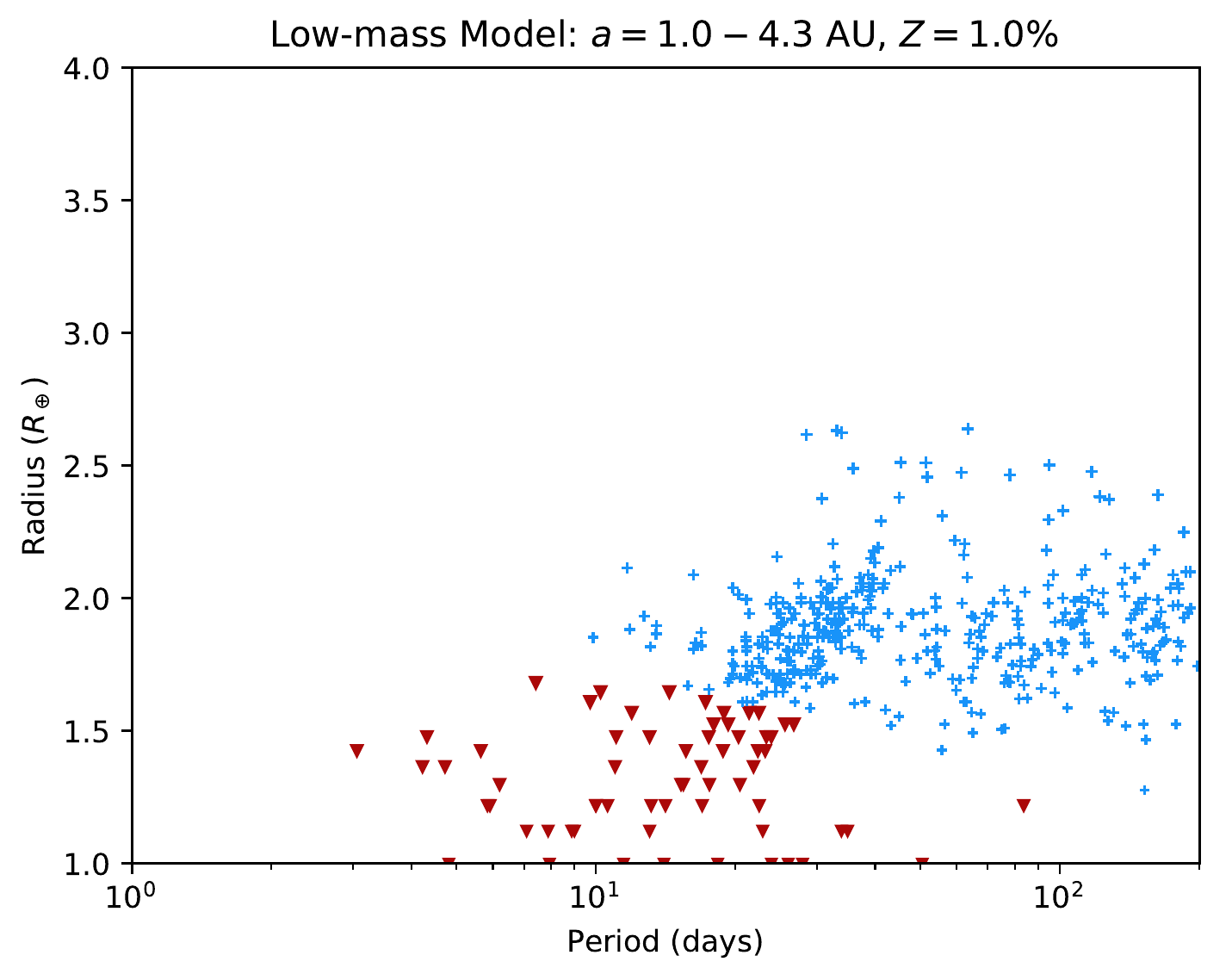}\\
  \caption{Radii and orbital periods of planets produced by our six models (Table \ref{tab:models}). In the baseline model (top left), planetary embryos form inside the ice line, starting at 1 AU. The disk metallicity is $Z = 1\%$, and the total mass in embryos is $50 M_\oplus$. In the ice-rich model (top right), embryos form as ice-rich bodies starting at the ice line at 5 AU. The metal-rich model (mid left) has a disk with $Z = 2\%$. The metal-poor model (mid right) has a disk with $Z = 0.5\%$. The high-mass model (bottom left) has $100 M_\oplus$ of embryos. The low-mass model (bottom right) has $25 M_\oplus$ of embryos. Photo-evaporated cores are marked as dark red triangles and planets with gaseous envelopes are marked as blue plus signs.}
  \label{fig:scatter}
\end{figure*}

Figure \ref{fig:scatter} shows the distribution of radii and orbital periods for the planets produced in each of our six models (Table \ref{tab:models}). In each case our simulations produce planets within a relatively narrow band of planet radii. However, changes in the disk properties --- especially metallicity and total mass of embryos --- can move the location of this band.

\subsection{Radius ratios}

\begin{figure*}[p!]
  \centering
  \includegraphics[width=0.46\textwidth]{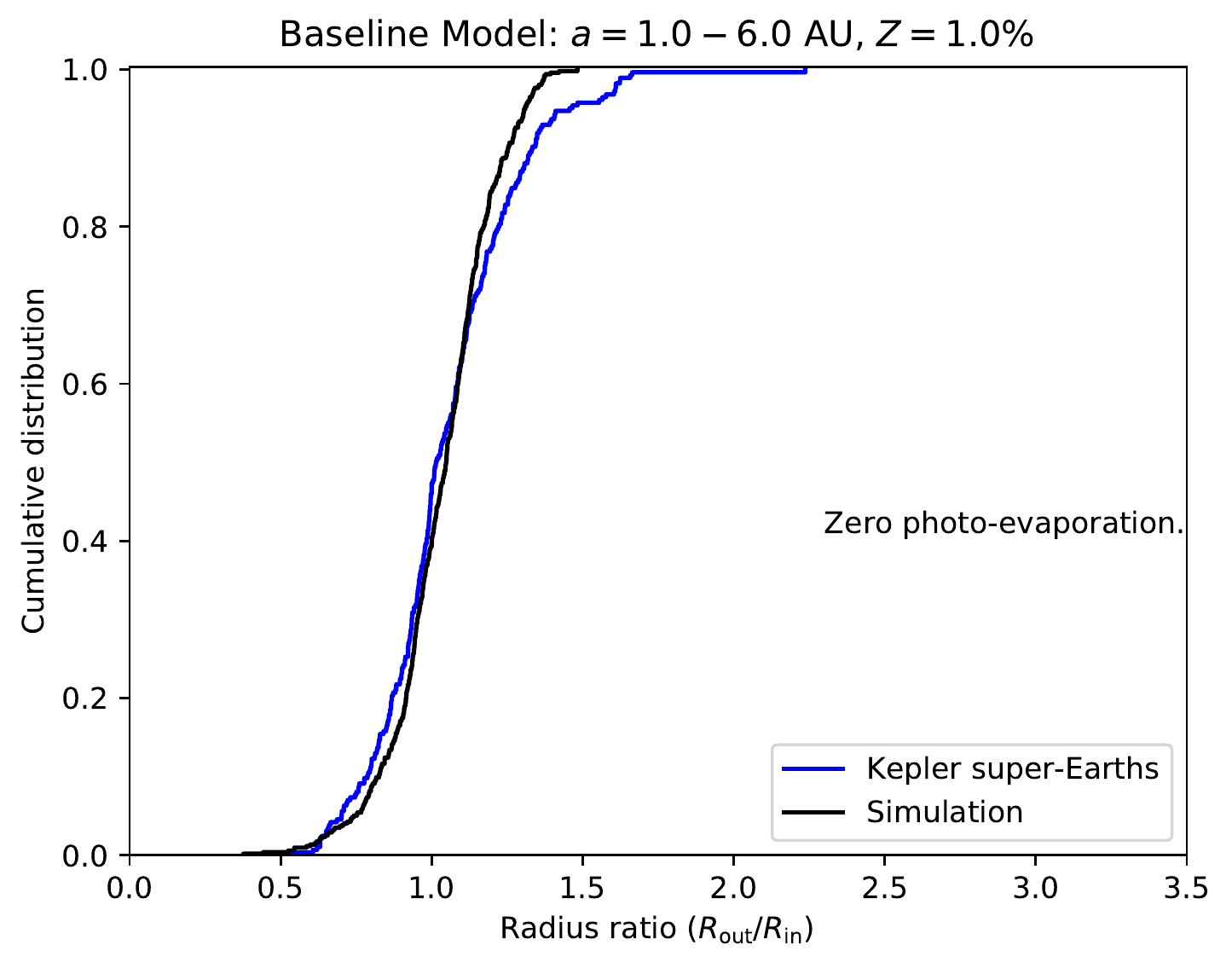}
  \includegraphics[width=0.46\textwidth]{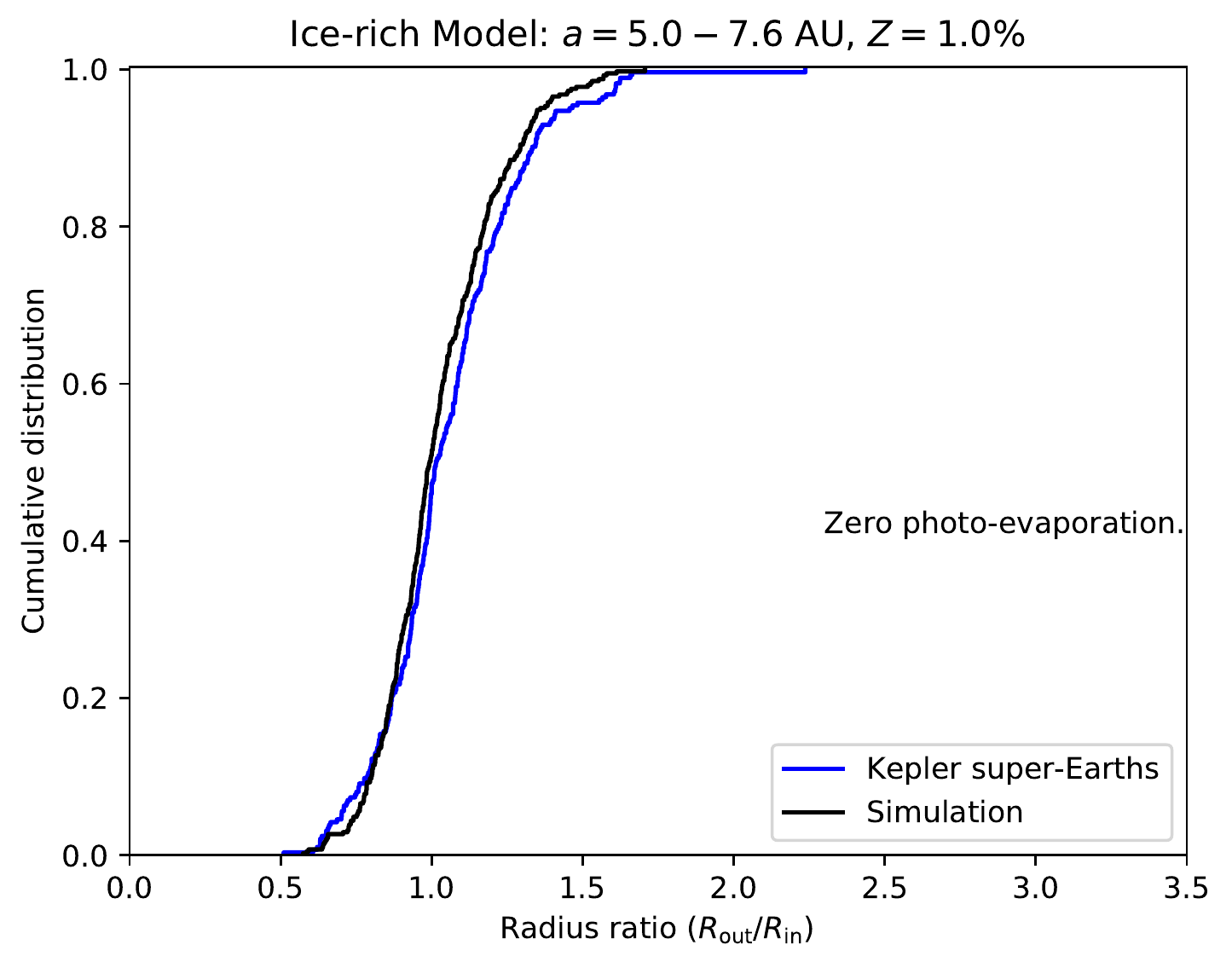}\\
  \includegraphics[width=0.46\textwidth]{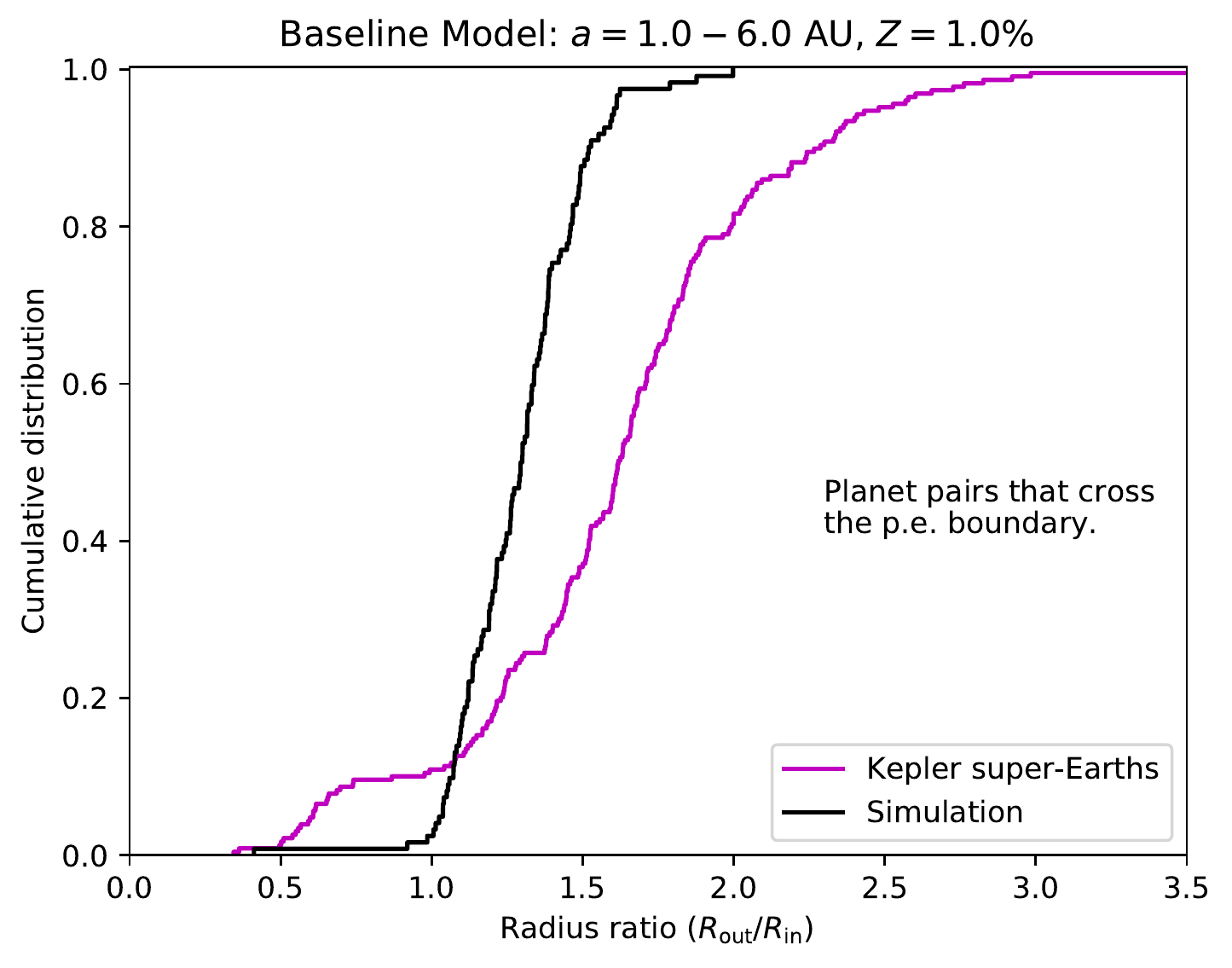}
  \includegraphics[width=0.46\textwidth]{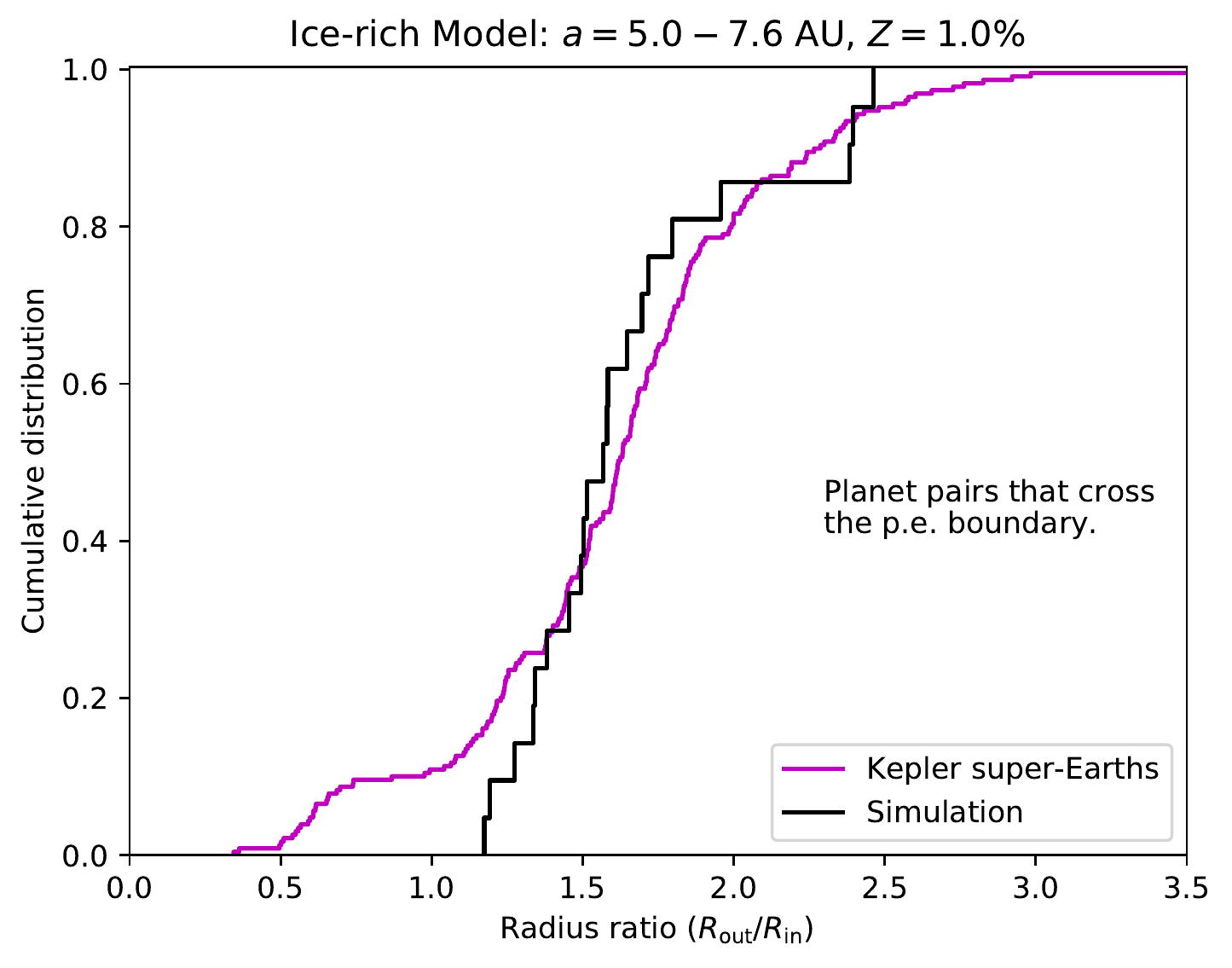}\\
  \includegraphics[width=0.46\textwidth]{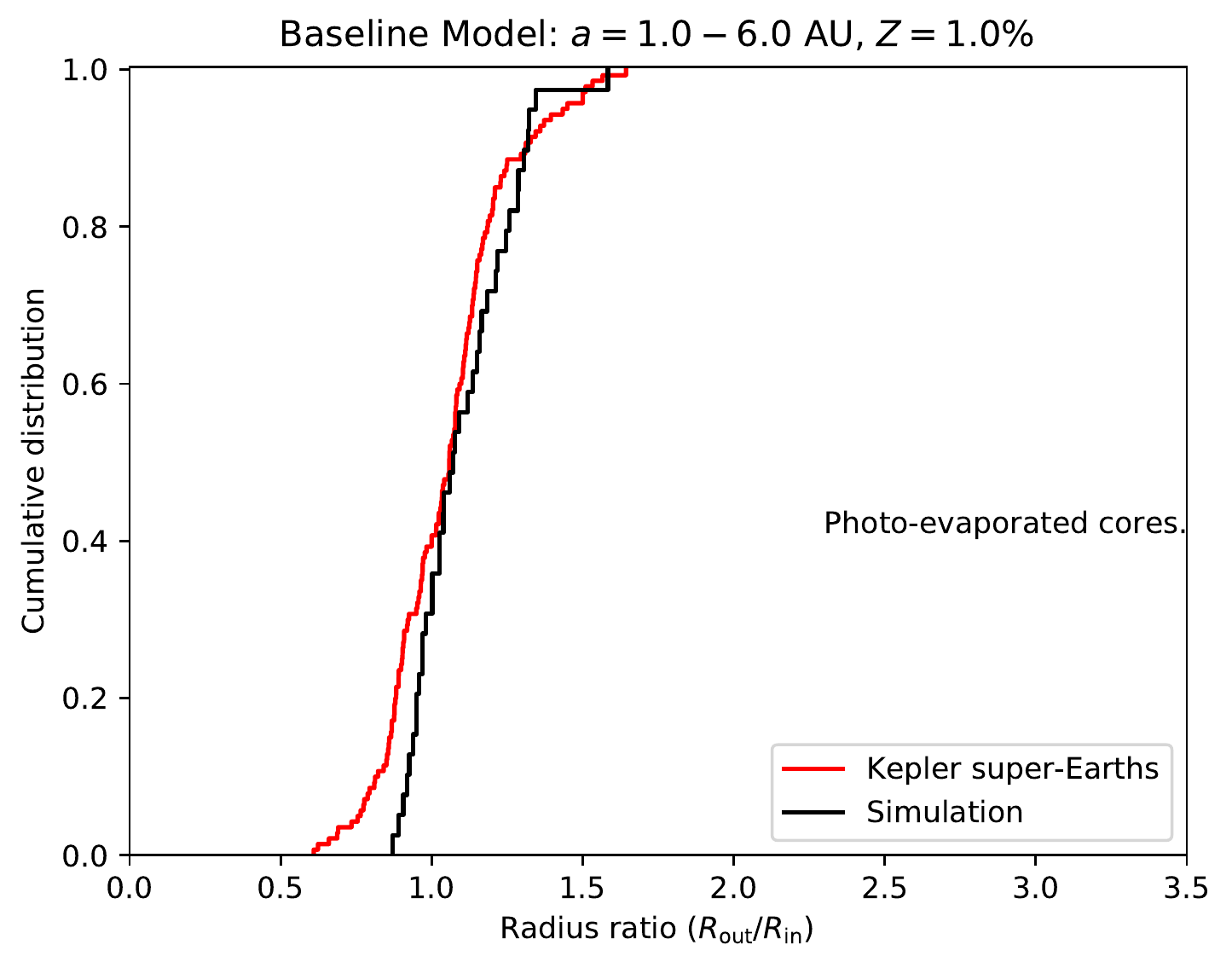}
  \includegraphics[width=0.46\textwidth]{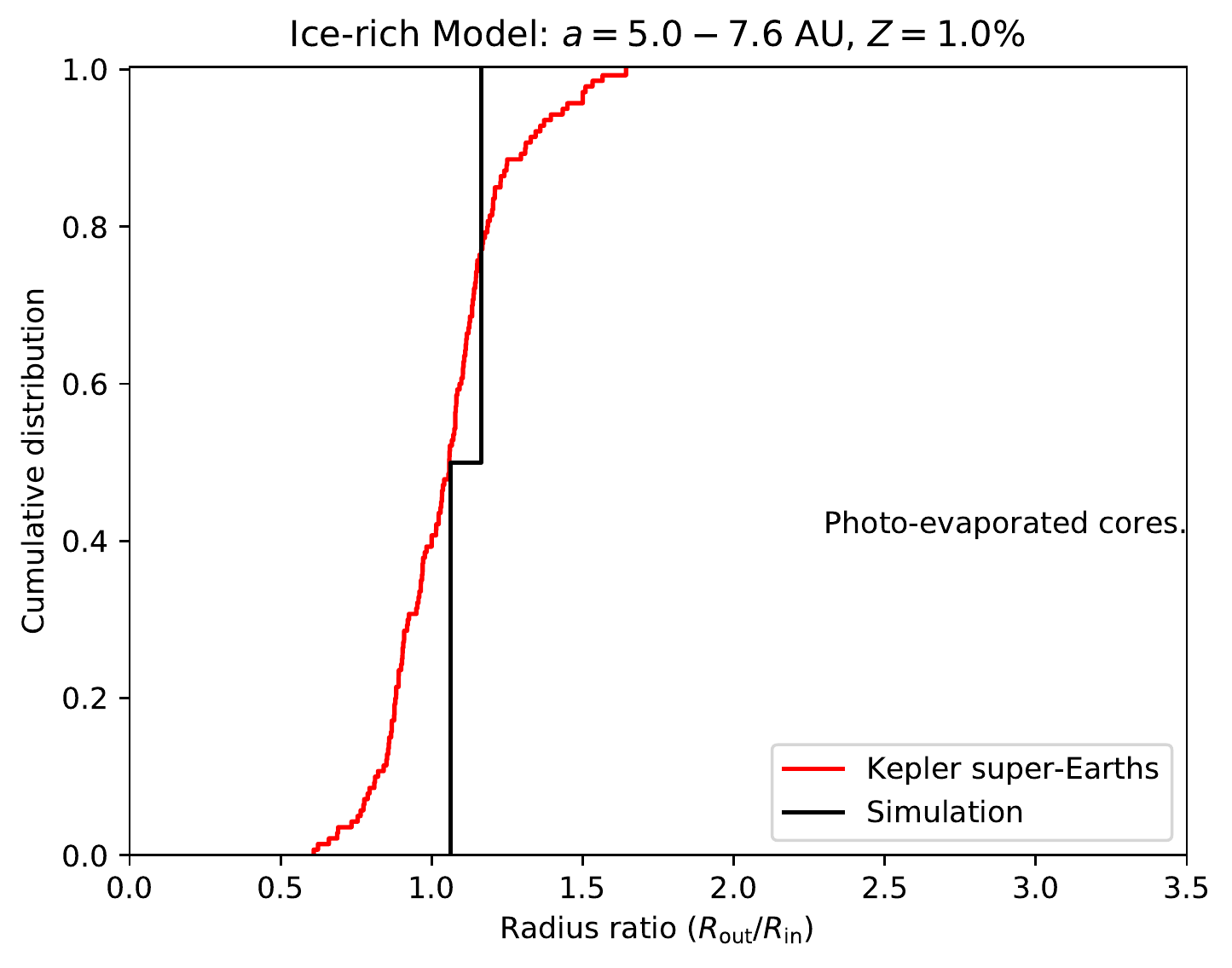}\\
  \caption{Cumulative distribution of the size ratio of neighboring planets ($R_{\rm out} / R_{\rm in}$) in our simulations (black) and the Kepler sample (see Figure \ref{fig:kepler}). The top plot shows the distribution of $R_{\rm out} / R_{\rm in}$ for planet pairs where both planets are above $R_{\rm trans}$. The middle plot shows $R_{\rm out} / R_{\rm in}$ when there is one planet on either side of $R_{\rm trans}$, and the bottom plot is for planet pairs where both planets are below the line. For planets with $R < R_{\rm trans}$ we replace $R$ with the core radius to simulate the effect of photo-evaporation. The left column shows the results for the baseline model and the right column shows the results for the ice-rich model.}
  \label{fig:R-ratios}
\end{figure*}

We compute the planet radii according to the method described in section \ref{sec:radius}. Here the planet radius is set by the height of the adiabatic atmosphere, following the model of \citet{Lopez_2014}. Figure \ref{fig:R-ratios} shows the distribution of $R_{\rm out} / R_{\rm in}$ for our baseline and ice-rich models (Table \ref{tab:models}) and for the Kepler sample. Figures for the other simulation sets are included in Appendix \ref{appendix:figures}. For planet pairs that are entirely above (top), or entirely below (bottom) the transition radius $R_{\rm trans}$, our simulations broadly reproduce the correct size ratios. However, for planet pairs that straddle the transition radius (middle), our model performs quite poorly.

\subsection{Inflated atmospheres}
\label{sec:results:inflated}

Finally, we examine the effect of atmospheres becoming inflated (or ``puffy'') when they are highly irradiated. In this section we compute the radii of all the planets above the transition line using the inflated atmosphere model (Equation \ref{eqn:R_fin}).

\begin{figure}[th!]
  \centering
  \includegraphics[width=0.46\textwidth]{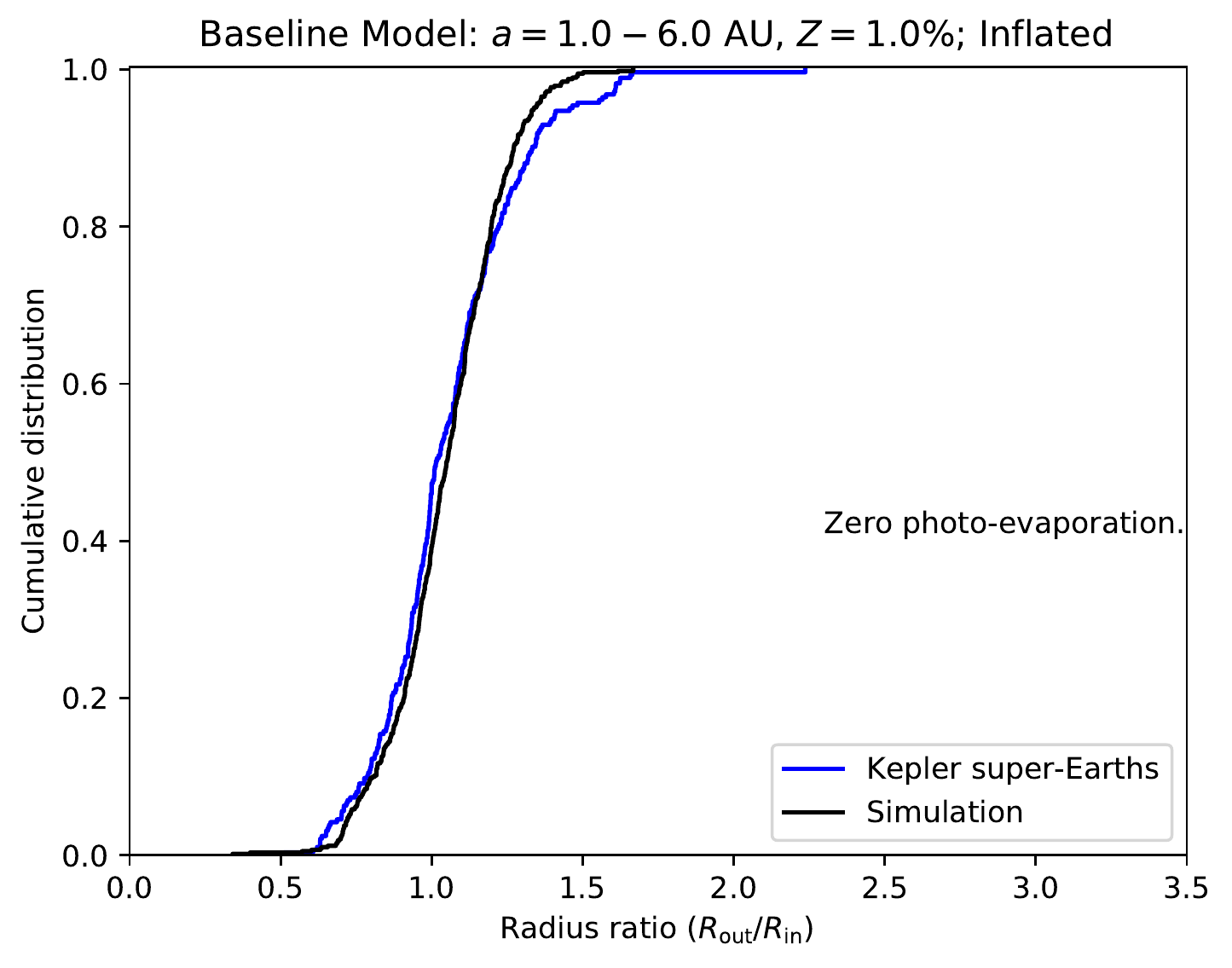}
  \includegraphics[width=0.46\textwidth]{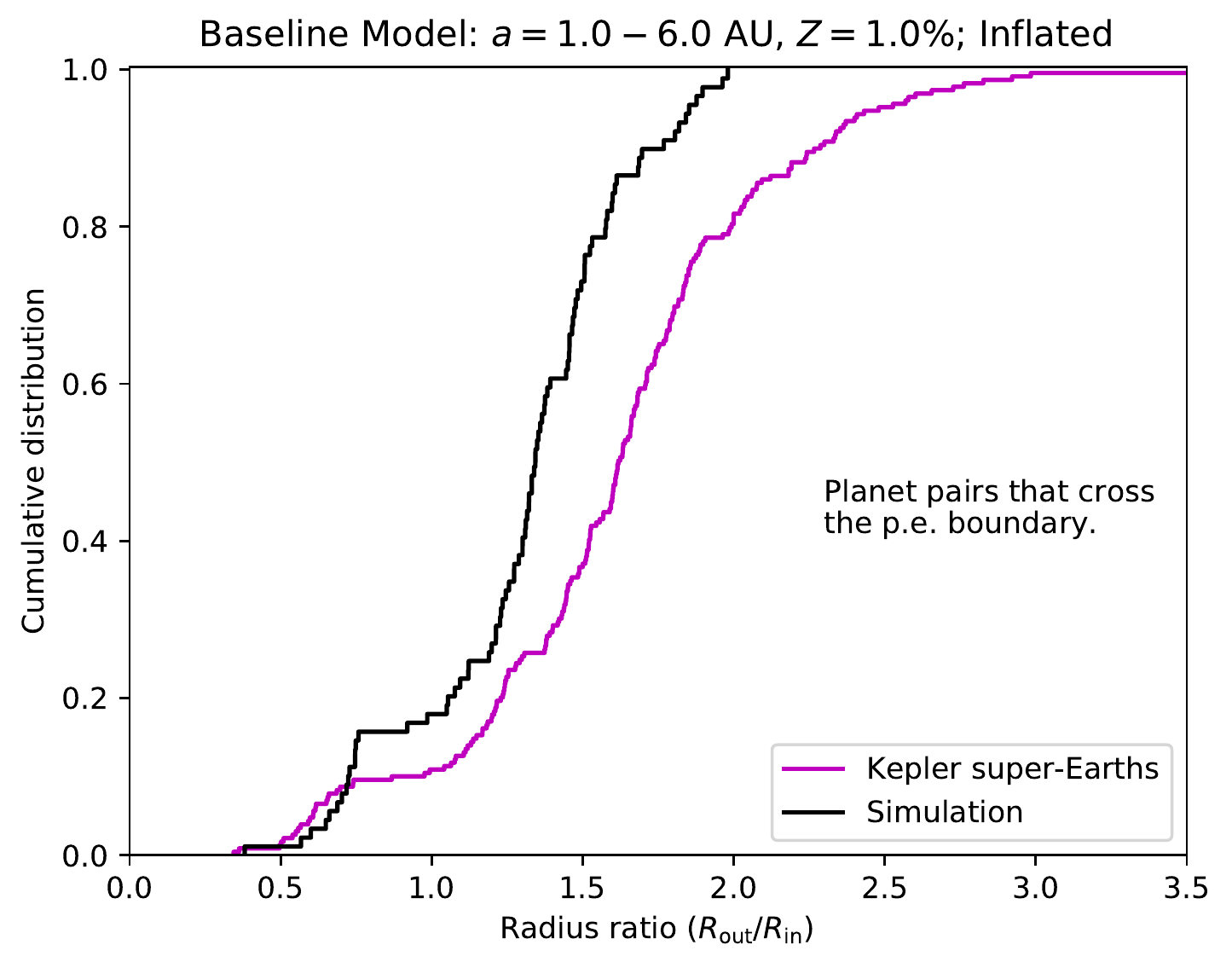}
  \includegraphics[width=0.46\textwidth]{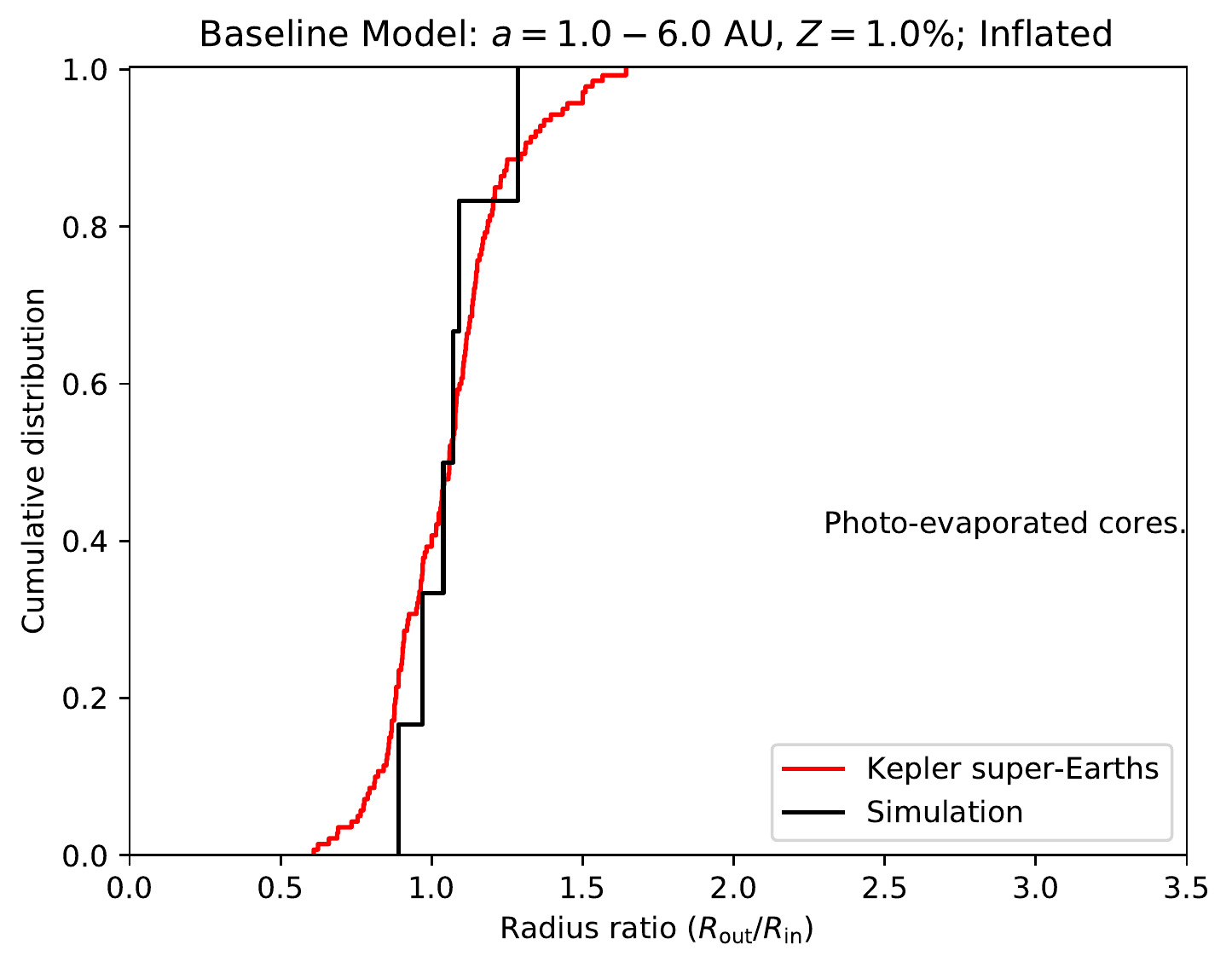}
  \caption{Size ratio distribution of neighboring planets ($R_{\rm out} / R_{\rm in}$) in our baseline model (black, green) and the Kepler sample. Compared to Figure \ref{fig:R-ratios}, Here we include the model for the isothermal layer of the atmosphere (Equation \ref{eqn:R_fin}) up to a maximum size of $R_{\rm max} = 0.2 \Rb$. The top plot shows the distribution for planet pairs where both planets are above $R/\Re = 1.05\,(F/\Fe)^{0.11}$. The middle plot shows $R_{\rm out} / R_{\rm in}$ when there is one planet on either side of the line, and the bottom plot is for planet pairs where both planets are below the line.}
  \label{fig:R-ratios-infl}
\end{figure}

We found that many of our simulated planets that lie just outside the photo-evaporation line are sufficiently inflated that $R_{\rm fin} > \Rb$ (Equation \ref{eqn:R_fin}), meaning that the planet radius reaches $\Rb$ before it reaches $P = 20$ mbar. That suggests that either these planets should have already photo-evaporated or might be actively out-gassing their atmospheres. Equation \ref{eqn:R_fin} has a singularity at $R_\rcb = \Rb/9$. However, all of our planet pairs have $R_\rcb < \Rb/9$. We set a maximum cut-off for the planet's atmosphere at $R_{\rm max} = c\,\Rb$. Figure \ref{fig:R-ratios-infl} shows the distribution of size ratios for $c = 0.2$ (black). We find that the shape of the plot depends only weakly on $c$. Despite its limitations, our model has clearly improved the fit for planet pairs that straddle the photo-evaporation threshold.

\begin{figure}[th!]
  \centering
  \includegraphics[width=0.48\textwidth]{scatter-radius-01a.pdf}\\
  \includegraphics[width=0.48\textwidth]{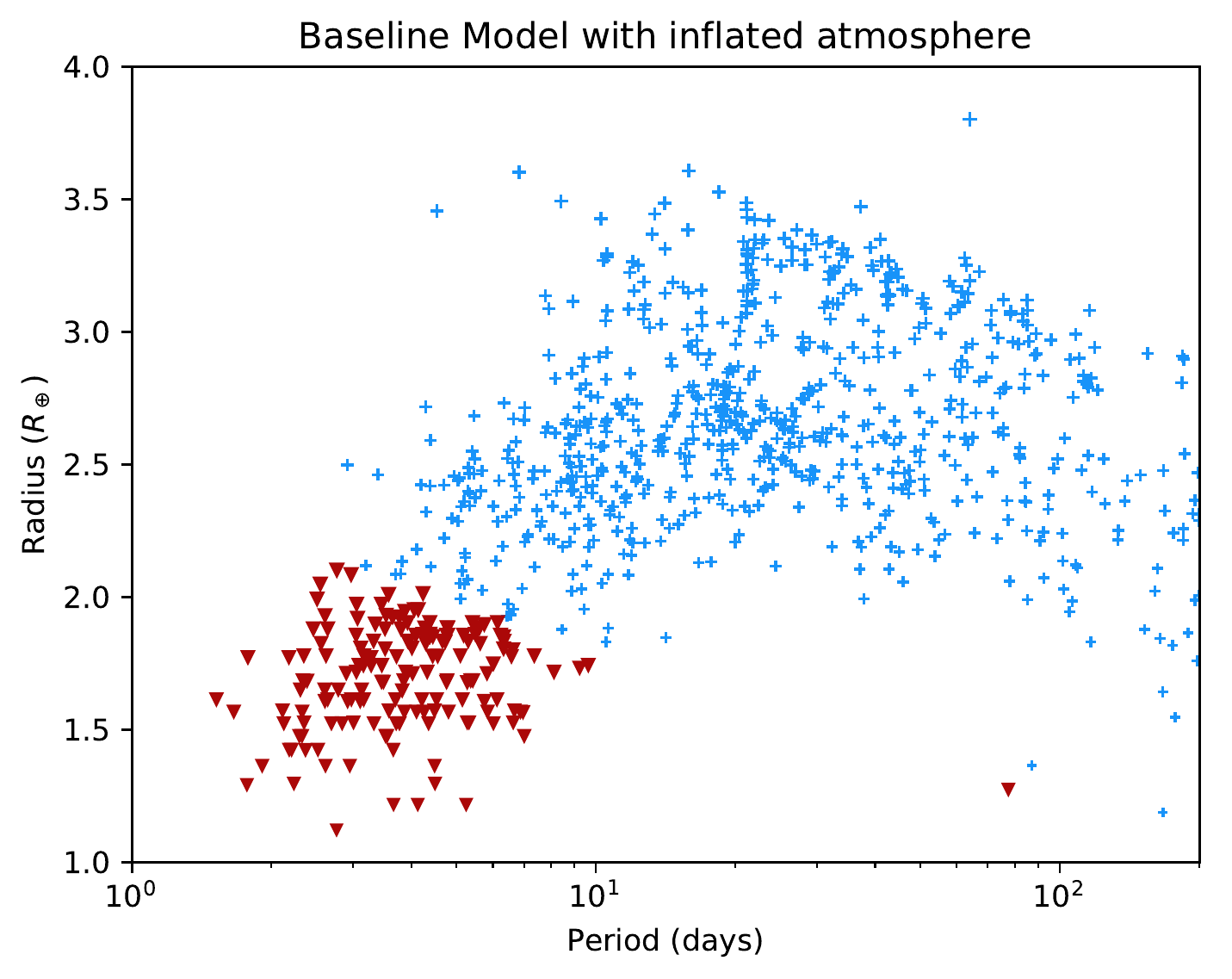}
  \caption{Radii and orbital periods of planets produced by the baseline model. A dark color indicates the the planet has been photo-evaporated, so that $R = R_{\rm core}$. The top plot includes only the adiabatic atmosphere, which contains most of the atmosphere mass. In the bottom plot we added a term that allows the atmosphere of the planet to become inflated up to a maximum size of $R_{\rm max} = 0.20 R_{\rm B}$ where $R_{\rm B}$ is the Bondi radius.}
  \label{fig:scatter-infl}
\end{figure}

Figure \ref{fig:scatter-infl} shows the period and size distribution for planets in our baseline model with and without atmosphere inflation, as described in this section. This figure illustrates how large some of the super-Earth atmospheres need to be in order to reproduce the extreme size ratios across the photo-evaporation valley.

%
%
\section{Discussion}
\label{sec:discussion}

In section \ref{sec:results} we presented the simulation results for the first planet formation model that includes full N-body dynamics, gas accretion, gas loss due to collisions, and atmospheric photo-evaporation. This model naturally reproduces many of the observed features of Kepler super-Earths. Most importantly, our model reproduces the distribution of relative sizes of super-Earths that either have mostly pristine atmospheres or have been fully photo-evaporated. This result seems to be a robust outcome of planet migration and accretion that does not require any special fine-tuning. Secondly, our model reproduces many but not all of the features of the distribution of period ratios (Figure \ref{fig:period-ratios}); and finally, some of our models predict the presence of ultra short-period planets ($P \sim 1$ day) and others do not. Despite the successes, in this discussion we will focus on how the model can be improved further, and we identify key areas where additional work is needed.

\subsection{Range of planet radii}

The planets produced by our model do not reproduce the radius-period distribution of super-Earths discovered by Kepler (compare Figure \ref{fig:scatter} vs Figure \ref{fig:kepler}). Adding our model for an isothermal layer does not fundamentally change that result (Figure \ref{fig:R-ratios-infl}). We propose two candidate explanations for the discrepancy:

\begin{itemize}
\item In Figure \ref{fig:scatter}, each set of disk properties produces planets within a relatively narrow size range. But changing the disk properties (including the location of embryos) clearly has a significant effect on planet sizes and periods. We propose that much of the scatter in super-Earth radii and periods may come from a scatter in the initial disk properties.

\item The atmosphere model in Equation \ref{eqn:adiabatic} is an analytic fit to a grid of simulations conducted by \citet{Lopez_2014}. It is probable that this fit hides much of the natural scatter in the model.
\end{itemize}

Future work should investigate both possibilities.


\subsection{Auto-correlation of planet radii}

Previous authors have noted that planets within the same Kepler system seem to have similar sizes \citep{Millholland_2017,Weiss_2018}. Our investigation confirms this result, and adds some nuance to the story:

\begin{itemize}
\item Away from the photo-evaporation valley (top and bottom of Figures \ref{fig:kepler}, \ref{fig:R-ratios}, and \ref{fig:R-ratios-infl}) the intra-system uniformity of planet sizes is more extreme that previously realized, and our planet formation model reproduces that uniformity.

\item The most extreme size ratios are concentrated on planet pairs that straddle the photo-evaporation valley. These size ratios require that the planet outside the photo-evaporation valley have a much larger radius than predicted by our baseline model.
\end{itemize}

\subsection{Ultra short period planets}

In Figure \ref{fig:scatter}, most of our models struggled to produce planets with orbital periods less than 2 days. In our simulations the period of the innermost planet is mostly determined by the inner edge of the disk, which is at 0.1 AU ($P \sim 10$ days). Planets with periods shorter than 10 days form when a resonant chain of planets migrates to the inner edge of the disk and the disk torques on the outer planets forces the resonant chain to push the inner planets past the edge of the disk. This can be seen in the example simulation in Figure \ref{fig:example}.

The models that most efficiently produced planets with $P < 2$ days were the metal-rich disk and the high-mass models (Figure \ref{fig:scatter}). Since the Kepler field is slightly above the galactic plane, Kepler systems are likely to be more metal poor than the Sun. While this seems to favor the high-mass model, that kind of assessment requires a careful look at the period distribution with better modeling of Kepler's detection biases. In this work we included the primary detection bias due to viewing geometry. Future work should also account for the higher integrated signal-to-noise for planets with very short orbital periods.

Lastly, stellar tides, or some variation in the location of the inner edge of the disk, may also play an important role in the frequency of ultra short period planets.

\subsection{Highly irradiated atmospheres}

Our models are generally quite successful at reproducing the size ratios of planet pairs away from the photo-evaporation boundary. However, the size ratios of planets that straddle the photo-evaporation valley present a significant challenge. For those planets, all our models fail to reproduce the largest size ratios (Figure \ref{fig:R-ratios}), and the group of planets with $R_{\rm out} / R_{\rm in} < 1$.

The ice-rich model can reproduce at least some of the larger size ratios, but it also fails to reproduce the $R_{\rm out} / R_{\rm in} < 1$ tail. It appears that the size ratios are being driven away from unity. This would be explained if the planet just outside the transition radius are being highly inflated:

\begin{itemize}
\item In most cases, the inner planet is photo-evaporated and the outer planet has an inflated atmosphere. When that happens, $R_{\rm out} / R_{\rm in}$ is driven to very large values.

\item In a few instances, the outer planet has a low-enough mass that it is inside the photo-evaporation region, and while the \textit{inner planet} is inflated. When that happens, $R_{\rm out} / R_{\rm in}$ is driven away from unity and toward values around $\sim 2/3$.
\end{itemize}

Our model with an isothermal layer (Equation \ref{eqn:R_fin}) supports this idea: In the middle plot of Figure \ref{fig:R-ratios-infl}, the baseline model with an isothermal atmosphere layer better reproduces the overall distribution of size ratios across the photo-evaporation boundary. Importantly, the model reproduces the excess of planet pairs with $R_{\rm out} / R_{\rm in} < 1$, which could not be reproduced with the ice-rich model.

The fact that many of our planets had $R_{\rm fin} > \Rb$ (Equation \ref{eqn:R_fin}) likely reflects the limitations of our model for the isothermal layer of the atmosphere. But it also raises the possibility that some of these highly-irradiated planets might be filling a large portion of their Bondi radius, or might be currently their atmospheres. It is also possible that a complete explanation for the observed radii requires both water-rich planets as well as a better atmosphere model.

In any case, we hope that future authors will investigate the behavior of super-Earth atmospheres in this extreme environment. Recent work by \citet{Chachan_2018} also shows that the dissolution of hydrogen in planetary mantles during the planet formation process, followed by out-gassing as the mantle cools, can significantly increase a planet's ability to retain an atmosphere over Gyr timescales. Therefore, future work should include the processes of hydrogen dissolution and out-gassing as well.

%
%
\section{Summary and conclusions}
\label{sec:conclusions}

Using Kepler observations and N-body simulations we have presented several different lines of evidence that point to a photo-evaporation threshold near $R/R_\oplus = 1.05\,(F/F_\oplus)^{0.11}$. Planets above that curve seem to have retained their gaseous envelopes, while planets below that curve are likely to be photo-evaporated cores. Some of our evidence is entirely empirical, and some builds upon theoretical modelling:

\begin{enumerate}
\item We showed, in Figure \ref{fig:kepler}, that the line $R/R_\oplus = 1.05\,(F/F_\oplus)^{0.11}$ separates the Kepler super-Earths into three distinct populations. Planet pairs that are entirely above or entirely below that line show a narrow distribution of size ratios. Planet pairs that straddle that line show extreme size ratios, including both very large ratios ($R_{\rm out} / R_{\rm in} \sim 1.5$) and very small ones ($R_{\rm out} / R_{\rm in} \sim 2/3$).

\item We used transit timing variations to constrain the mass ratios of three of the planets with the most extreme size ratios (section \ref{sec:obs:ttv}). All of them have extreme density ratios, and two of them are very difficult to explain without a significant amount of atmosphere inflation.

\item We developed a sophisticated planet formation model that included disk migration, atmosphere accretion, and atmosphere loss through giant impacts. For each simulated planet we calculated the size of the core and the H$_2$ and He envelope. When both planets are are on the same size of the photo-evaporation line, we successfully reproduce the observed size ratios. However, we cannot reproduce the extreme size ratios of planet pairs that straddle the photo-evaporation line. Planets outside the photo-evaporation line appear oversized atmospheres.

\item Finally, we derived a simple expression to model an inflated radiative envelope. With this expression, we significantly improved the match with the observed planet ratios across the photo-evaporation boundary. Furthermore, adding this inflation term did not damage the good fits for the other two populations.
\end{enumerate}

Taken together, this points to a distinct photo-evaporation threshold, and a newly identified population of super-Earths with inflated atmospheres.

While it seems inevitable that any planet that formed in the protoplanetary disk experienced disk migration, our results cannot distinguish between an ice-rich model in which planets migrate from beyond the snow line, and a ``rocky'' model, in which planets form and migrate inside the snow line. In the absence of atmosphere inflation, the ice-rich model is favored because it can produce planet pairs with larger size ratios, and it better approximates the observed distribution of planet size ratios. However, both models seem to require inflated atmospheres. We hope that future work will develop a better model for highly irradiated super-Earth atmospheres. When such a model becomes available, it might be possible to use observed size ratios to distinguish between formation scenarios that start beyond the snow line and models that don't.

Finally, this work has an important implication for target selection with the up-coming James Webb Space Telescope: For planets pairs that straddle the photo-evaporation boundary, the planet that lies outside the photo-evaporation is more likely to be a good target for transit spectroscopy. If the stellar properties are not known sufficiently well to constrain the incident flux, observations should target the outer planet in planet pairs with $R_{\rm out} / R_{\rm in} > 2$.

%
%
\section*{Acknowledgements}
D.C.\,developed the model for gas accretion and gas loss, conducted the simulations, and performed the main analysis. D.J-H.\,and A.W.\,contributed the TTV analysis. A.I.\,and S.R.\,provided an adapted version of the {\sc Mercury} code that includes disk migration and disk evolution. E.B.F.\,contributed to the scientific analysis and provided advice throughout the project.

D.C.\, acknowledges Hilke Schlichting, Eric Lopez, and Jonathan Fortney for discussions on modelling super-Earth atmospheres. D.~C.~'s research was supported by an appointment to the NASA Postdoctoral Program within NASA's Nexus for Exoplanet System Science (NExSS), administered by Universities Space Research Association under contract with NASA. This work benefited from the 2018 Exoplanet Summer Program in the Other Worlds Laboratory (OWL) at the University of California, Santa Cruz, a program funded by the Heising-Simons Foundation. 

A.~I.\,thanks financial support from FAPESP via proc. 16/19556-7 and 16/12686-2.

E.~B.~F.\,acknowledges support from NASA Exoplanet Research Program award NNX15AE21G. The results reported herein benefitted from collaborations and/or information exchange within NASA's Nexus for Exoplanet System Science (NExSS) research coordination network sponsored by NASA's Science Mission Directorate. The Center for Exoplanets and Habitable Worlds is supported by the Pennsylvania State University, the Eberly College of Science, and the Pennsylvania Space Grant Consortium. We gratefully acknowledge support from NSF grant MRI-1626251. This research or portions of this research were conducted with Advanced CyberInfrastructure computational resources provided by The Institute for CyberScience at The Pennsylvania State University (\texttt{http://ics.psu.edu}), including the CyberLAMP cluster supported by NSF grant MRI-1626251. A.I.\,thanks financial support from FAPESP via proc. 16/19556-7 and 16/12686-2.

S.~N.~R.\,thanks the Agence Nationale pour la Recherche via grant ANR-13-BS05-0003-002 (grant MOJO) and NASA Astrobiology Institute's Virtual Planetary Laboratory Lead Team, funded under solicitation NNH12ZDA002C and cooperative agreement no. NNA13AA93A.

%
%

\bibliographystyle{aasjournal}
\bibliography{references}

%
%
\appendix
\section{Disk torques}
\label{appendix:torques}
Super-Earths and mini-Neptunes experience Type-I migration. In Type-I migration, a planet experiences a negative Lindblad torque $\Gamma_{\rm L}$ and a positive co-rotation torque $\Gamma_{\rm C}$. The total torque on the planet is given by

\begin{equation}
	\Gamma_{\rm tot} = \Gamma_{\rm L} \Delta_{\rm L}
    				 + \Gamma_{\rm C} \Delta_{\rm C}
\end{equation}
where $\Delta_{\rm L}$ and $\Delta_{\rm C}$ are corrections of order unity. \citet{Paardekooper_2010,Paardekooper_2011} derived the expressions for $\Gamma_{\rm L}$ and $\Delta_{\rm L}$, while $\Delta_{\rm L}$ and $\Delta_{\rm C}$ were calculated by \citet{Cresswell_2008,Coleman_2014,Fendyke_2014}. The full set of equations was gathered together by \citet{Izidoro_2017} and are reproduced again here for convenience (\textbf{Note:} Their paper has some typos which have been corrected here). Following \citet{Paardekooper_2010,Paardekooper_2011} and \citet{Izidoro_2017}, the formulas below assume a smoothing length of the planet potential of $b = 0.4h$ where $h = H/r \approx 0.05$ is the disk aspect ratio, $r$ is the orbital distance, and $H$ is the pressure scale height. We start with the formulas for the two corrections, $\Delta_{\rm L}$ and $\Delta_{\rm C}$. The correction for the Lindblad torque is given by

\begin{equation}
 \Delta_{\rm L}
 	=
    \left[
    P_{\rm e} + \frac{P_{\rm e}}{|P_{\rm e}|} \times
    \left\lbrace
    	0.07 \left( \frac{i}{h}\right)
        + 0.085\left( \frac{i}{h}\right)^4
        - 0.08\left(  \frac{e}{h} \right) \left( \frac{i}{h} \right)^2
	\right\rbrace
    \right]^{-1},
\end{equation}
where $e$ and $i$ are the planet orbital eccentricity and inclination, and 

\begin{equation}
	P_{\rm e}
    = \frac{
    	1 + \left( \frac{e}{2.25h}\right)^{1.2} + \left( \frac{e}{2.84h}\right)^6
	}{
    	1 - \left( \frac{e}{2.02h}\right)^4
	}.
\end{equation}

The correction for the co-rotation torque is

\begin{equation}
	\Delta_{\rm C}
    =	\exp\left( \frac{e}{e_{\rm f}} \right)
    	\left\lbrace
        	1 - \tanh\left(\frac{i}{h} \right)
        \right\rbrace,
\end{equation}
where $e_{\rm f} = 0.5h + 0.01$ \citep{Fendyke_2014}. The formulas for the torques $\Gamma_{\rm L}$ and $\Gamma_{\rm C}$ are

\begin{eqnarray}
	\Gamma_{\rm L}
    &~=~& (-2.5 -1.7\beta + 0.1x)~\frac{\Gamma_0}{\gamma_{\rm eff}},\\
	\Gamma_{\rm C}
    &~=~&  \Gamma_{\rm c,hs,baro}~F(p_\nu)~G(p_\nu) \\
	&~+~&  \Gamma_{{\rm c,lin,baro}}~(1 - K(p_\nu)   \nonumber\\
    &~+~&  \Gamma_{\rm c,hs,ent}~F(p_\nu)~F(p_\chi)~
    	   \sqrt{G(p_{\rm \nu})~G(p_\chi)} \nonumber  \\
	&~+~&  \Gamma_{\rm c,lin,ent}~
           \sqrt{(1 - K(p_\nu))~(1 - K(p_\chi)},\nonumber\\
	\Gamma_{\rm c,hs,baro}
    &~=~& 	1.1\left( \frac{3}{2}-x\right) \frac{\Gamma_0}{\gamma_{\rm eff}},\\
	\Gamma_{\rm c,lin,baro}
    &~=~& 	0.7\left( \frac{3}{2}-x\right) \frac{\Gamma_0}{\gamma_{\rm eff}},\\
	\Gamma_{\rm c,hs,ent}
    &~=~& 	7.9~\xi~\frac{\Gamma_0}{\gamma_{\rm eff}^2},\\
	\Gamma_{\rm c,lin,ent}
    &~=~& 	\left( 2.2 - \frac{1.4}{\gamma_{\rm eff}}\right)\xi
    		\frac{\Gamma_0}{\gamma_{\rm eff}}.
\end{eqnarray}
where $\xi = \beta - (\gamma -1)x$ is the negative of the entropy slope, $\gamma = 1.4$ is the adiabatic index, $x$ is the negative negative of the surface density profile, and $\beta$ is the temperature gradient,

\begin{equation}
	x = - \frac{\partial {\rm ln} ~\Sigma_{\rm gas}}{\partial {\rm ln}~r},
	~~~
    \beta = - \frac{\partial {\rm ln} ~T}{\partial {\rm ln}~r}.
\end{equation}

We have also used the scaling factor $\Gamma_0 = (q/h)^2 \Sigma_{\rm gas} r^4 \Omega_k^2$, where $q$ is the planet-star mass ratio $h = H/r \approx 0.05$ is, as before, the disk aspect ratio, $\Sigma_{\rm gas}$ is the surface density, and  $\Omega_k$ is the planet's Keplerian orbital frequency. The terms that we haven't defined yet describe thermal and viscous diffusion effects that contribute differently to the different components of the co-rotation torque. We begin with the thermal diffusion coefficient,

\begin{equation}
	\chi = 
    \frac{16~\gamma~(\gamma -1)~\sigma T^4}{3 \kappa~\rho^2~(hr)^2~\Omega_k^2},
\end{equation}
where $\rho$ is the gas volume density, $\kappa$ is the opacity and $\sigma$ is the Stefan-Boltzmann constant. Using $\chi$ we can define the effective adiabatic index $\gamma_{\rm eff}$ as

\begin{eqnarray}
	\gamma_{\rm eff}
    &~=~&
    \frac{2~Q \gamma}{
    	\gamma Q + \frac{1}{2}
        \sqrt{2 \sqrt{( \gamma^2Q^2+1)^2-16 Q^2(\gamma-1)}+2 \gamma^2 Q^2 - 2}
    },\\
	Q
    &~=~& \frac{2 \chi}{3 h^3 r^2 \Omega_k}.
\end{eqnarray}

Next, $p_\nu$ and $p_\chi$ are parameters that govern the viscous saturation and the thermal saturation respectively. They are defined in terms of the non-dimensional half-width of the horseshoe region ${x_{\rm s}}$,

\begin{eqnarray}
	p_\nu
    &~=~& \frac{2}{3}\sqrt{\frac{r^2\Omega_k}{2\pi\nu}x_s^3},\\
	p_\chi
    &~=~& \sqrt{\frac{r^2\Omega_k}{2\pi\chi}x_s^3},\\
	x_{\rm s}
    &~=~& \frac{1.1}{\gamma_{\rm eff}^{1/4}}\sqrt{\frac{q}{h}}.
\end{eqnarray}

Finally, we give the expression for functions $F$, $G$, and $K$,

\begin{eqnarray}
  F(p)
  &~=~& \frac{1}{1+ \left( \frac{p}{1.3}\right)^2},\\
  G(p)
  &~=~&
  \begin{cases}
    \frac{16}{25}
    	\left( \frac{45\pi}{8}\right)^{\frac{3}{4}}
    	p^{\frac{3}{2}},
    	& {\rm if}~~ p < \sqrt{\frac{8}{45\pi}} \\
    1 - \frac{9}{25}
    	\left( \frac{8}{45\pi}\right)^{\frac{4}{3}}
        p^{-\frac{8}{3}},
        & {\rm otherwise}.
  \end{cases},\\
  K(p)
  &~=~&
  \begin{cases}
    \frac{16}{25}
    	\left( \frac{45\pi}{28} \right)^{\frac{3}{4}}
        p^{\frac{3}{2}},
        & {\rm if}~~ p < \sqrt{\frac{28}{45\pi}} \\
    1 - \frac{9}{25}
    	\left( \frac{28}{45\pi} \right)^{\frac{4}{3}}
        p^{-\frac{8}{3}},
        & {\rm otherwise}.
  \end{cases}.
\end{eqnarray}

Now that we have an expression for the total torque $\Gamma_{\rm tot}$, we need to implement it as an additional force term in the N-body code. To do this, we follow \citet{Papaloizou_2000} and \citet{Cresswell_2008} in defining the planet migration timescale as $t_{\rm mig} =- L / \Gamma_{\rm tot}$, where $L$ is the planet's orbital angular momentum. With this definition, the timescale for the planet to reach the star is $t_m/2$. To implement orbital migration we add the force term

\begin{equation}\label{eqn:a_mig}
	\mathbf{a}_{\rm mig} = -\frac{\mathbf{v}}{t_{\rm mig}},
\end{equation}
where $\mathbf{v}$ is the planet's instantaneous velocity. In addition to orbital migration, planet-disk interactions also dampen the orbital eccentricity. To implement these, we also need the eccentricity and inclination damping timescales,  ${\rm t_e}$ and ${\rm t_i}$. Expressions for these timescales have been derived by \citet{Papaloizou_2000} and \citet{Tanaka_2004}, and were later modified by \citet{Cresswell_2006,Cresswell_2008}. The timescales are

\begin{eqnarray}
	t_e
    &~=~&
    \frac{t_{\rm wave}}{0.780}
    \left(
    	1-0.14\left(\frac{e}{h}\right)^2 
        + 0.06\left(\frac{e}{h}\right)^3
        + 0.18\left(\frac{e}{h}\right)\left(\frac{i}{h}\right)^2
    \right),\\
	t_i
    &~=~&
    \frac{t_{\rm wave}}{0.544}
    \left(
    	1-0.3\left(\frac{i}{h}\right)^2 
        + 0.24\left(\frac{i}{h}\right)^3
        + 0.14\left(\frac{e}{h}\right)^2\left(\frac{i}{h}\right)
    \right),\\
	t_{\rm wave}
    &~=~&
    \left(\frac{M_{\odot}}{m}\right)
    \left(\frac{M_{\odot}}{\Sigma_{\rm gas} a^2}\right)
    h^4 \Omega_k^{-1},
\end{eqnarray}
where $M_{\odot}$ is the mass of the Sun, and $m$ and $a$ are the planet's mass and semimajor axis. Similar to Equation \ref{eqn:a_mig}, we implement eccentricity damping and inclination damping, respectively, as the forces

\begin{eqnarray}
	\mathbf{a}_e\label{eqn:a_e}
    &~=~& -2~\frac{(\mathbf{v \cdot r})\mathbf{r}}{r^2 t_e},\\
	\mathbf{a}_i\label{eqn:a_i}
    &~=~& -\frac{v_z}{t_i}\mathbf{k},
\end{eqnarray}
where $\mathbf{r}$ and $\mathbf{v}$ are the position and velocity vectors of the planet, $v_z$ is the $z$ component of the planet's velocity, and $\mathbf{k}$ is the unit vector in the $z$ direction. Equations \ref{eqn:a_mig}, \ref{eqn:a_e}, and \ref{eqn:a_i} are given in \citet{Paardekooper_2010} and \citet{Cresswell_2006,Cresswell_2008}.

\section{Gas accretion}
\label{appendix:gas-accretion}

We adapted the gas accretion model of \citet{Ginzburg_2016}. Once the protoplanet forms, it rapidly accretes an initial atmosphere from the surrounding nebula. This atmosphere extends to either the Bondi radius, $\Rb$, or the Hill radius, $\Rh$ (whichever is smaller).

\begin{eqnarray}
	\Rb &=& \frac{G \Mc \mu}{\kb \Td} \\
    \Rh &=& a \left( \frac{\Mc}{3 \Ms} \right)^{1/3}
\end{eqnarray}
where $\Mc$ is the core mass, $\Td$ is the local disk temperature, $\mu$ is the molecular weight, $a$ is the semimajor axis, and $\kb$ is the Boltzmann constant. A planet is in the Bondi regime whenever 

\begin{equation}
	\Rb < \Rh \; \Rightarrow \; \frac{\Mc}{\Me} < 50
    			\left( \frac{a}{\au} \cdot \frac{T}{10^3\K} \right)^{3/2}.
\end{equation}
For our baseline disk model (section \ref{sec:model:init}) with metallicity $Z = 1\%$, $\Rb < \Rh$ at 0.1 AU whenever the core mass is $\Mc < 25.5\Me$. As the disk cools, the transition mass between the Bondi and Hill radius occurs at lower masses. Near the end of the disk lifetime, at 4 Myr, the transition occurs at $\Mc = 1.5\Me$, with less massive and more distant cores having $\Rb < \Rh$. Therefore, most or nearly all of the accretion occurs in the Bondi regime. If a very massive core forms very late, very close to the star (i.e. a giant impact occurs close to the end of the disk lifetime) the formulas of \citet{Ginzburg_2016} might slightly overestimate the mass of the tenuous accreted envelope.

The initial atmosphere is accreted on a dynamical timescale. Since it has had no time to cool, it follows an adiabatic density profile. This initial atmosphere has a mass of

\begin{equation}\label{eqn:Matm}
	M_{\rm atm,0} \sim \rhod \; \Rb^3
\end{equation}
where $\rhod$ is the local disk density. As in \citet{Ginzburg_2016}, we adopt an adiabatic index of $\gamma = 7/5$ (i.e.\,diatomic gas). As the atmosphere cools, it forms an outer radiative layer while the lower layer remains convective. At the radiative-convective boundary (RCB), the temperature is in equilibrium with the disk, $T_{\rm rcb} \approx \Td$. The radius of the RCB is

\begin{equation}
	R_\rcb = \frac{\Rb}{1 + \ln(\rho_\rcb / \rhod)}
\end{equation}
For $\Matm$ a few percent of $\Mc$,  \citet{Ginzburg_2016} estimate that $\rho_\rcb / \rhod \sim 10^1 - 10^2$, so that $R_\rcb$ is smaller than $\Rb$ a factor of a few. After the atmosphere cools, it contracts, allowing more gas to enter the Bondi radius. \citet{Ginzburg_2016} compute the mass and energy contained in the envelope, as well as the cooling rate. For adiabatic index $\gamma = 7/5$ the values are

\begin{eqnarray}
	\Matm &=& 
        \frac{5\pi^2}{4} \left( \frac{2}{7} \right)^{5/2}\;
        R_\rcb^3 \; \rho_\rcb
    	\left( \frac{\Rb}{R_\rcb} \right)^{1 / (\gamma - 1)} \\
	E &=&
    	\frac{-64}{35\pi}
    	\frac{G\,\Mc\,\Matm}{\Rc}
    	\left( \frac{R_\rcb}{\Rc} \right)^{-(3\gamma-4)/(\gamma-1)} \\
	L &=& \frac{64\pi}{3} \left( \frac{\gamma-1}{\gamma} \right)
    	\frac{\sigma \Td^4}{\kappa \rho_\rcb} \, \Rb
\end{eqnarray}
where $\sigma$ is the Stephan-Boltzmann constant, $\kappa$ is the opacity. Setting $L = - \dot{E}$ we obtain the gas accretion rate,

\begin{equation}\label{eqn:Mdot}
\dot{M}_{\rm atm} =
        \frac{25\pi^4}{6} \left( \frac{2}{7} \right)^{5/2}
		\frac{\sigma \Td^4}{\kappa \Matm} \; 
		\left( \frac{R_\rcb}{\Rb} \right) \frac{\Rb^{9/2} \Rc^{1/2}}{G\,\Mc} \\
\end{equation}

As in \citet{Ginzburg_2016}, we use the approximation $R_\rcb \approx \Rb$. This means that in the late stages of accretion we will overestimate $\dot{M}_{\rm atm}$ by a factor of order unity. The authors use $\Rc \propto \Mc^{1/4}$ to allow for gravitational compression. This is close to the $\Rc \propto \Mc^{1/3.7}$ estimated by \citet{Zeng_2016}. In any case, the weak dependence of $\dot{M}_{\rm atm}$ on $\Rc$ makes the difference between the two power laws insignificant. The authors also adopt $\kappa \approx 0.1 \; {\rm cm}^2 \, {\rm g}^{-1}$ \citep{Allard_2001,Freedman_2008}. Therefore we find

\begin{equation}
	\frac{\dot{M}_{\rm atm}}{\Me/\Myr}
    = 0.045
    \left( \frac{\Matm}{10^{-2}\Me} \right)^{-1}
    \left( \frac{\Mc}{\Me} \right)^{3.625}
    \left( \frac{\Td}{10^3\K} \right)^{-0.5}
\end{equation}

\section{Additional results}
\label{appendix:figures}

\begin{figure*}[th!]
  \centering
  \includegraphics[width=0.45\textwidth]{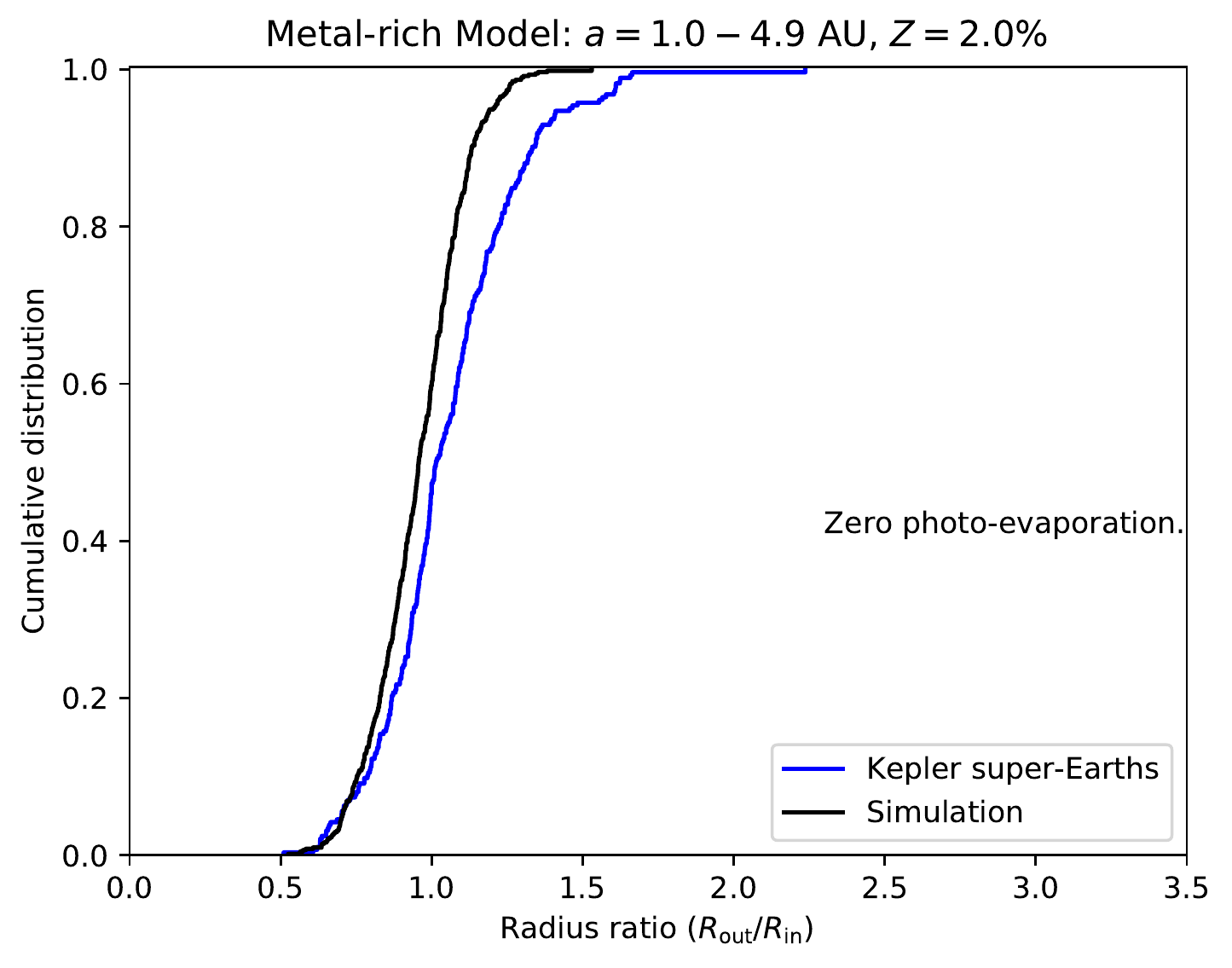}
  \includegraphics[width=0.45\textwidth]{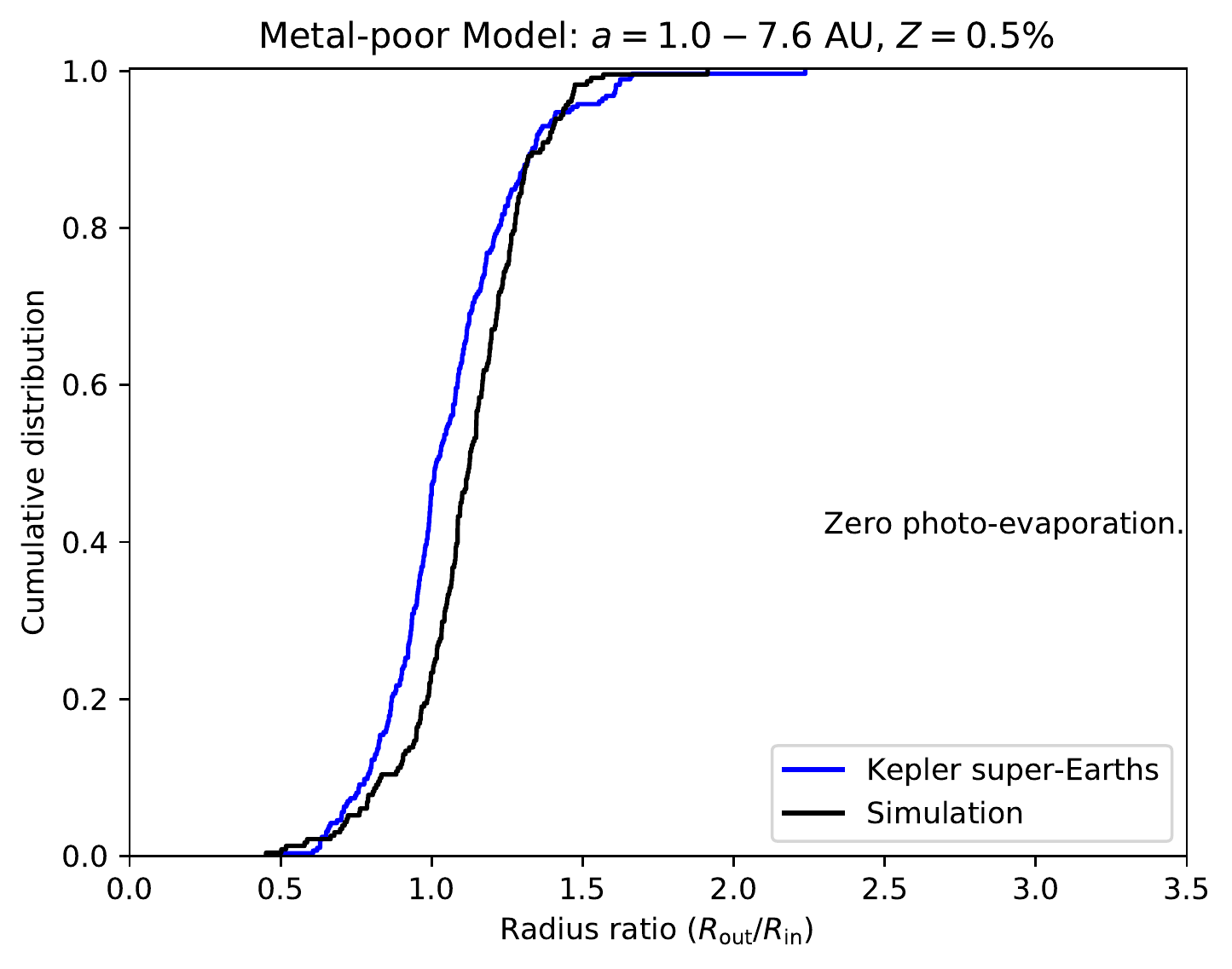}\\
  \includegraphics[width=0.45\textwidth]{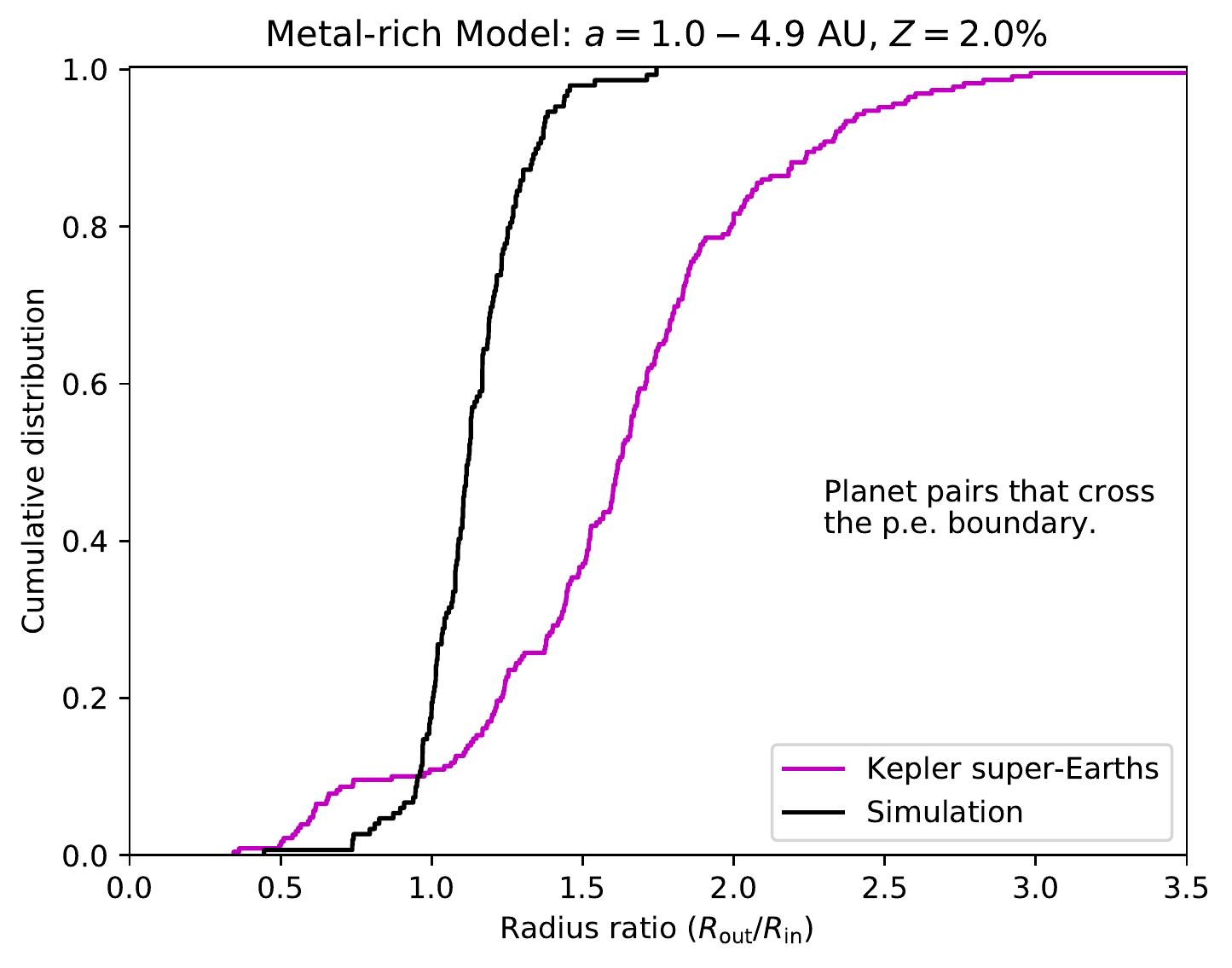}
  \includegraphics[width=0.45\textwidth]{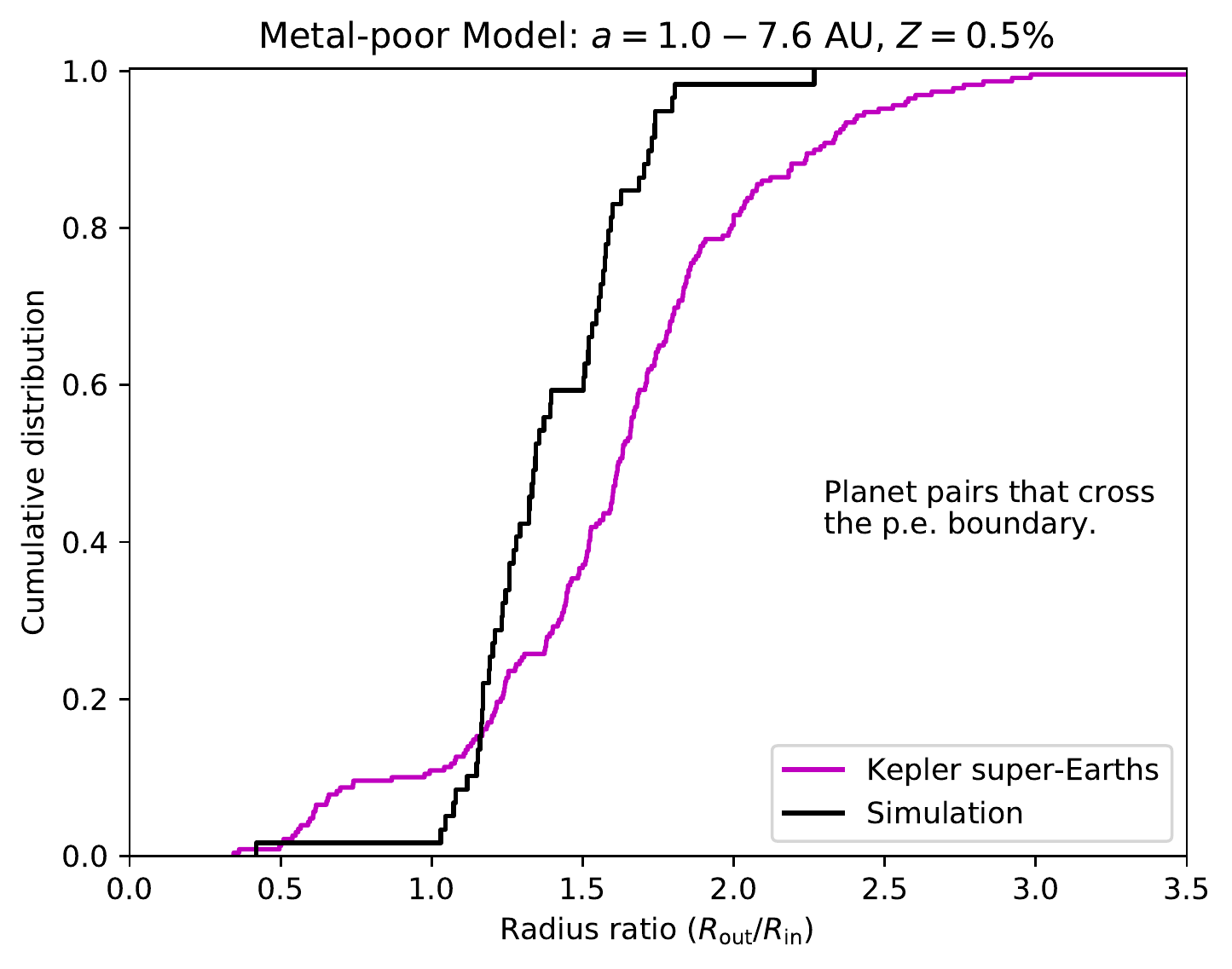}\\
  \includegraphics[width=0.45\textwidth]{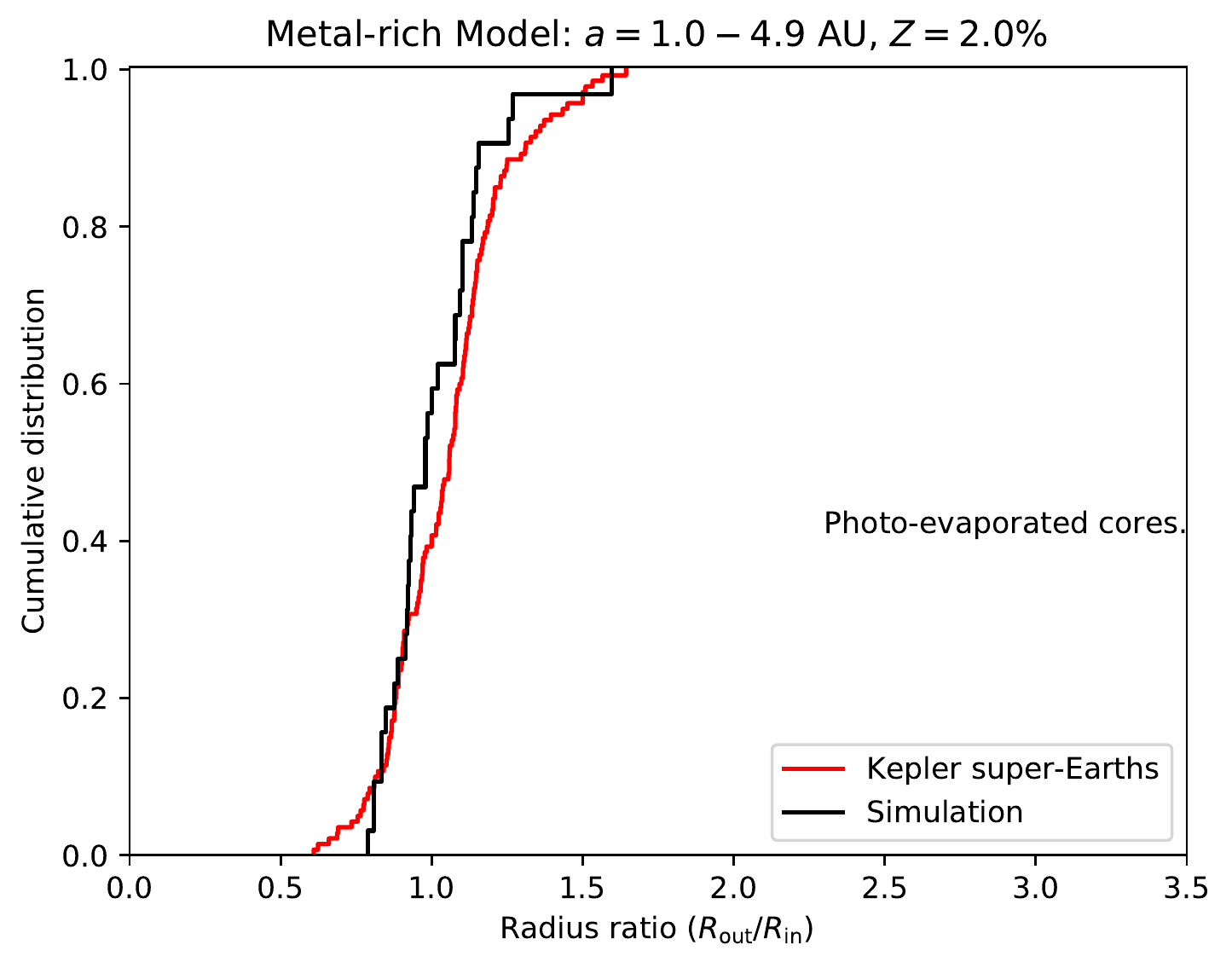}
  \includegraphics[width=0.45\textwidth]{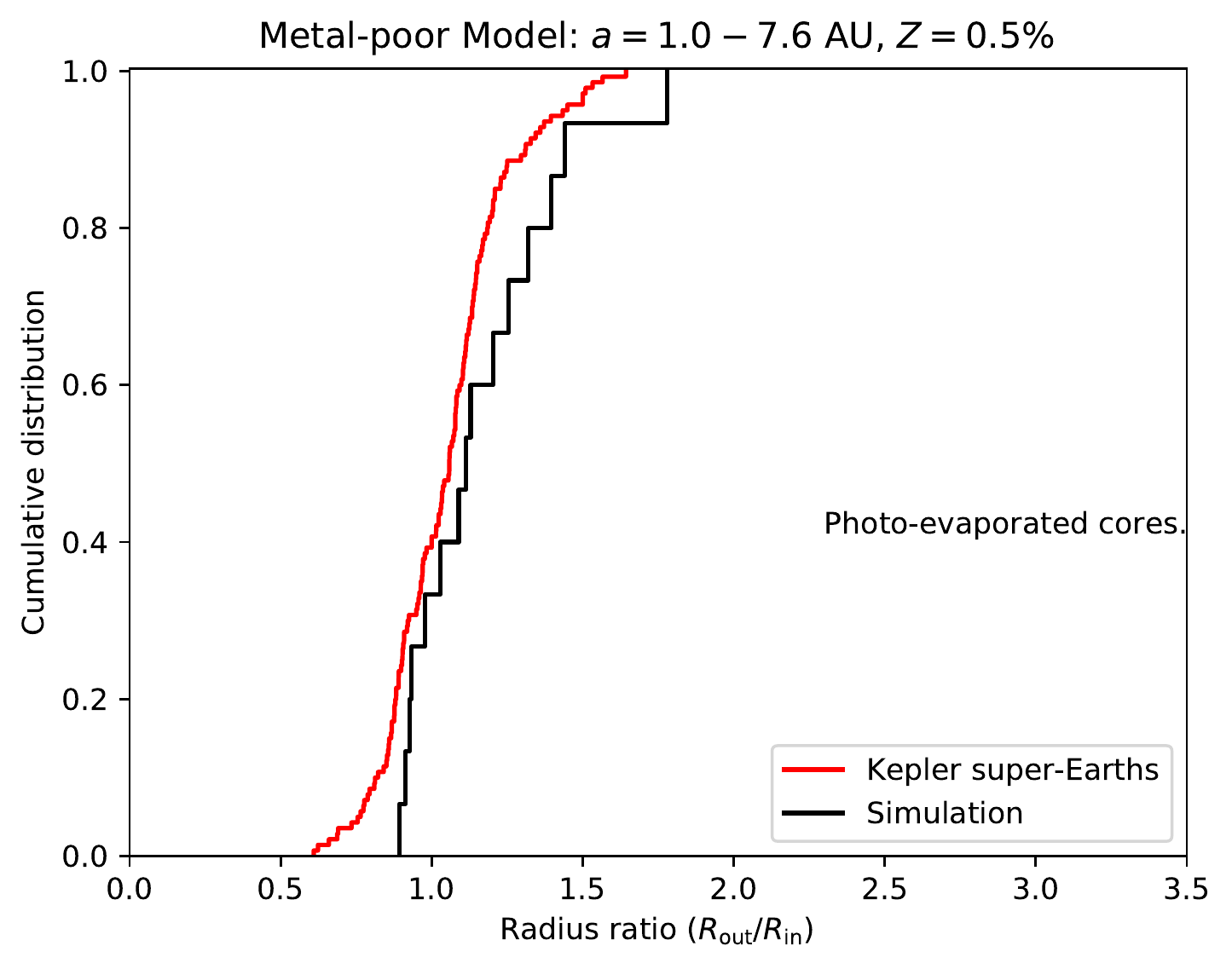}\\
  \caption{Cumulative distribution of the size ratio of neighboring planets ($R_{\rm out} / R_{\rm in}$) in our simulations (black) and the Kepler sample (see Figure \ref{fig:kepler}). The top plot shows the distribution of $R_{\rm out} / R_{\rm in}$ for planet pairs where both planets are above $R_{\rm trans}$. The middle plot shows $R_{\rm out} / R_{\rm in}$ when there is one planet on either side of $R_{\rm trans}$, and the bottom plot is for planet pairs where both planets are below the line. For planets with $R < R_{\rm trans}$ we replace $R$ with the core radius to simulate the effect of photo-evaporation. The left column shows the results for the metal-rich model and the right column shows the results for the metal-poor model.}
  \label{fig:R-ratios-metal}
\end{figure*}

\begin{figure*}[th!]
  \centering
  \includegraphics[width=0.45\textwidth]{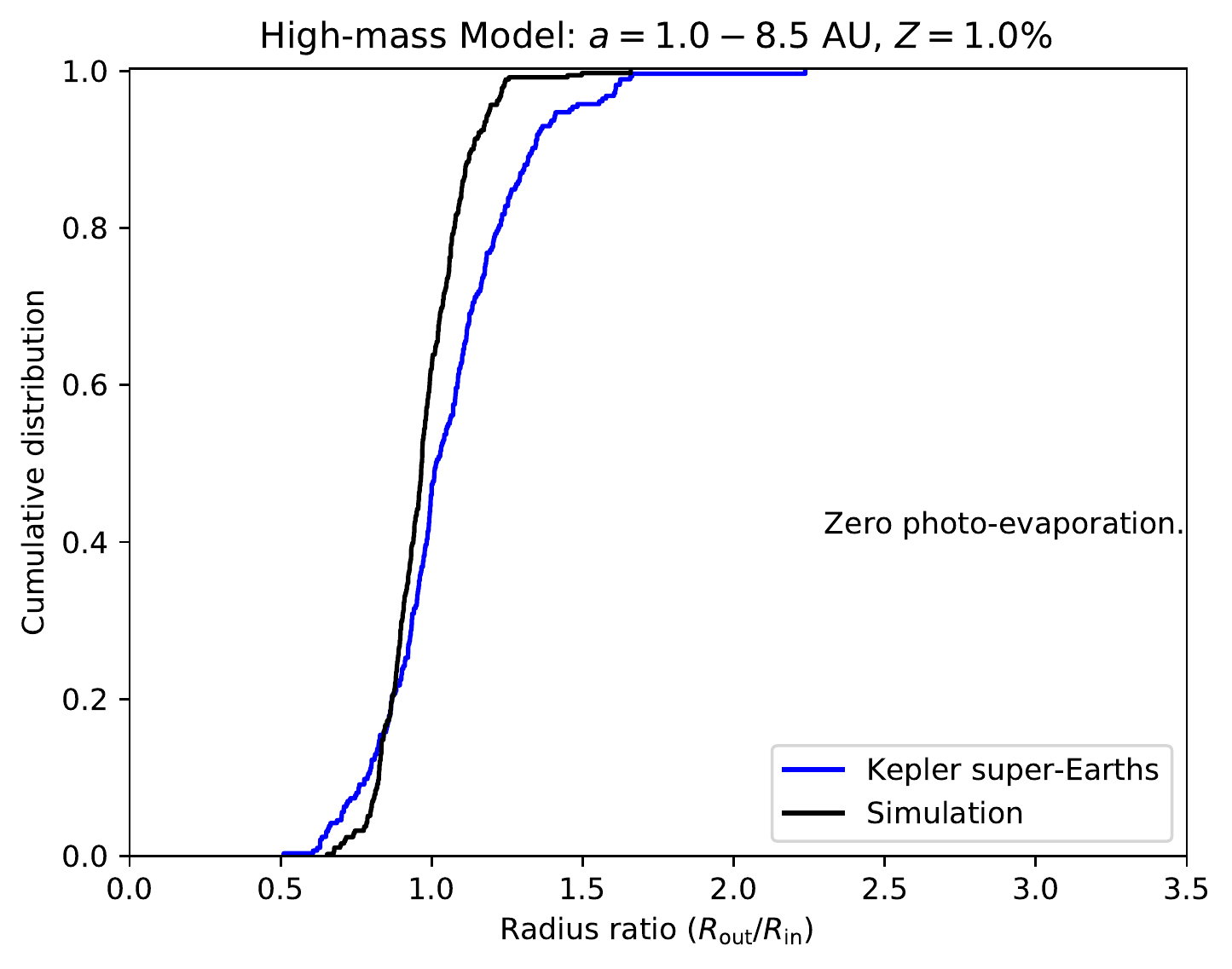}
  \includegraphics[width=0.45\textwidth]{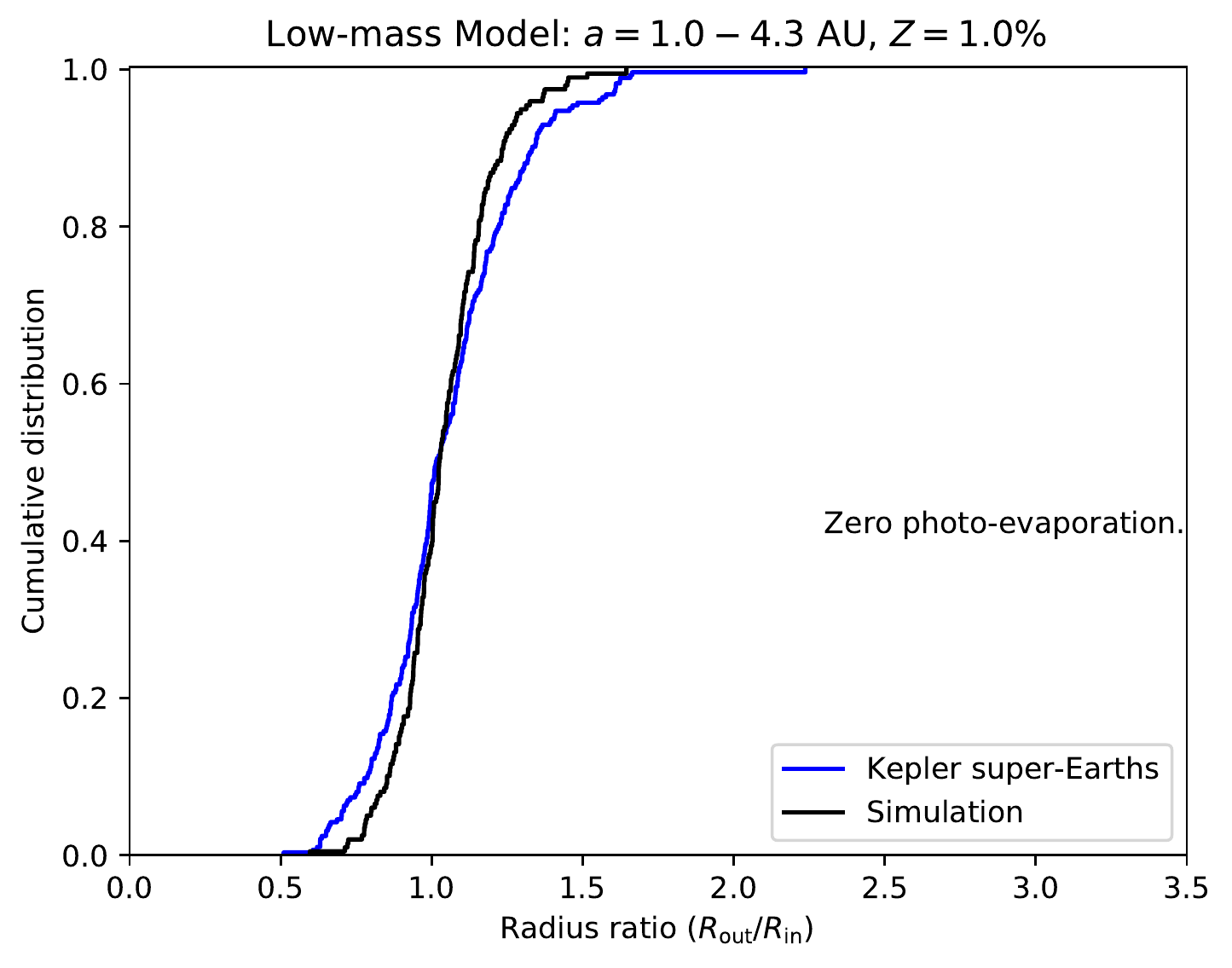}\\
  \includegraphics[width=0.45\textwidth]{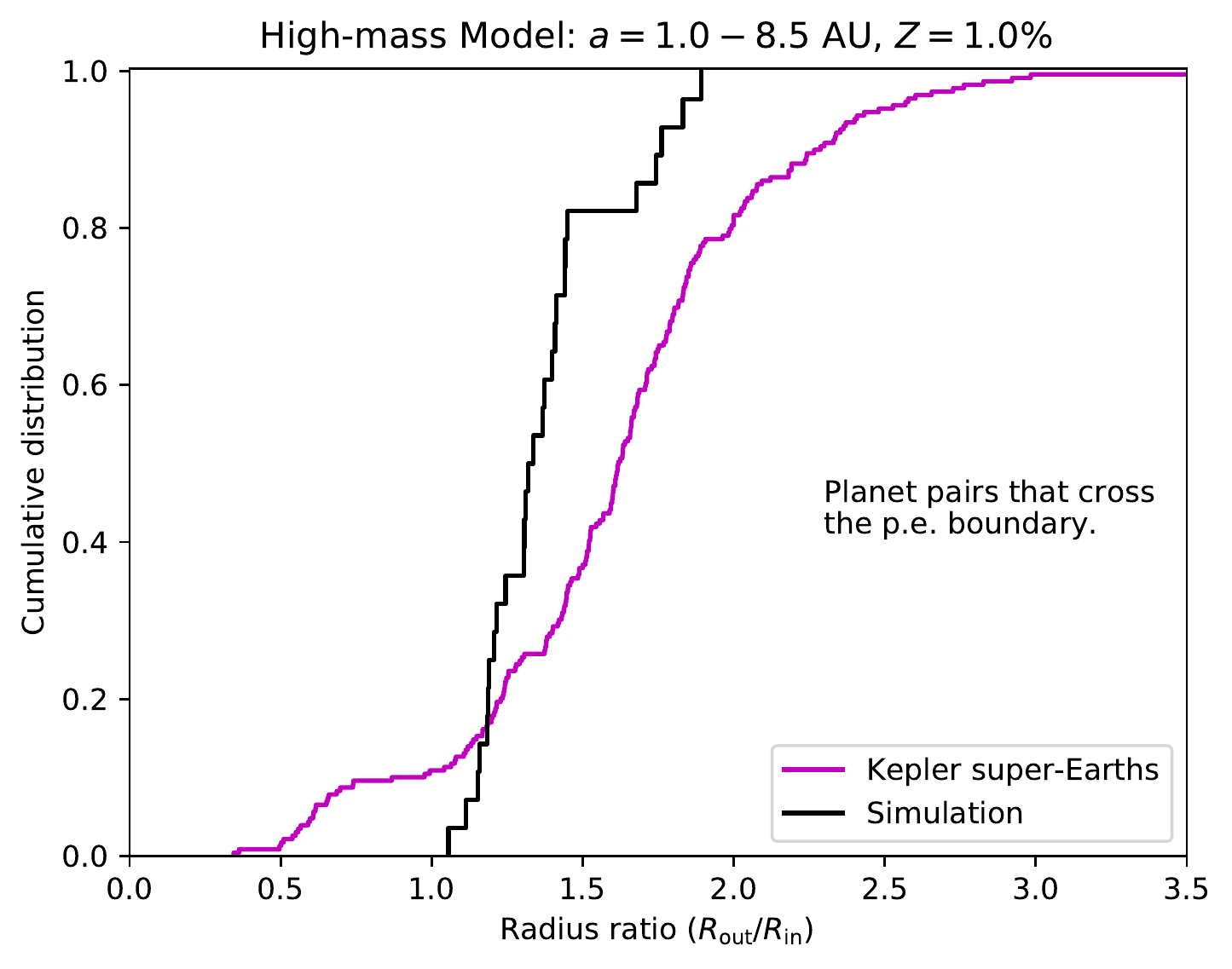}
  \includegraphics[width=0.45\textwidth]{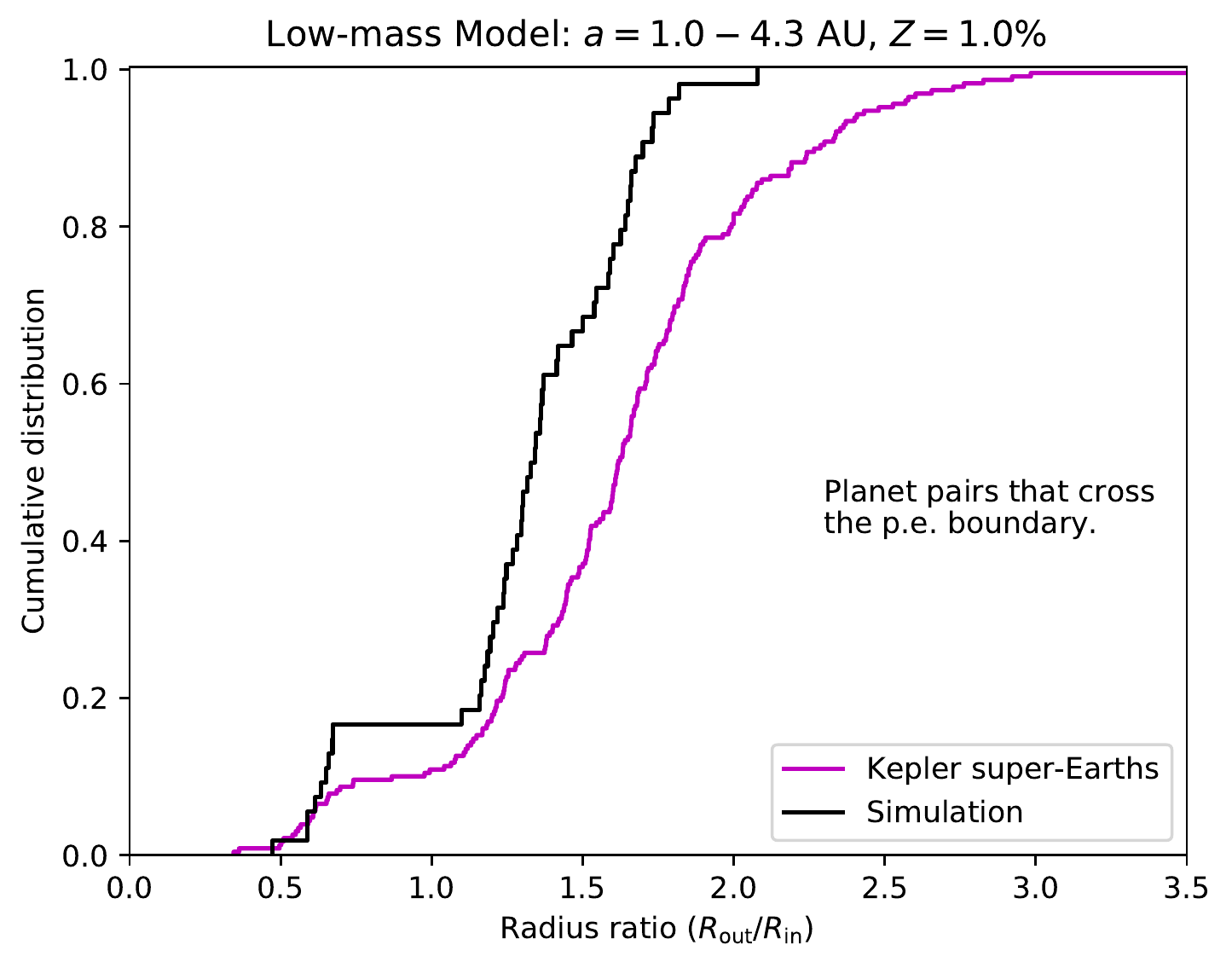}\\
  \includegraphics[width=0.45\textwidth]{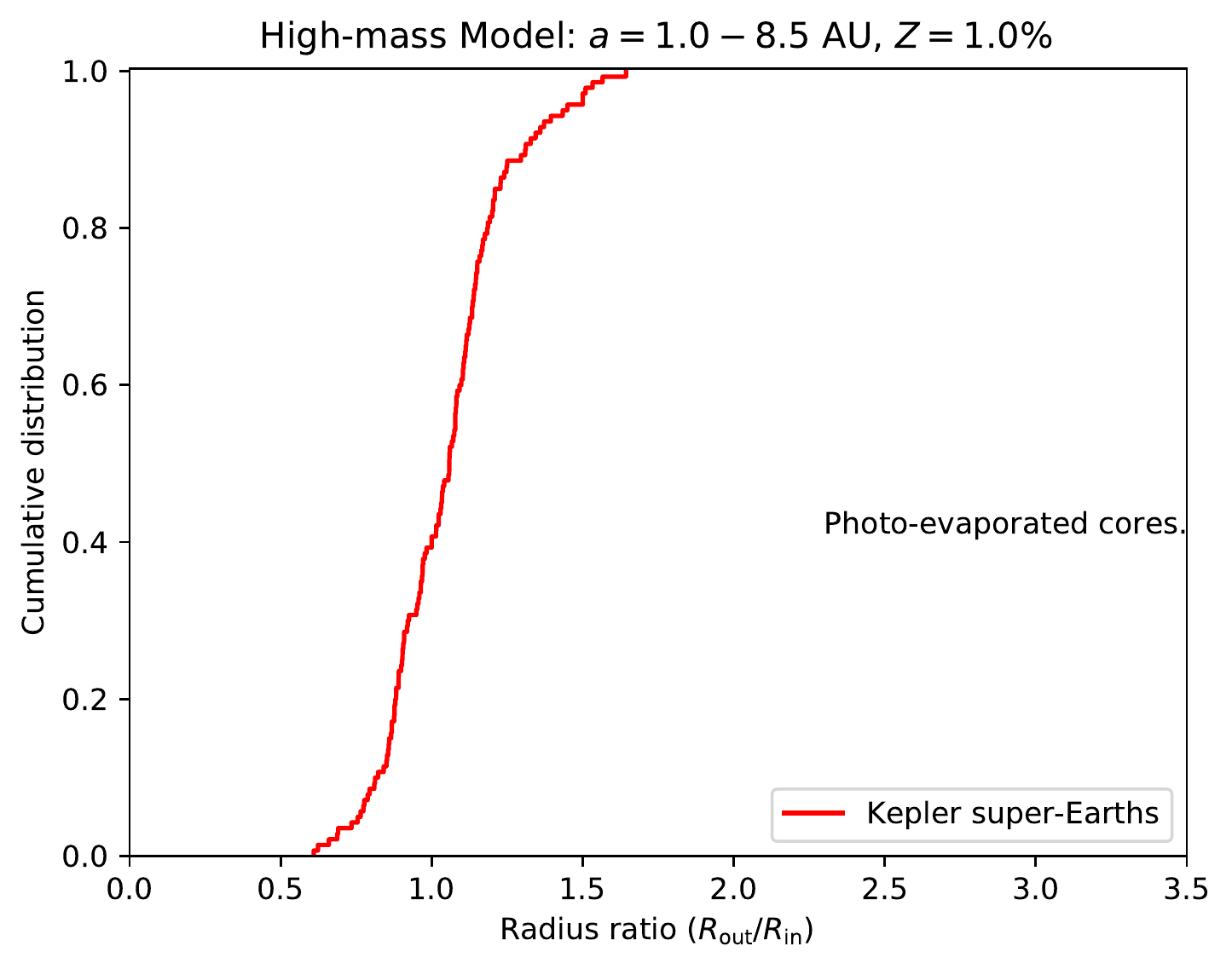}
  \includegraphics[width=0.45\textwidth]{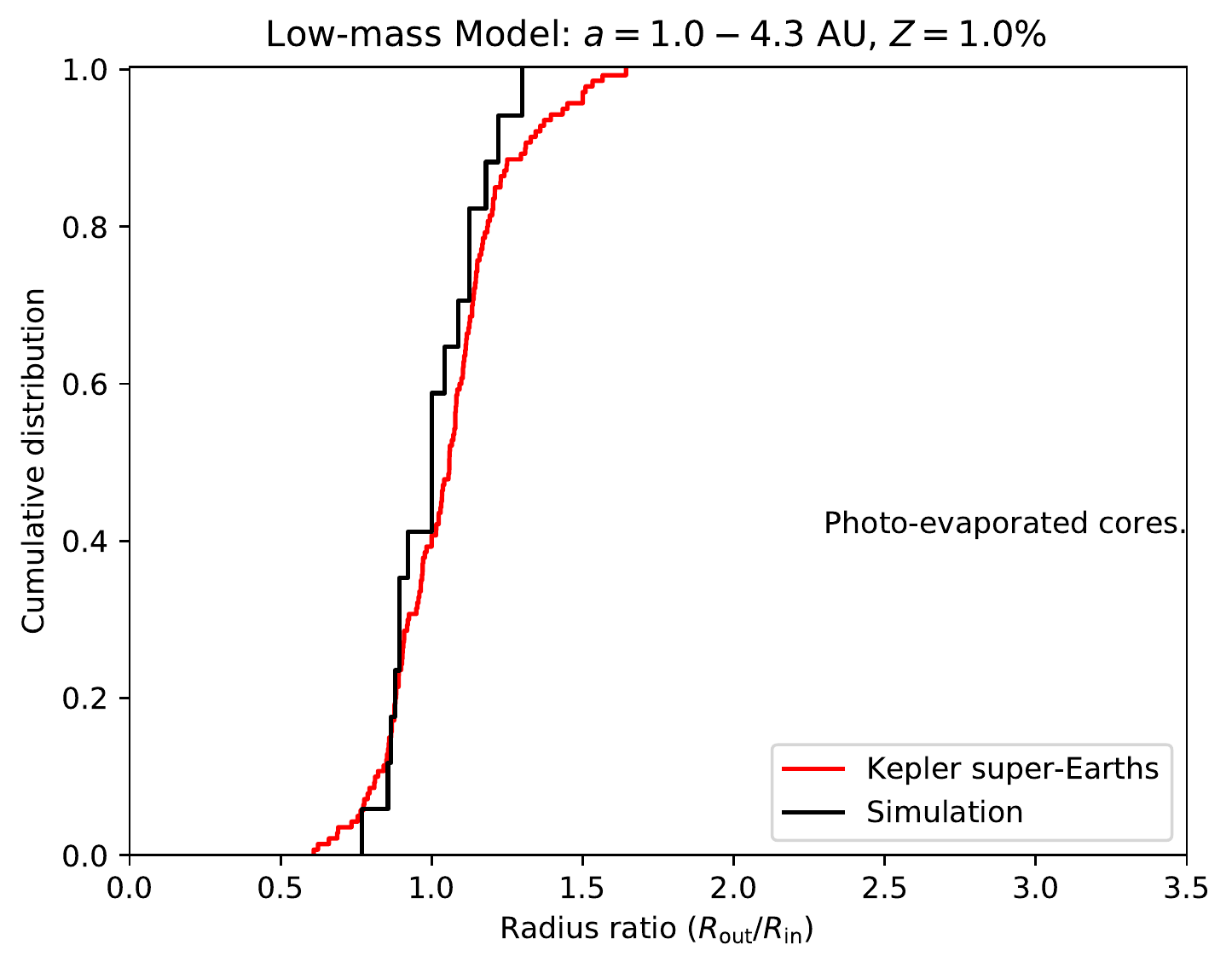}\\
  \caption{Cumulative distribution of the size ratio of neighboring planets ($R_{\rm out} / R_{\rm in}$) in our simulations (black) and the Kepler sample (see Figure \ref{fig:kepler}). The top plot shows the distribution of $R_{\rm out} / R_{\rm in}$ for planet pairs where both planets are above $R_{\rm trans}$. The middle plot shows $R_{\rm out} / R_{\rm in}$ when there is one planet on either side of $R_{\rm trans}$, and the bottom plot is for planet pairs where both planets are below the line. For planets with $R < R_{\rm trans}$ we replace $R$ with the core radius to simulate the effect of photo-evaporation. The left column shows the results for the high-mass model and the right column shows the results for the low-mass model. For the high-mass model, there were no simulations where two planets were photo-evaporated. Hence, the bottom-left plot does not have a model prediction.}
  \label{fig:R-ratios-mass}
\end{figure*}

Here we include the size ratios of the four sets of simulation where we vary the disk metallicity and the total mass in embryos (Table \ref{tab:models}). Figures \ref{fig:R-ratios-metal} and \ref{fig:R-ratios-mass} are complementary to Figure \ref{fig:R-ratios}.

\end{document}